\documentclass[twocolumn,superscriptaddress]{revtex4-1}
\usepackage{amssymb}
\usepackage{bm,amsmath}
\usepackage{graphicx} 
\usepackage{array}
\usepackage{setspace}
\usepackage{xcolor}
\usepackage{hyperref}
\usepackage{gensymb}
\usepackage{soul}
\hypersetup{
	colorlinks,
	linkcolor={red!90!black},
	citecolor={black!10!blue},
	urlcolor={blue!80!black}
}
\newcommand{\f}{\frac}

\newcommand{\pmin}{p_\text{min}}
\newcommand{\pref}{p_0^\text{ref}}

\newcommand{\Dreff}{D_r^\text{eff}}
\newcommand{\nhat}{\hat{\mathbf{n}}_i}
\newcommand{\shat}{\hat{\mathbf{s}}_i}
\renewcommand{\b}{\textcolor{black}}
\newcommand{\tpeff}{\tau_p^\text{eff}}

\begin{document}

	\title{Motility driven glassy dynamics in confluent epithelial monolayers}
	
	\author{Souvik Sadhukhan}
	\email{ssadhukhan@tifrh.res.in}
	\affiliation{Tata Institute of Fundamental Research, 36/P Gopanpally Village, Hyderabad - 500046, India}
	
	\author{Manoj Kumar Nandi} 
	\affiliation{Universit{\'{e}} Claude Bernard Lyon 1, Institut National de la Sant{\'{e}} et de la Recherche M{\'{e}}dicale, Stem Cell and Brain Research Institute, Bron 69500, France.}
	
	\author{Satyam Pandey}
	\affiliation{Tata Institute of Fundamental Research, 36/P Gopanpally Village, Hyderabad - 500046, India}
	
	\author{Matteo Paoluzzi}
	\affiliation{Istituto per le Applicazioni del Calcolo del Consiglio Nazionale delle Ricerche, Via Pietro Castellino 111 80131 Napoli, Italy}
	
	\author{Chandan Dasgupta}
	\affiliation{Department of Physics, Indian Institute of Science, Bangalore 560012, India}
	\affiliation{International Centre for Theoretical Sciences, TIFR, Bangalore 560089, India}

	\author{Nir Gov}
	\affiliation{Department of Chemical and Biological Physics, Weizmann Institute of Science, Rehovot 7610001, Israel}
	
	\author{Saroj Kumar Nandi}
	\email{saroj@tifrh.res.in}
	\affiliation{Tata Institute of Fundamental Research, 36/P Gopanpally Village, Hyderabad - 500046, India}
	
\begin{abstract}
As wounds heal, embryos develop, cancer spreads, or asthma progresses, the cellular monolayer undergoes a glass transition between solid-like jammed and fluid-like flowing states. During some of these processes, the cells undergo an epithelial-to-mesenchymal transition (EMT): they acquire in-plane polarity and become motile. Thus, how motility drives the glassy dynamics in epithelial systems is critical for the EMT process. However, no analytical framework that is indispensable for deeper insights exists. Here, we develop such a theory inspired by a well-known glass theory. One crucial result of this work is that the confluency affects the effective persistence time-scale of active force, described by its rotational diffusivity, $D_r^{\text{eff}}$. $\Dreff$ differs from the bare rotational diffusivity, $D_r$, of the motile force due to cell shape dynamics, which acts to rectify the force dynamics: $D_r^{\text{eff}}$ is equal to $D_r$ when $D_r$ is small and saturates when $D_r$ is large. We test the theoretical prediction of $\Dreff$ and how it affects the relaxation dynamics in our simulations of the active Vertex model. This novel effect of $\Dreff$ is crucial to understanding the new and previously published simulation data of active glassy dynamics in epithelial monolayers. 
\end{abstract} 
	
\maketitle
	
\section{Introduction} 
	
The glassy nature of collective cellular dynamics in epithelial monolayers is crucial for many biological processes, such as embryogenesis  \cite{Tambe2011,Friedl2009b,Kakkada2018,Schotz2013}, wound healing \cite{poujade2007,Das2015,Brugues2014,malinverno2017}, and cancer progression \cite{Streitberger2020,Friedl2003a}. Glass transition refers to the drastic change of dynamics, from a solid-like jammed to a fluid-like flowing state, without much apparent change in the static properties \cite{Berthier2011,Berthier2019c,pareek2023,Atia2021,activereview}. However, there are several dynamical signatures of glassiness, and they also appear in these biological systems. For example, complex non-exponential relaxation \cite{Angelini2011,Park2015}, dynamical heterogeneity \cite{Nnetu2012,Park2015}, caging \cite{Szabo2006}, non-Gaussian particle displacements \cite{Trepat2009a,giavazzi2018,Atia2021}, etc. Nevertheless, the biological systems are distinctive from the non-living glassy systems: the cells can divide and die \cite{jacques2015,ranft2010,silke2017}, differentiate \cite{alvarado2014}, be confluent \cite{Farhadifar2007,Garcia2015}, be motile and self-propel \cite{sriramreview,sriramrmp}, etc. The epithelial monolayers are confluent, where the cells cover the entire space and remain sedentary under healthy conditions \cite{Farhadifar2007,Park2015,Park2016}. On the other hand, during several of these biological processes, the cells undergo an epithelial-to-mesenchymal (EMT) transition, where the cells assume a migratory nature with an internal motility \cite{thiery2002,Bi2016,mitchel2020}. Thus, it is imperative to understand how motility affects the glass transition in confluent epithelial systems.

The effects of motility on the glassy dynamics of particulate systems are relatively well-understood \cite{Janssen2019,Berthier2019c,paoluzzi2022}. These systems comprise self-propelled particles (SPPs) with a self-propulsion force, $f_0$, and a persistence time, $\tau_p$ \cite{sriramreview,sriramrmp,activereview}. $\tau_p$ leads to a separation of time scales between the thermal and active dynamics. When $\tau_p$ is not too large, an equilibrium-like description with an extended fluctuation-dissipation relation applies \cite{Palacci2010,Szamel2014,Han2017,fodor2016,parisi2005}. Due to the inherent complexities of these systems, simulation studies \cite{Berthier2014,Mandal2016,Flenner2016,Ni2013} and experiments on synthetic systems \cite{Klongvessa2019a,Klongvessa2019b,Arora2022} have provided crucial insights. Theories of equilibrium glasses, such as the mode-coupling theory (MCT) \cite{Berthier2013,Szamel2016,Flenner2016,Liluashvili2017,Feng2017,Nandi2017} or random first-order transition (RFOT) theory \cite{Nandi2018,Mandal2022}, have been extended for active SPP systems.
However, the constraint of confluency leads to new control parameters and behaviors for the epithelial systems, and these theories are not directly applicable. With no analytical theory, simulation studies of model confluent systems have played vital roles.

Several computational models exist to study the static and dynamic properties of confluent epithelial monolayers: for example, the cellular Potts model (CPM) \cite{Graner1992,Glazier1993,Hogeweg2000,hirashima2017,Chiang2016,Sadhukhan2021}, the Vertex model \cite{Farhadifar2007,Honda1980,Marder1987,Fletcher2014}, the Voronoi model \cite{Bi2016,Li2021,czajkowski2019}, the phase-field model \cite{nonomura2012,palmieri2015,loewe2020}, etc. These models essentially vary in the details of implementations. Recent works have shown that all these models are similar from the perspective of the glassy dynamics, exhibiting a jamming transition from fluid-like fast to solid-like slow dynamics \cite{bi2014,Bi2016,Sussman2018,Li2021,Sadhukhan2021}. Note that this `jamming transition' is distinct from the zero-temperature zero-activity jamming transition found in the physics of disordered systems \cite{Berthier2019c,Atia2021,sadhukhan2022}; here, the jamming phenomenon is synonymous with glassy dynamics. In confluent systems, the glassy dynamics have several unusual properties.

Compared to most particulate systems, confluent systems readily show sub-Arrhenius relaxation dynamics \cite{Sussman2018,Li2021,Sadhukhan2021}, and cellular shape is a crucial control parameter for these systems \cite{Bi2016,sadhukhan2022}. Moreover, the same system also shows super-Arrhenius behavior with changing control parameters. In a recent work, some of us have phenomenologically extended one of the most popular theories of glassy dynamics, the random first-order transition (RFOT) theory \cite{lubchenko2007,kirkpatrick2015,Biroli2012}, for a confluent monolayer of cells (without self-propulsion) \cite{Sadhukhan2021}. The theoretical predictions agree well with the simulation results of the CPM in the absence of motility. However, as discussed above, EMT plays a crucial role in several biological processes where the cells assume in-plane polarity and motile character. 

Simulation studies show nontrivial effects of $f_0$ and $\tau_p$ on the glassy dynamics in confluent systems \cite{Bi2016,mitchel2020,henkes2020,li2024}. However, how motility couples with confluency and what is its precise role in controlling the glassy dynamics remain unclear. In this work, we have combined active Vertex model simulations and RFOT-inspired analytical theory and discovered a novel effect of motility in confluent systems and how it affects the glassy dynamics. The main results of the work are as follows: (1) We show that confluency has a nontrivial effect on the self-propulsion; specifically, it modifies the effective persistence time of the active forces, such that they differ from the intrinsic $\tau_p$. This effect leads to the modified rotational diffusivity $D_r^{\text{eff}}$ arising from the coupling between shape and active force relaxation. (2) We extend the RFOT theory of particulate glasses for confluent systems of motile cells and show that the theory agrees well with new and previously published simulation results of active confluent systems. (3) We demonstrate that the novel effect of confluency on the bare rotational diffusivity, $D_r$, of active forces is crucial to understanding the effects of motility on the relaxation dynamics of confluent systems.
Our theory provides a simple analytical framework to analyse the active glassy dynamics in confluent systems.

\section{Models for confluent cell monolayer}
\label{model}
To set the notations, we first briefly describe the confluent models representing an epithelial monolayer. Experiments show that the height of a monolayer remains nearly constant \cite{Farhadifar2007}. Thus, a two-dimensional description, representing cells as polygons, is possible \cite{Marder1987,Weaire2001,albert2016}. The computational models have two parts: the energy function and the way in which cells are defined in the model. The energy function, $\mathcal{H}$, describing a cellular monolayer is 
\begin{equation}\label{hamiltonian}
	\mathcal{H} = \sum_{i=1}^{N} \Big[ \Lambda_A (A_i - A_0)^2 + \Lambda_P (P_i - P_0)^2 \Big ],
\end{equation} 
where $N$ is the total number of cells in the monolayer, $A_0$ and $P_0$ are the target area and target perimeter, respectively. $A_i$ and $P_i$ are the instantaneous area and perimeter of the $i$th cell. $\Lambda_A$ and $\Lambda_P$ are area and perimeter moduli; they determine the strength with which the area and perimeter constraints in Eq.~(\ref{hamiltonian}) are satisfied. The physical motivation behind the energy function, $\mathcal{H}$, is the following: We can treat the cell cytoplasm as an incompressible fluid \cite{jacques2015}; therefore, the total cell volume is constant. Since the cellular heights in the monolayer remain nearly the same, cells want to have a preferred area, $A_0$. $A_0$ can vary for different cells, but we have assumed it to be uniform for simplicity. On the other hand, $P_0$ is a coarse-grained variable containing several effects. For most practical purposes, a thin layer of cytoplasm known as the cellular cortex determines the mechanical properties of a cell. Moreover, various junctional proteins, such as E-Cadherin, $\alpha$-Catenin, $\beta$-Catenin, tight-junction proteins, etc, determine cell-cell adhesion \cite{Farhadifar2007,Barton2017,jacques2015}. The properties of all these different proteins and the effect of the cell cortex lead to the perimeter term with a preferred perimeter, which is a balance of cortical contractility and cellular adhesion.

Multiplying the Hamiltonian with a constant does not affect any system property. Therefore, we can rescale length by $\sqrt{A_0}$, and write Eq.~(\ref{hamiltonian}) as
\begin{equation}\label{energyfunction}
	\mathcal{H}=\sum_{i=1}^N\bigg[ \lambda_A(a_i-1)^2+\lambda_P(p_i-p_0)^2\bigg],
\end{equation}
where $\lambda_A=\Lambda_A$, $a_i=A_i/A_0$, $\lambda_P=\Lambda_P/A_0$, $p_i=P_i/\sqrt{A_0}$, and $p_0=P_0/\sqrt{A_0}$. $A_0$ is the average area when we consider poly-disperse systems. One can also have a temperature $T$ that represents various active processes as well as the equilibrium $T$ \cite{sadhukhan2022} (see Appendix \ref{simulationdetails} for the details of implementation). $p_0$ and $T$ are the main control parameters of the system. In this work, we have used $T=0$ and study the athermal self-propelled system.

Using the energy function $\mathcal{H}$, we can calculate the force on a cell, $\mathbf{F}_i=-\nabla_i\mathcal{H}$ (see Appendix \ref{simulationdetails} for more details of the implementation). For self-propulsion, we assign a polarity vector, $\nhat=(\cos\theta_i,\sin\theta_i)$, where $\theta_i$ is the angle with the $x$-axis. We obtain the active force, $\mathbf{f}_a=f_0\nhat=\mu v_0\nhat$. We have set the friction coefficient $\mu$ to unity. $\theta_i$ performs rotational diffusion and is governed by the equation \cite{Bi2016}, 
\begin{equation}\label{thetaeq}
	\partial_t \theta_i(t) =\sqrt{2D_r} \eta_i(t)
\end{equation}
where $\eta_i$ is a Gaussian white noise, with zero mean and a correlation $\langle \eta_i(t)\eta_j(t^\prime)\rangle = \delta(t-t^\prime)\delta_{ij}$ and we have set the rotational friction coefficient to unity. $D_r$ is the rotational diffusion coefficient that is related to a persistence time, $\tau_p=1/D_r$, for the motility director of the cells.

Theoretical studies of the confluent cellular monolayers comprise analyzing the evolution of the system, with or without self-propulsion via various confluent models, either at zero or non-zero $T$. Some of the models are lattice-based discrete models, such as the Cellular Potts Model (CPM) \cite{Glazier1993,Graner1992,Hogeweg2000}, others are continuum models, such as the Vertex \cite{Honda1980,Farhadifar2007,Fletcher2014,Barton2017} or the Voronoi model \cite{yang2017,Bi2016,Li2021}, and then there are models which have some elements of both, for example, the phase field models \cite{nonomura2012,palmieri2015}. One cellular process, the $T1$ transition, is crucial for dynamics in these systems. In a $T1$ transition, two neighboring cells move away, and other two cells become neighbors. It is also known as the ``neighbor-switching" process \cite{Farhadifar2007,Fletcher2014,bi2014}. In this work, we have simulated the Vertex model, the Voronoi model, and the CPM to test the theoretical predictions. \b{For the Vertex and Voronoi model, $\sqrt{A_0}$ is the unit of length and $1/(\mu\lambda_AA_0)$ gives the unit of time.}

{\it The Vertex model:} Within this model, the vertices of the polygons representing the cells are the degrees of freedom \cite{Farhadifar2007,Fletcher2014}. The cell perimeter connecting adjacent vertices is a line, either straight or with a constant curvature. One can simulate the model via either molecular dynamics (MD) \cite{Farhadifar2007,Fletcher2014,Barton2017} or Monte Carlo (MC) \cite{wolff2019}; we have used MD in this work (see Appendix \ref{simulationdetails} for details). Within the Vertex model, $T1$ transition is implemented externally: when an edge length becomes smaller than a predefined cutoff value, $l_c$, a T1 transition is performed \cite{Fletcher2014,bi2014}. After a successful T1 transition, we increase the new length to $\lambda_\text{T1}$ times $l_c$ such that the same edge does not undergo immediate T1 transition. We have used $\lambda_\text{T1}=2.0$.
Unless otherwise mentioned, for the results presented in this work, we have used a binary system, $A_{0\alpha} = 1.2$ and $A_{0\beta} = 0.8$, with a $50:50$ ratio such that the average area remains $1.0$, $l_c = 0.04$. We start with a random initial configuration of cells and equilibrate the system for a time $3\times 10^5$ before collecting data. \b{We have used $N=256$ cells for all the simulations, except for the data presented in Fig. \ref{elongation_time}, where we have used $N=100$ cells such that the simulation time is small and we obtain better averaging over time origins and initial conditions.} We averaged each curve for $100$ time origins and $32$ ensembles.   
	
	{\it Voronoi model:} In the Voronoi model, the degrees of freedom are the Voronoi centers of the polygons \cite{Bi2016,yang2017,Sussman2018}. The tessellation of these Voronoi centers gives the polygons representing the cells. The T1 transitions naturally occur in the Voronoi model. We can use both MD and MC for the dynamics of the model; we have used the MC algorithm in this work \cite{paoluzzi2021}. The cell centers are moved stochastically by an amount $dr$ via the MC algorithm using Eq.~(\ref{energyfunction}). We perform the Voronoi tessellation at each time to construct the cells and calculate the updated area and perimeter. These updated values give the energy for the next step. Unless otherwise specified, we have used $N = 100$ cells, $dr = 0.25$, and a random initial configuration. We equilibrate the system for a time of $10^5$ before collecting data. We averaged each curve over $50$ time origins and $32$ ensembles. 
	
\b{	{\it The CPM:} The CPM is a lattice-based model for confluent epithelial monolayers \cite{Glazier1993,Graner1992,Sadhukhan2021,Chiang2016}. This model represents the system by a $L\times L$ square lattice. Each lattice site has an integer Potts variable, $\sigma \in [1,N]$, where $N$ is the total number of cells. $\sigma = 0$ is reserved for the extracellular medium. A collection of sites with the same Potts variable designates a cell. Dynamics proceeds via the interchange of lattice sites between neighboring cells using the Monte-Carlo (MC) algorithm with the energy function, Eq. \ref{hamiltonian}. We have included activity following Ref. \cite{Chiang2016}. The lattice side gives the unit of length and $L\times L$ MC attempts define unit of time. In our simulations, we have kept $A_0=100$, $P_0=40$, $\Lambda_A=1$, and $\Lambda_P=0.5$. For convenience, we have used $T=0$ in our simulations (Fig. \ref{elongation_time}d).
}

	\section{Results}
	
	\subsection{Effects of confluency on rotational diffusivity of self-propulsion}
	\label{effectiveDrconfluency}

\begin{figure}
		\includegraphics[width=8.6cm]{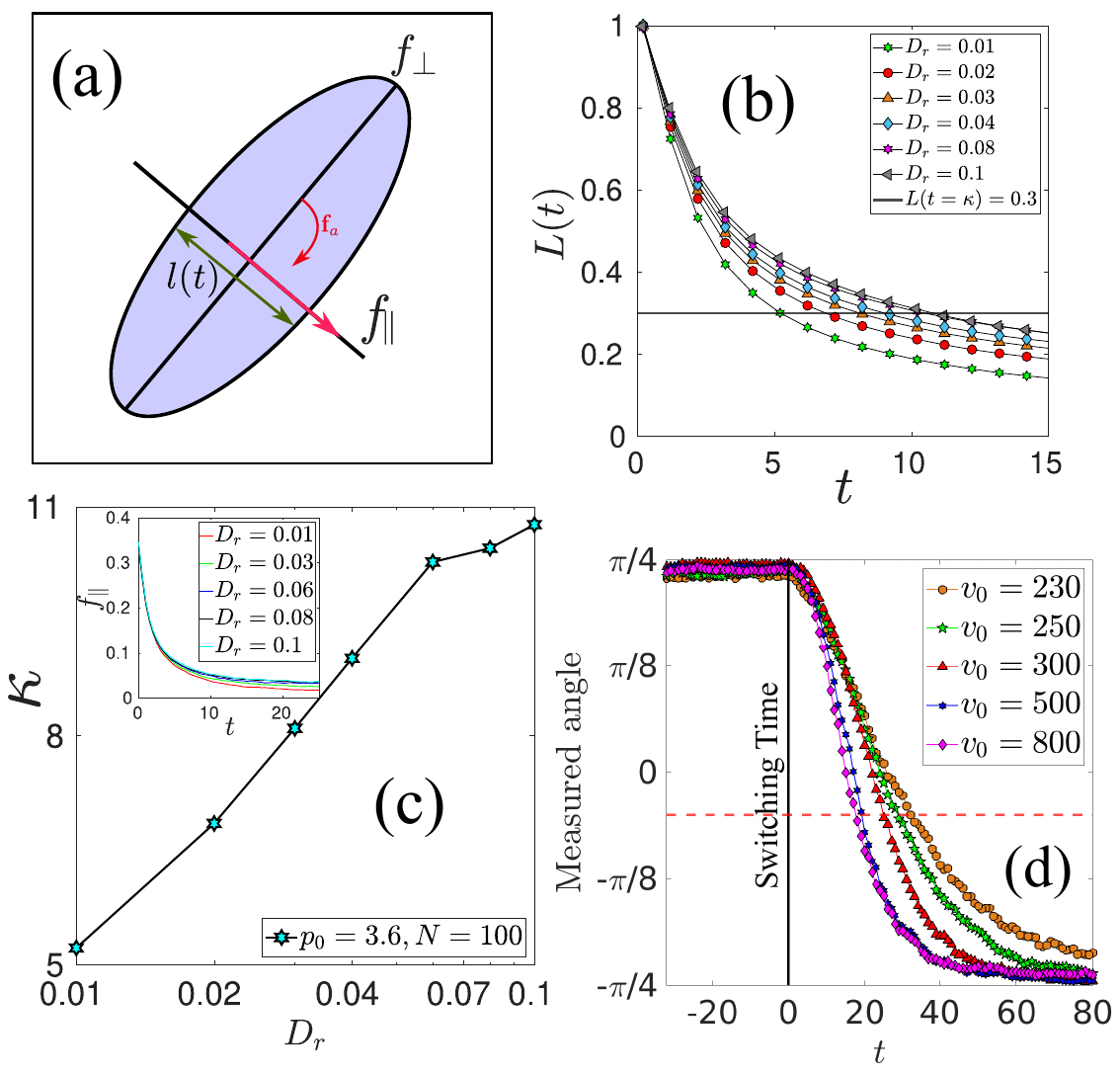}
		\caption{Demonstration of the cellular elongation time, $\kappa$. (a) For a particular cell, we freeze the direction of active force, $\mathbf{f}_a$, and make its magnitude $16$ times higher for better statistics. The cell gets elongated along that direction. We then change the direction $\mathbf{f}_a$ by $90 \degree$ and measure the cellular length, $l(t)$, along that direction and the magnitude of forces, $f_{\parallel}(t)$ and $f_\perp(t)$, as denoted in the figure. \b{(b) $L(t)$ for various values of $D_r$. We obtain the time scale $\kappa$ when $L(t)$ reaches 0.3 (indicated by the line). Note that $L(t)$ goes to zero at long-times by definition, however, the decay of $L(t)$ is slower than exponential. We have highlighted the regime from where we have obtained $\kappa$. (c) $\kappa$ extracted from $L(t=\kappa)=0.3$ as a function of $D_r$. {\bf Inset:} The force, $f_{\parallel}(t)$, along the direction of $\mathbf{f}_a$ after the switch of direction decays to the steady state value on a similar time-scale as $\kappa$. For these measurements, we simulated $100$ cells for better averaging, used $v_0=0.5$ and $p_0=3.6$, equilibrated the system for $3 \times 10^5$ {(time is measured in natural units)} before collecting data, and performed $32$ ensembles and $1000$ initial time averaging. (d) The angle of the major axis with the x-axis as a function of MC time in CPM. We equilibrate the system and then change the motility direction of the active cell at $t = 0$ from $\pi/4$ to $-\pi/4$ (vertical line). The cell takes some time to align its major axis along the motility direction. The dashed line shows the $70\%$ of the total change of the angle. Parameters used for this simulation: $\lambda_A$ = 1, $A_0 = 100$, $P_0$ = 40.0, $\lambda_P = 0.5$, $T = 0$.}}
		\label{elongation_time}
	\end{figure}

	In confluent systems, the cells are extended objects with no gap between neighboring cells. In these systems, cells can move only through shape changes of their boundaries. \b{For the concreteness of our arguments, we first consider the Vertex model, although the discussion below holds in general (Fig. \ref{elongation_time}(d) shows results for the CPM)}. Within the Vertex model, cells move via the movement of the vertices. \b{Persistent cellular movement will elongate the cell in that direction as it leads to lesser frictional force}. Cellular elongation takes time and depends on the properties of the entire system since the movement of one cell must coordinate with that of other cells to keep the system confluent. We show below that this leads to an effective rotational diffusivity, $\Dreff$, of the self-propulsion. Active particles are associated with a director along which the active propulsion force, $\mathbf{f}_a$, acts. In confluent systems, when the internal propulsion force $\mathbf{f}_a$ changes its direction, the forces exerted on the neighboring cells evolve as the cell deforms and elongates along the new force direction, which takes time. We now demonstrate the role of this elongation time, $\kappa$, and show that it leads to a $\Dreff$, different from the bare $D_r$.
	
	We simulate a system for a long time to reach a steady state. We then freeze the direction of $\mathbf{f}_a$ for a chosen cell and again reach a steady state. This particular cell will elongate along the motile force direction. We then suddenly change the direction of $\mathbf{f}_a$ of this cell. The cell then starts elongating along the new motile force direction (see Supplementary Movie). Immediately after switching the motile force direction, we measure the cell length, $l(t)$, as a function of time $t$, along the direction of $\mathbf{f}_a$, and going through the center of mass of the cell (Fig. \ref{elongation_time}a). We also make the motile force magnitude of the chosen cell 16 times higher than the rest of the cells for better statistics. The qualitative results do not depend on the higher magnitude of the motile force of this cell. \b{$l(t)$ starts from a low value and then saturates. The saturation value, $l_s$, depends on the system parameters. To obtain a measure of the elongation time-scale, $\kappa$, we define $L(t)=[l_s-l(t)]/[l_s-l(0)]$: $L(t)$ starts at one and goes to 0 at long times. Therefore, we can define $L(t=\kappa)=0.3$. Figure \ref{elongation_time}(b) shows the behavior of $L(t)$ for various $D_r$. $\kappa$ increases as $D_r$ increases when the system becomes more glassy (Fig. \ref{elongation_time}c). When we suddenly switch the direction of $\mathbf{f}_a$, the other cells require time to respond, and the force along that direction will be higher. This large force will slowly relax with time as the system reaches the new steady state, other cells move, and the particular cell elongates. The inset of Fig. \ref{elongation_time}(c) shows that the average force parallel to $\mathbf{f}_a$ on this cell relaxes towards the steady-state value on a similar time scale as $\kappa$. This elongation time leads to a modified rotational diffusivity.
}

\b{	To test whether the result of the modified rotational diffusivity is a generic feature of confluent systems, we have also simulated the CPM, making one of the cells active and treating the rest of the system as passive. We first equilibrate the system with the direction of $\mathbf{f}_a$ along $\pi/4$ and then switch the direction along $-\pi/4$. We show the angle of the major axis of the cell as a function of time in Fig. \ref{elongation_time}(d). We can again define $\kappa$ when the major axis angle relaxes by $70\%$ of its total change. The cell takes some time to elongate along the direction of $\mathbf{f}_a$. Moreover, similar to the simulations in the Vertex model, we see that the relaxation becomes faster, and therefore $\kappa$ decreases, with increasing self-propulsion speed $v_0$. Thus, we believe that the change in rotational diffusivity is a generic feature of confluent systems and not limited to Vertex models.
}
	
\b{Confluent systems relax via T1 transitions. Therefore, the time scale of T1 transitions must set the cellular elongation time. This is consistent with the result that $\kappa$ depends on $D_r$ that changes the time scale of T1 transitions (Fig. \ref{elongation_time}c).
Let us consider a scenario where we change the direction of the motile force for a particular cell every $\tau_p$ time duration. When $\tau_p$ is much larger compared to $\kappa$, the cell will take time $\kappa$ to elongate along the direction of the active force. Albeit this delay, the effective persistence time $\tpeff$ will be given by $\tau_p$. Conversely, when $\tau_p$ is small compared to $\kappa$, the cell cannot follow the rapid change of motile force direction. Nevertheless, the cellular shape and major axis will change on the time-scale of cellular reorganization (driven by T1 processes). Therefore, we can phenomenologically write}
	\begin{equation}
		\tau_p^\text{eff}=\tau_p+\kappa,
	\end{equation}
	\b{where $\tpeff$ is dominated by the larger of the two time-scales.} Writing the time-scales in terms of diffusivities, $\tau_p^\text{eff}=1/\Dreff$ and $\tau_p=1/D_r$, we obtain
	\begin{equation}\label{effectiveDr}
		\Dreff=\frac{D_r}{\kappa D_r+1}.
	\end{equation}
	Since $\Dreff$ is the effective diffusivity of the active forces (and cell elongations) that give the structural relaxation, in the formalism of extended RFOT theory for self-propulsion \cite{Nandi2018}, Eq.~(\ref{rfotgeneral}) below, we must use $D_r^\text{eff}$ instead of $D_r$ for a confluent system. As we show below, this modification of $D_r$ due to the constraint of confluency and finite cell elongation time is crucial for the dynamics of confluent systems in the presence of self-propulsion.
	
			\begin{figure*}
		\centering
		\includegraphics[width=16.6cm]{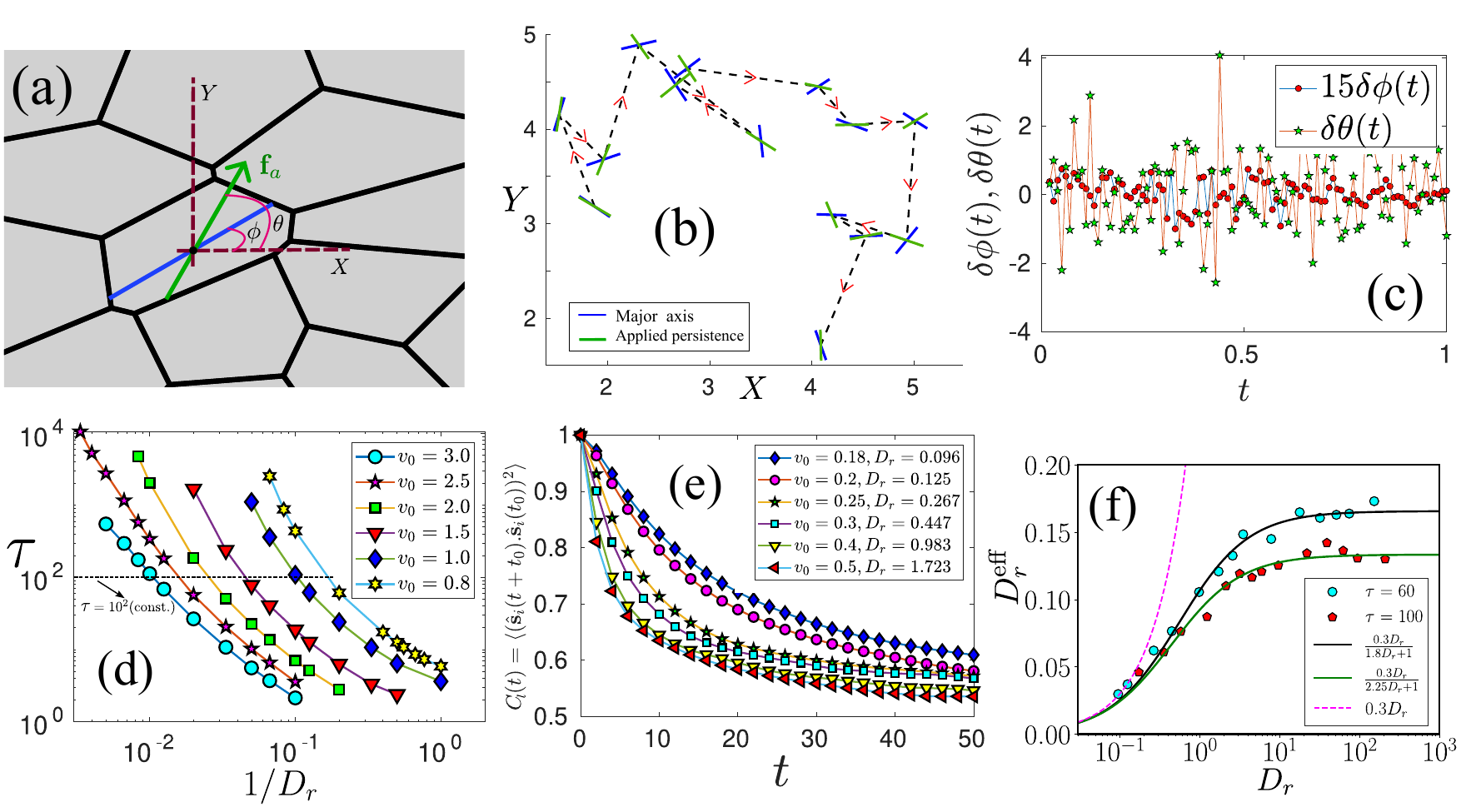}
		\caption{  Modification of rotational diffusivity in confluent systems. (a) A configuration of the cells in a dense confluent monolayer. Cells are always in contact with their neighbors. The blue line represents the major axis of the cell and the green arrow is the motile force direction. We schematically show the angles $\theta$ and $\phi$. (b) We show the major axis (blue) and the sense of the active force (green) along the trajectory of a typical cell in the monolayer. The arrow indicates the direction of time. \b{We have shown the configurations at a time-separation of $10$. The parameters for this simulation: $p_0 = 3.75$, $v_0 = 0.5$, $D_r = 1.0$} (c) $\delta \theta$ (green pentagram) corresponds to the \b{instantaneous} angular variation of the active force, and $\delta \phi$ (red circles) corresponds to that of the major axis. Note that we have multiplied $\delta\phi$ by a factor of 15 to show in the same scale. $\delta\phi$ varies much slower than $\delta\theta$. \b{We have used $p_0=3.72$, and larger values of activity parameters to highlight the effect: $v_0 = 3.0$, $D_r = 74$.} (d) $\tau$ as a function of $D_r^{-1}$ for different $v_0$ and fixed $p_0 = 3.72$. From such plots, we obtain the values of different parameters that give a specific value of $\tau$. We have used $N = 100$ cells for these simulations. (e) The auto-correlation function for the nematic order parameter $\shat$ (major axis). We keep $p_0 = 3.72$ fixed, and use $v_0$ and $D_r$ such that $\tau=100$ (obtained from Fig. d). We define a cut-off as $C_l(t=\tau_p^\text{eff})=0.65$ and obtain ${D_r^\text{eff}}=1/\tau_p^\text{eff} $. (f) $D_r^\text{eff}$ as a function of $D_r$ obtained at two different $\tau$. The lines are the plot of our analytical argument [Eq.~(\ref{effectiveDr})]. $D_r^\text{eff}$ is proportional to $D_r$ at small $D_r$ and saturates at large $D_r$. The dashed line shows the behavior without the modification due to confluency.}
		\label{effectiveDrFig}
	\end{figure*}

	We now present our simulation results supporting this physical picture and analytical expression, Eq.~(\ref{effectiveDr}). To follow the cell elongation direction, characterized by $\phi(t)$, we first calculate the gyration tensor for each cell and diagonalize it \cite{sadhukhan2022} to obtain two eigenvalues. The eigenvector corresponding to the major eigenvalue, i.e., the major axis direction $\shat$, gives $\phi$ [Fig.~\ref{effectiveDrFig}(a)]. Note that $\shat$ is {\em nematic} by nature as it corresponds to the cell elongation axis. We show in Fig.~\ref{effectiveDrFig}(b) the direction of the internal active force and that of $\shat$ along a trajectory. Fig.~\ref{effectiveDrFig}(c) shows the direction change of the internal active force, $\delta\theta$, and that of $\shat$, $\delta \phi$. Note that the magnitude of $\delta\theta$ is much higher than that of $\delta\phi$. We now discuss how to obtain $D_r^\text{eff}$ from the auto-correlation function of $\delta\phi$.

	Since $\kappa$ represents a time-scale for cell shape changes, \b{it should depend on the properties of the system, i.e., on the control parameters.
The dynamical properties of the system is characterized by the structural relaxation time $\tau$ of the system (see Appendix \ref{simulationdetails} for definition).} For various $p_0$ and $v_0$, we have first calculated $\tau$ as a function of $D_r$ (Fig. \ref{effectiveDrFig}d). We then take a line of constant $\tau$, and for this set of parameters, we calculate the auto-correlation function, $C_l(t)$, given as
	\begin{equation}\label{nematicautocor}
		C_{l}(t)=\langle [\shat(t+t_0)\cdot \shat(t_0)]^2  \rangle,
	\end{equation}
	where $\langle\ldots\rangle$ denotes averages over various cells, initial times $t_0$, and ensembles \cite{rivas2020}. Note that the definition of $C_l(t)$ in Eq.~(\ref{nematicautocor}) accounts for the nematic nature of the major axis and $C_l(t\to\infty)=1/2$. Figure \ref{effectiveDrFig}(e) shows the typical behavior of $C_l(t)$. From these plots, we get a time-scale, $\tau_p^\text{eff}$, by defining a cut-off value of  {$C_l(t=\tau_p^\text{eff})=0.65$, i.e., when $C_l(t)$ decays by $70\%$.} We obtain $D_r^\text{eff}=1/\tau_p^\text{eff}$. Note that the definition of the cut-off value is arbitrary, and this introduces a scale in $D_r^\text{eff}$. We show the simulation data for $D_r^\text{eff}$ as a function of $D_r$ for two different values of $\tau$ in Fig.~\ref{effectiveDrFig}(f). We fit the data with $D_r^\text{eff}=\Gamma D_r/(\kappa D_r+1)$, where $\Gamma$ accounts for the arbitrary constant introduced by the cut-off value of $C_l(t)$. We have also used various other cut-off values in our analysis: $\Gamma$ remains constant for a particular cut-off value. On the other hand, $\kappa$ depends on the value of $\tau$: \b{$\kappa$ increases as $\tau$ becomes larger}. However, this dependence is not strong: \b{within the range of parameters we explored, $\kappa$ seems to logarithmically vary with $\tau$. This suggests a linear variation of $\kappa$ with the changing control parameters (Sec. \ref{comparisonwithsim}).}
	The solid lines in Fig.~\ref{effectiveDrFig}(f) show the comparison of the proposed analytical form (Eq. \ref{effectiveDr}), and the dashed line shows the behavior when we do not consider the modification due to confluency: the excellent agreement with the simulation data justifies the proposed analytical form. We use $D_r^\text{eff}$ for the contribution coming from self-propulsion in the extended RFOT theory below.

	\subsection{RFOT for the active confluent cell monolayer}
	\label{activerfotmodel1}
	The confluent models in the absence of self-propulsion are equilibrium systems. Although $T$ in these models actually represents various energy-consuming active processes as well as equilibrium temperature, within the models it has the status of an equilibrium $T$. Such a description agrees well with experiments on systems where self-propulsion is small \cite{Park2015,Atia2018,Sadhukhan2021,Graner1992,Farhadifar2007}. On the other hand, self-propulsion drives the system out of equilibrium. However, in the regime of small $\tau_p$, linear response still applies and we can define a generalized fluctuation-dissipation relation \cite{parisi2005,fodor2016,sadhukhan2022}. \b{On the other hand, when $\tau_p$ is much larger compared to the natural time-scale of the system, various other effects, such as intermittency and jamming, will appear \cite{mandal2020,keta2022}.} We consider a regime where $\tau_p$ is not very large such that linear response remains applicable. In this regime, the relaxation dynamics remain equilibrium-like at an effective temperature $T_\text{eff}$ \cite{Berthier2013,Nandi2017,paul2023}. Thus, we treat activity as a small perturbation \cite{Nandi2018,Sadhukhan2021,Paul2021b}. The control parameters for the system are $T$ and the three parameters of activity: $p_0$ for confluency and $v_0$ and $\tau_p$ for self-propulsion.

	RFOT theory posits that a glassy system consists of mosaics of glassy domains \cite{Kirkpatrick1987,Kirkpatrick1989,kirkpatrick2015,lubchenko2007,Biroli2012}. The typical length scale of these mosaics comes from a nucleation-like argument. The free energy cost for rearranging a mosaic of length $R$ is
	\begin{align}\label{freeenergy}
		\Delta F=-\Omega_d R^d f+S_d R^{\tilde{\theta}} \Upsilon
	\end{align}
	where $\Omega_d$ and $S_d$ are the volume and surface area of a unit hypersphere in dimension $d$, $f$ is the free energy gain per unit volume, and $\Upsilon$ is the surface free energy cost per unit area. \b{$\tilde{\theta}$ is the surface area exponent}.  The first term in Eq.~(\ref{freeenergy}) gives the free energy gain in the bulk, while the second term gives the cost at its surface due to mismatch. Within RFOT theory, configurational entropy, $s_c(T)$, is the driving force for the reconfiguration of mosaics and $f=Ts_c(T)$ at temperature $T$. Minimizing Eq.~(\ref{freeenergy}) yields the domain of length-scale, $\xi=\left[{\tilde{\theta} S_d\Upsilon}/{d\Omega_dTs_c}\right]^{1/(d-\tilde{\theta})}$. We now discuss the effects of self-propulsion and confluency on $\Upsilon$ and $s_c$.
	
	We first consider the effects of activity on the configurational entropy, $s_c[\Phi,T]$, which is governed by the interaction potential $\Phi$ and $T$ \cite{wolynesbook,parisi2010,Nandi2018}. We denote the potential energy of the passive system as $\Phi_p$, and that due to self-propulsion and confluency as $\Phi_s$ and $\Phi_c$, respectively, and obtain $s_c=s_c[\Phi_p+\Phi_s+\Phi_c,T]$. \b{Here, $\Phi_p$ includes the potential due to the area term in Eq. (\ref{energyfunction}), and any other possible interactions coming from the definition of the models}. As discussed above, $\Phi_s$ is a small perturbation in the regime of our interest. \b{Besides, the inter-cellular interaction potential, $\Phi_c$, in the confluent systems is represented by the perimeter term in Eq. (\ref{energyfunction}) and governed by the parameter $p_0$}. Although $\Phi_c$ is not small for a confluent system, we can still use a perturbative expansion around a reference value. 
	
	The constraint of confluency leads to a geometric transition point along $p_0$. For a polygon of unit area, there is a specific minimum value of the perimeter, $p_\text{min}$, but there is no limit on the maximum value. 
	The ground state of a confluent system consists of hexagons, for which $p_\text{min}\simeq3.712$. In the case of disordered systems showing glassy dynamics, $p_\text{min}$ becomes slightly higher. $p_\text{min}$ has significant role in the static and dynamic properties of confluent systems. When $p_0>p_\text{min}$, the system can satisfy the perimeter constraint of Eq.~(\ref{energyfunction}), the cells have irregular sides represented by mostly concave polygons, and the system properties do not depend on $p_0$: this is known as the large-$p_0$ regime \cite{Sadhukhan2021}. On the other hand, when $p_0<p_\text{min}$, the system cannot satisfy the perimeter constraint of Eq.~(\ref{energyfunction}), the cells have regular sides (straight or with constant curvature) represented by mostly convex polygons, and the system properties strongly depend on $p_0$: this is known as the low $p_0$ regime \cite{Sadhukhan2021}. We take a reference value of $p_0$, $\pref$, close to $\pmin$ and determine its value via fitting to the simulation data. We take the potential corresponding to $\pref$ as a reference state and treat the potential around this value as a small parameter, i.e., $\Phi_c=\Phi_c^\text{ref}+\delta\Phi_c$ (see Appendix \ref{altarg} for an alternative RFOT argument based on this $p_0^\text{ref}$). Reference \cite{Sadhukhan2021} has shown that such a description works well for confluent systems in the absence of self-propulsion.
	As shown in the Appendix, $\delta\Phi_c\propto (p_0-p_0^\text{ref})$, and we have
	\begin{equation}
		s_c[\Phi,T]\simeq \frac{\Delta C_p(T-T_K)}{T_K}+\chi_c(p_0-p_0^\text{ref})+\kappa_s\Phi_s,
	\end{equation} 
	where $\Delta C_p$ is the specific heat difference between the liquid and crystalline phase, $T_K$ is the Kauzmann temperature \cite{Kauzmann1948}, $\chi_c$ and $\kappa_s$ are two constants.
	
	Let us now consider the surface term. The temperature dependence of $\Upsilon$ is linear \cite{wolynesbook}. Therefore, $\Upsilon=\Xi(\Phi) T$, where the interaction potential also governs \b{the $T$-independent part of the surface reconfiguration energy, $\Xi(\Phi)$}. For a system of SPPs close to the glass transition, the dominant contribution of self-propulsion comes from the bulk and enters the RFOT theory via $s_c$ \cite{Nandi2018}. Assuming $\Xi$ to be independent of $v_0$ and $\tau_p$ provides a good description of the behavior in such systems \cite{Nandi2018}. For simplicity, we keep this assumption also for active cell monolayer. On the other hand, confluency results in strong inter-cellular interaction at cell boundaries: this will affect $\Xi$. As detailed in the Appendix \ref{rfotdetails} \cite{Sadhukhan2021}, we have $\Xi(\Phi_p,\Phi_c)=[B-C(p_0-p_0^{\text{ref}})]$, where $B$ and $C$ are two constants.

	Within RFOT theory, relaxation dynamics occurs via the relaxation of individual mosaics \cite{lubchenko2007,Biroli2012,Nandi2018}. The energy barrier associated with a region of length scale $\xi$ is $\Delta=\Delta_0 \xi^\psi$, where $\Delta_0$ is an energy scale, and $\psi$ is another exponent. Considering a barrier-crossing scenario, we obtain the relaxation time as $\tau=\tau_0\exp[\Delta/k_BT]$, where $\tau_0$ is the relaxation time at high $T$. As detailed in the Appendix \ref{rfotdetails}, using $\psi=\tilde{\theta}=d/2$ \cite{wolynesbook,lubchenko2007,Biroli2012}, we obtain
	\begin{equation}\label{rfotgeneral}
		\ln\left(\frac{\tau}{\tau_0}\right)=\frac{E[1 - F(p_0 - p_0^\text{ref})]}{(T-T_K)+\chi(p_0 - p_0^\text{ref})+\tilde{\kappa}_s \Phi_s}
	\end{equation}
	where, $\tilde{\kappa}_s = \frac{\kappa_sT_K}{\Delta C_p}$, $\chi=\chi_c T_K/\Delta C_p$, $F=C/B$, and $E = \frac{k\tilde{\theta} S_d T_K B}{k_B d \Omega_d \Delta C_p}$ are all constants that we can obtain by fitting the equation with simulation or experimental data. $\Phi_s$ is the effective mean potential energy due to self-propulsion, and we provide its detailed form below. We now confront our theory with new and existing simulation data.

	\subsection{Comparison with simulation results of active Vertex model}
	\label{comparisonwithsim}
	
	\begin{figure*}
		\includegraphics[width=17cm]{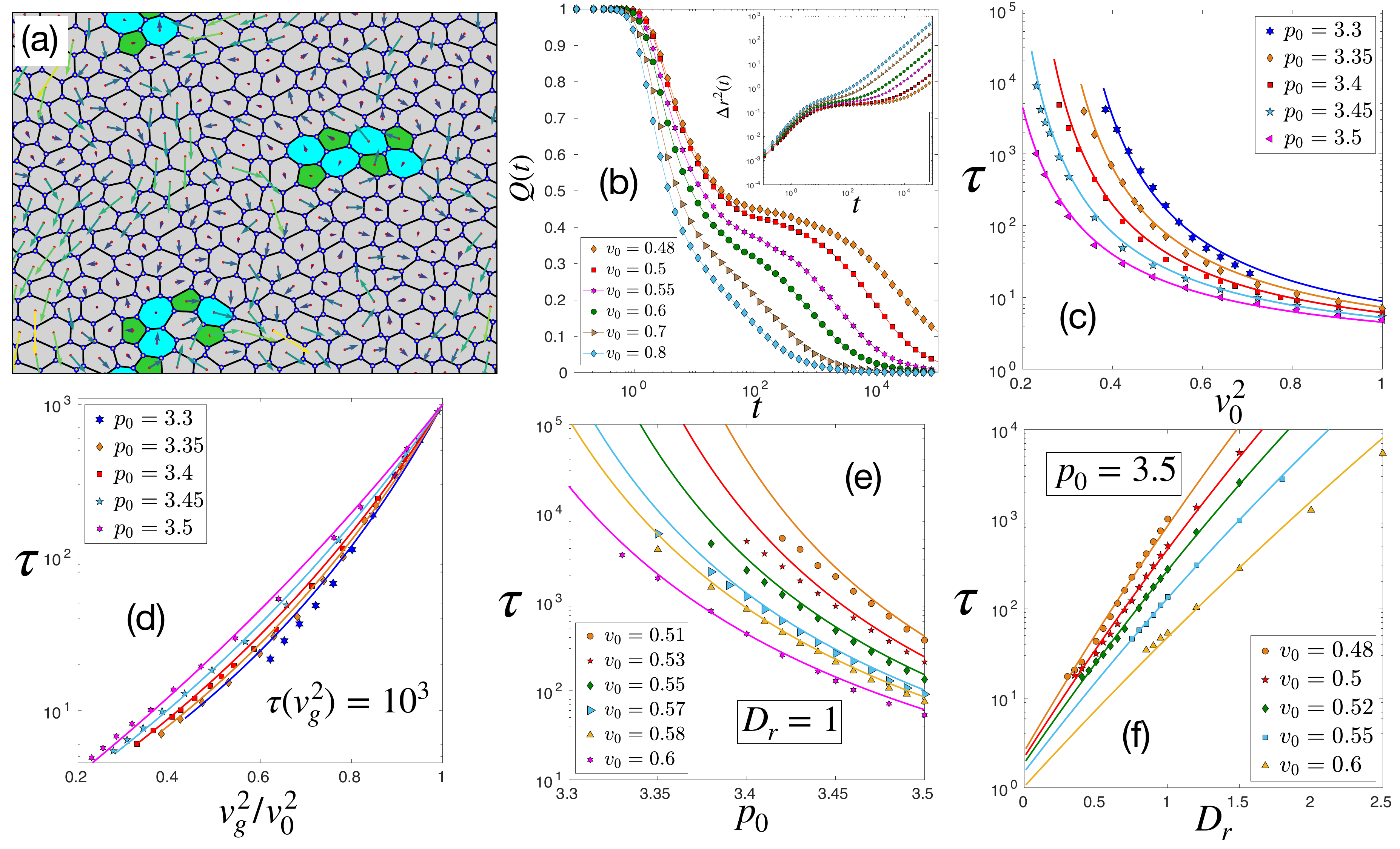}
		\caption{Comparison of theoretical predictions with simulation results of active Vertex model. (a) A typical configuration of the active Vertex model with $p_0 = 3.5$, $v_0 =0.52$, and $D_r = 2.0$. The arrowheads denote the velocity directions, and their lengths denote the magnitudes. Different colors indicate the number of vertices of the cells: green correspond to 5, gray to 6, and cyan to 7. (b) Self-overlap function, $Q(t)$, for different $v_0$ with a fixed $p_0 = 3.45$, and $D_r = 1$. {\bf Inset:} Mean square displacement for the same set of parameters as in the main figure. (c) $\tau$ as a function of $v_0^2$: points are simulation data for different values of $p_0$ and lines are the plot of Eq.~(\ref{rfotgeneral}) with the parameters given in the main text. (d) Angell plot representation of the data of Fig. (c). The system shows super-Arrhenius relaxation for this set of parameters. (e) $\tau$ as a function of $p_0$: symbols are simulation data and lines are the theory. (f) $\tau$ as a function of $D_r$: symbols represent simulation data and lines are theory.} 
		\label{simcomp}
	\end{figure*}
	
	We now compare the theoretical predictions with our simulation results of the active Vertex model (see Appendix \ref{simulationdetails} for simulation details). Fig.~\ref{simcomp}(a) shows a snapshot of the cells with the arrows indicating the self-propulsion direction. To characterize the dynamics, we compute the self-overlap function, $Q(t)$, as defined in Appendix \ref{simulationdetails}. We show the behavior of $Q(t)$ for different values of $v_0$ and fixed $p_0$ and $D_r$ in Fig.~\ref{simcomp}(b). The inset of Fig.~\ref{simcomp}(b) shows the behavior of the mean-square displacement, $\Delta r^2(t)$ (defined in Appendix \ref{simulationdetails}). The plateau in $Q(t)$ and sub-diffusive behavior in $\Delta r^2(t)$ \b{at intermediate times} are typical \b{characteristics} of glassy systems \cite{Berthier2011,Berthier2019c,activereview}. We define the relaxation time $\tau$ as $Q(t=\tau)=0.3$. Fig.~\ref{simcomp}(c) shows the simulation results for $\tau$ (symbols) as a function of $v_0^2$ for different values of $p_0$ and $D_r=1$.
	To compare with the RFOT theory prediction, we first fix the parameters via fitting Eq.~(\ref{rfotgeneral}) with one set of data. Once these parameters are determined, there exist no other free parameters in the theory and we can then compare the model with the rest of the simulation data. Our simulation results suggest $\pref=3.81$ which we have kept constant. As detailed in Ref. \cite{Nandi2018}, considering a one-dimensional model for the dynamics of a single self-propelled particle in a confining potential with strength $k$ and friction $\gamma$, we obtain
	\begin{equation}\label{phisform}
		\Phi_s=\frac{Hv_0^2}{D_r^\text{eff}+G},
	\end{equation}
	where ${H}=1/\gamma$ and $G=k/\gamma$. Note that, as discussed in Sec. \ref{effectiveDrconfluency}, we have used $D_r^\text{eff}$ [Eq.~(\ref{effectiveDr})] instead of $D_r$ in the expression of $\Phi_s$ for a confluent system.

	{\bf Comparison for varying $v_0$:}
	Using the expression of $D_r^\text{eff}$ in Eq.~(\ref{rfotgeneral}), for fixed $p_0$ and $D_r$ we obtain
	\begin{equation}
		\ln\Big(\frac{\tau}{\tau_0}\Big) =\frac{E^\prime - F^\prime(p_0 - 3.81)}{1+\chi^\prime(p_0 - 3.81)+ Kv_0^2},
		\label{Eq:lntaufinal1}
	\end{equation}
	where, $E^\prime = E/(T-T_K)$, $F^\prime = EF/(T-T_K)$, $\chi^\prime = \chi/(T-T_K)$, $K = \tilde{\kappa}_sH/[\{\Gamma D_r/(\kappa D_r+1)+G\}(T-T_K)]$. \b{$\kappa$ has a weak dependence on the control parameters (see Sec. \ref{effectiveDrconfluency}), however, this dependence is extremely weak, and we can ignore this variation with respect to 1. Thus, we treat $K$ as a constant.} Dividing the numerator and denominator of the right-hand side of Eq.~(\ref{Eq:lntaufinal1}) by $K$, we obtain
	\begin{equation}
		\ln\Big(\frac{\tau}{\tau_0}\Big) =\frac{\tilde{E} - \tilde{F}(p_0 - 3.81)}{\tilde{K}+\tilde{\chi}(p_0 - 3.81)+ v_0^2},
		\label{Eq:lntaufinal2}
	\end{equation}
	where, $\tilde{E} = E^\prime/K$, $\tilde{F} = F^\prime/K$, $\tilde{\chi}=\chi^\prime/K$, and $\tilde{K} = 1/K$.
	The simulation results suggest a weak $p_0$-dependence in $\tau_0$. However, for simplicity, we take $\tau_0 = 0.265$ as a constant. We obtain the rest of the parameters via fit with one particular set of data for $p_0 = 3.35$: $\tilde{F} = 1.664$, $\tilde{E} = 0.65$, $\tilde{\chi} = 0.735$, and $\tilde{K} = 0.17$. There are no other free parameters in the theory and we can compare the theoretical predictions with the simulation data, as shown in Fig.~\ref{simcomp}(c).

	We now present the same data as in Fig.~\ref{simcomp}(c) in a way that is analogous to the well-known Angell plot \cite{angell1991a,angell1995b}. We define the glass transition point for the self-propulsion velocity, $v_g$, as $\tau(v_g^2)=10^3$. Fig.~\ref{simcomp}(d) shows $\log\tau$ as a function of $v_g^2/v_0^2$: the lines are the RFOT theory plot, obtained via Eq.~(\ref{rfotgeneral}) with the same parameters used in Fig.~\ref{simcomp}(c) and the symbols are simulation data. All the curves meet at $v_0=v_g$ by definition. The motivation behind plotting the data as a function of $v_0^2$ instead of $v_0$ is that $v_0^2$ has the scale of temperature when $D_r$ is constant. In the Angell plot representation, the diagonal line represents the Arrhenius relaxation. All the curves falling below this diagonal line signify super-Arrhenius relaxation. 
	
	{\bf Comparison for varying $p_0$:}
	Next, we look at the dynamics as a function of $p_0$, keeping $D_r$ and $v_0$ fixed. For the comparison with the RFOT theory, we rewrite Eq.~(\ref{rfotgeneral}) as
	\begin{align}
		\ln\Big(\frac{\tau}{\tau_0}\Big) =\frac{(\tilde{E} + 3.81\tilde{F}) - \tilde{F}p_0}{(\tilde{K}-3.81\tilde{\chi} + v_0^2) + \tilde{\chi}p_0}
		=C'\frac{A^\prime - p_0}{B^\prime + p_0},
		\label{Eq:lntaufinal3}
	\end{align}
	where, $C'= {\tilde{F}}/{\tilde{\chi}}$, $A^\prime = (\tilde{E} + 3.81\tilde{F})$, and $B' =  (\tilde{K}-3.81\tilde{\chi} + v_0^2)$. 
	{ We can obtain the values of the new parameters using the previously determined values of $\tilde{F}$, $\tilde{\chi}$, $\tilde{E}$ and $\tilde{K}$. Thus,} we obtain $C'=2.264$, $A'=71.34$, and $B'=v_0^2-2.63$. We show the comparison of the theory, Eq. (\ref{Eq:lntaufinal3}), with the simulation data for varying $p_0$ at different values of $v_0$.
	
	{\bf Comparison for varying $D_r$:}
	We finally study the dynamics at varying $D_r$, keeping $p_0$ and $v_0$ fixed, and compare it with the RFOT theory. To compare the simulation data from the RFOT theory prediction, we write Eq.~(\ref{rfotgeneral}) as
	\begin{equation}
		\ln\Big(\frac{\tau}{\tau_0}\Big) =\frac{\tilde{E} - \tilde{F}(p_0 - 3.81)}{\tilde{K}+\tilde{\chi}(p_0 - 3.81)+ \frac{v_0^2}{\frac{\tilde{\Gamma} D_r}{\kappa D_r + 1} + \tilde{G}}}.
		\label{Eq:lntaufinal4}
	\end{equation}
	\b{Note that we now have to explicitly keep the $D_r$-dependence in $\Phi_s$ (Eq. \ref{rfotgeneral}). We have written Eq. (\ref{Eq:lntaufinal4}) by dividing the numerator and denominator by the constant $K$, such that $\tilde{E}$, $\tilde{F}$, $\tilde{K}$, and $\tilde{\chi}$ remain the same as before and absorbed a constant factor in $\tilde{\Gamma}$ and $\tilde{G}$. Thus, we now need to obtain the new constants $\kappa$, $\tilde{\Gamma}$ and $\tilde{G}$.
	After straightforward algebra, we write Eq. (\ref{Eq:lntaufinal4}) as}
	\begin{equation}
		\ln\Big(\frac{\tau}{\tau_0}\Big) =\frac{\tilde{E} - \tilde{F}(p_0 - 3.81)}{\tilde{K}+\tilde{\chi}(p_0 - 3.81)+ \frac{v_0^2(\kappa D_r + 1)}{\tilde{N}D_r + \tilde{G}}}
		\label{Eq:lntaufinal5}
	\end{equation}
	where, $\tilde{N} = (\tilde{\Gamma} + \tilde{G} \kappa)$. Thus, we now need to determine the constants $\kappa$, $\tilde{N}$, and $\tilde{G}$. \b{Furthermore, as in the particulate systems \cite{Flenner2016,Nandi2018}, $\tau_0$ also depends on activity when we vary $D_r$. Analyzing our data, we find $\tau_0\simeq 4.57-7.57v_0$. As discussed in Sec. \ref{effectiveDrconfluency}, $\kappa$ is nearly around 1, with a weak dependence on the control parameters. Since we are varying $v_0$ here, we expect $\kappa$ to be a function of $v_0$. From the simulation data, we find $\kappa\simeq 1-5v_0/4$. We obtain the other two parameters via fits with one set of data, for $v_0 = 0.5$: $\tilde{N}\simeq 1.245$ and $\tilde{G}=0.001$.}
	With this set of parameters, Fig. \ref{simcomp}(f) shows the comparison of simulation data (symbols) with the RFOT theory predictions (lines). Despite all these approximations involved in attaining Eq.~(\ref{rfotgeneral}), the agreement with simulation is remarkable.

	\subsection{Comparison with existing simulation results}
	To demonstrate the broad applicability of our theory, we now show that it rationalizes the previously published simulation results on active Voronoi model. For this purpose, we chose the simulation data from Ref. \cite{Bi2016}. Specifically, we compare the theory with the glass transition phase diagram demarcating the fluid-like to solid-like behavior.  Reference \cite{Bi2016} defined this transition when the translational diffusion coefficient, $D_\text{eff}$, goes below $10^{-3}$. However, such a definition is not unique, we can equivalently define the glass transition via $\tau$. We define the glass transition when $\tau/\tau_0=10^6$. As we show below, the functional form governing the phase diagram becomes independent of this specific form of the definition.
	
\begin{figure}
		\centering
		\includegraphics[width=8.6cm]{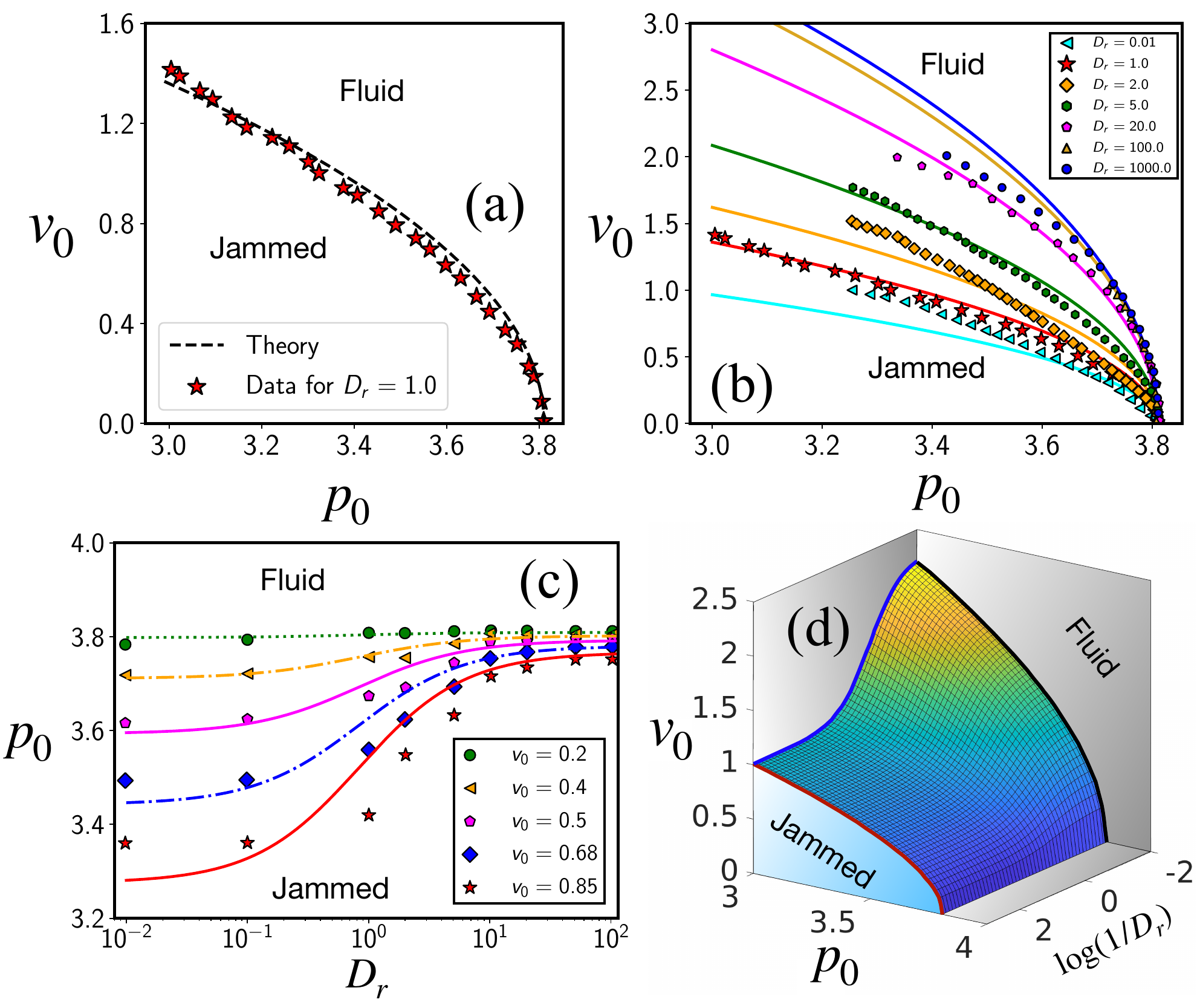}
		\caption{Comparison of theoretical predictions with existing simulation data for the glass transition phase diagram. (a) We fit our theory, Eq.~(\ref{phasediaRFOT_model1}), to the simulation data of Ref. \cite{Bi2016} and obtain the constants $r=1.138$, $a=0.798$ and $b=0.077$. Symbols represent data from Ref. \cite{Bi2016}; the dashed line is the fit. (b) Lines are the plots of Eq.~(\ref{phasediaRFOT_model1}), with the constants obtained from the fit shown in (a), for different values of $D_r$. Symbols are the corresponding simulation data taken from Ref. \cite{Bi2016}. (c) Theoretical predictions for phase diagram in $p_0-D_r$ plane for different values of $v_0$ and symbols are the corresponding simulation data taken from Ref. \cite{Bi2016}. Note that the values of $v_0$ used here are {\em different} from those in Ref. \cite{Bi2016} (see text). Lines are the plot of Eq.~(\ref{phasediaRFOT_model1}). (d) Theoretical 3$d$ phase diagram (Eq. \ref{phasediaRFOT_model1}) for the active cell monolayer.}
		\label{model1}
	\end{figure}

	We use the expressions for $\Phi_s$ [Eq.~(\ref{phisform})] and $D_r^\text{eff}$ [Eq.~(\ref{effectiveDr})] in Eq.~(\ref{rfotgeneral}), and obtain
	\begin{equation}
		\ln\left(\frac{\tau}{\tau_0}\right)=\frac{E[1 - F(p_0 - p_0^\text{ref})]}{(T-T_K)+\chi(p_0 - p_0^\text{ref})+ \frac{\tilde{\kappa}_s{H}v_0^2}{\frac{\Gamma D_r}{\kappa D_r+1}+G}}.
	\end{equation}
	To compare the theory with the data of Ref. \cite{Bi2016}, we are interested in the phase diagram as functions of $D_r$, $v_0$, and $p_0$ at a fixed $T$. Therefore, we can treat $T-T_K$ as a constant and write
	\begin{equation}\label{eqmodel1}
		\ln\left(\frac{\tau}{\tau_0}\right)=\frac{\bar{E}[1 - F(p_0 - p_0^\text{ref})]}{1+\bar{\chi}(p_0 - p_0^\text{ref})+ \frac{\bar{H}v_0^2}{\frac{\Gamma D_r}{\kappa D_r+1}+G}},
	\end{equation}
	where $\bar{E}=E/(T-T_K)$, $\bar{\chi}=\chi/(T-T_K)$ and $\bar{H}=\tilde{\kappa}_s H/(T-T_K)$. Now, the definition of the glass transition, $\tau/\tau_0=10^6$, is independent of any control parameter value. Therefore, we can take $p_0=p_0^\text{ref}$ and $v_0=0$ and obtain $\bar{E}=6\ln 10$. Therefore, from Eq.~(\ref{eqmodel1}), we obtain
	\begin{equation}
		\bar{K}(p_0^\text{ref}-p_0)= \frac{\bar{H}v_0^2}{\frac{\Gamma D_r}{\kappa D_r+1}+G},
	\end{equation}
	where $\bar{K}=\tilde{\chi}+F$. Therefore, the phase boundary governing the solid-like to fluid-like jamming transition is determined by the following equation:
	\begin{equation}\label{phasediaRFOT_model1}
		(\pref-p_0)= \frac{v_0^2}{r+\frac{D_r}{bD_r+a}},
	\end{equation}
	where we have redefined the constants as $r=G\bar{K}/\bar{H}$, $a=\bar{H}/\Gamma \bar{K}$, and $b=\kappa \bar{H}/\Gamma\bar{K}$. The analysis of the data from Ref. \cite{Bi2016} suggests $p_0^\text{ref}=3.81$. We obtain the three parameters, $r$, $a$, and $b$, by fitting Eq.~(\ref{phasediaRFOT_model1}) with one set of simulation data (Fig. 2(a) of Ref. \cite{Bi2016} for $D_r = 1$): $a=0.798$, $b=0.077$, and $r=1.138$. Fig.~\ref{model1}(a) shows the corresponding plot. Once these constants are determined, there are no other free parameters. We can now compare our theory with the other sets of data from Ref. \cite{Bi2016}. 
	
	We show in Fig.~\ref{model1}(b) the theoretical predictions (lines) for the solid-like to fluid-like phase boundary in the $v_0-p_0$ plane for different values of $D_r$ and also the simulation data by symbols taken from Fig. 3(a) of Ref. \cite{Bi2016}. Here, we emphasize two points: first, the theoretical curves are {\em not} fits to the simulation data; they are the {\em predictions} of the theory given the values of the three constants obtained from the fit in Fig.~\ref{model1}(a). Similar phase diagrams in the $v_0-p_0$ plane have been reported in Refs. \cite{Barton2017,yang2017} for different realizations of active cell monolayer. Second, such an agreement with the simulation data is impossible (see Fig. \ref{Drmodification}) without including the effects of confluency on the self-propulsion via Eq.~(\ref{effectiveDr}). 
	
	Next, we look at the behavior of the critical value of $p_0$ as a function of $D_r$ for different values of $v_0$ and compare the theory with the simulation data presented in Fig. (3b) of Ref. \cite{Bi2016}. We plot our theoretical {\em predictions} as lines in Fig.~\ref{model1}(c), and the corresponding simulation data (symbols) are from Ref. \cite{Bi2016}. We emphasize that the values of $v_0$ used in this plot are {\em different} from those presented in Ref. \cite{Bi2016}. We obtained the values of $v_0$ analyzing the other figures in the same paper \cite{Bi2016}, and then the theoretical curves agree with the simulation data. The critical value of $p_0$ saturates in both limits when $D_r\to 0$ and $D_r\to\infty$ and agrees well with Eq.~(\ref{phasediaRFOT_model1}); in the former case, it saturates to $3.81- v_0^2/r$, and in the latter case, it saturates to a value $3.81 -{v_0^2}/{(r+1/b)}$. Summarizing all these results, we can present the solid-like to fluid-like transition in a $3d$ phase diagram from Eq.~(\ref{phasediaRFOT_model1}). We show this phase diagram in Fig.~\ref{model1}(d), and find that it is remarkably similar to the phase diagram of Bi {\it et al.} (Fig. 3(c) of Ref. \cite{Bi2016}). Thus, our work provides the theoretical foundation for the previously published simulation results.
	
	\section{Discussion}
The glassy dynamics in epithelial monolayers plays a crucial role during many biological processes, for example, healing of wounds, progression of asthma, spreading of cancer, etc. During several of these processes, the cell monolayer undergoes EMT, when the cells change their nature from sedentary to motile \cite{thiery2002}. Thus, understanding the effect of motility on the glassy dynamics of confluent systems is imperative. We have developed an analytical framework for the glassy dynamics in active epithelial monolayers via the extension of the RFOT theory of glasses. We have compared our theoretical predictions with new and existing simulation data and show that our theory adequately describes the primary characteristics of the glass transition from a solid-like jammed to a fluid-like flowing state in such systems. Our work provides a simple analytical framework to analyze the glassy relaxation dynamics in active epithelial systems. \b{It will be interesting to extend our theoretical framework to include additional effects such as active stresses or inter-cellular friction \cite{zhang2020,chiang2023}.}

The principal result of this work is that confluency modifies the effective persistence time of self-propulsion. This modification appears as an effective rotational diffusivity, $D_r^\text{eff}$, of the active propulsion forces. In cellular systems, a cell needs time to elongate along the changing direction of the active force, and this elongation time, $\kappa$, should depend on both $f_0$ (or $v_0$) and $\tau$ of the system. We have shown in our simulations that $\kappa$ relates to the time scale of active force relaxation as well as that of cellular elongation. Using $\kappa$, we have phenomenologically obtained an analytical expression for the modified diffusivity of the active force $\Dreff$ and verified it in our simulation.

Analysis of the relaxation dynamics of existing and new simulation results suggest that $\Dreff$ is crucial for confluent cellular systems. We have shown in Fig. \ref{Drmodification} that one cannot explain the simulation results for the relaxation dynamics without considering $\Dreff$. In the experimental context, one can vary the cellular activity through different oncogenes, such as the Ras \cite{gysin2011}, or via growth factors, such as the protein kinase C \cite{rodrigues2019}. For example, Merlin, a tumor suppressor protein, increases directional persistence and reduces motility force in Madin–Darby Canine kidney (MDCK) dog epithelial cells \cite{Das2015}. However, the change in the dynamics should be less sensitive to the variation in $\tau_p$ in the large $D_r$ limit, where $\Dreff$ becomes nearly constant due to the effect of confluency.

Within the theoretical descriptions of confluent systems via Eq.~(\ref{energyfunction}), $p_0$ parameterizes inter-cellular interaction. The phase diagrams in Fig.~\ref{model1} suggest that the system remains fluid beyond a specific value of $p_0$. The parameter $\pref$ in Eq.~(\ref{phasediaRFOT_model1}) is crucial in obtaining this behavior. Our theory does not predict $\pref$; it is an input. We provide an alternate argument in Appendix \ref{altarg} for the extended RFOT theory for confluent systems based on $\pref$. The value of $\pref$ seems to vary for different systems and depends on the details of the models. It is possibly related to the geometric constraint leading to $\pmin$. It is also unclear whether it is related to the rigidity transition of Ref. \cite{bi2015}. $\pref$ seems to control the glassy dynamics in confluent systems (however, see Refs. \cite{saraswathibhatla2020,saraswathibhatla2021}), and it is critical to understand what leads to $\pref$ and how it affects the dynamical behavior.

The glassy behavior with changing self-propulsion velocity, $v_0$, seems similar to that in particulate systems where increasing $v_0$ fluidizes the system. However, the behavior with varying $D_r$ seems non-trivial, where the plateau height in $Q(t)$ seems to vary. This behavior generally signifies a variation in caging length. We will explore this aspect in more detail later. The other aspect, specific to confluent systems, is the fluidization with increasing $p_0$ even when $p_0$ is much lower than $\pref$. As shown in Fig.~\ref{simcomp}(e), $\tau$ decreases with increasing $p_0$. Experiments seem to agree with this result \cite{Garcia2015,bazellieres2015}. For example, Ref. \cite{malinverno2017} has shown that expression of RAB5A, a key endocytic protein, produces large-scale motion and fluidizes a solid-like jammed tissue of human mammary epithelial MCF-10A cells. RAB5A enhances endosomal trafficking and macropinocytic internalization and, thus, facilitates the dynamics of cellular junctional proteins, reducing inter-cellular interaction. This effect implies an increase in $p_0$ that leads to fluidization, much like the results of Fig.~\ref{simcomp}(e).

Apart from their biological significance, the glassy dynamics in confluent systems is also interesting from the perspective of theories of glassy systems. The confluent systems provide a rich testing ground for various theories of glassy dynamics. For example, the sub-Arrhenius relaxation in these systems seems to result in a negative $T_K^\text{eff}$, implying a finite relaxation time even at zero $T$. Based on this result, we can define a $\pref$ and develop the RFOT arguments as in Appendix \ref{altarg}. These arguments show that the relaxation should become super-Arrhenius at lower values of $p_0$ \cite{Sadhukhan2021}. We have verified this prediction for the cellular Potts model in Ref. \cite{Sadhukhan2021} and for the Voronoi and the Vertex models here (Appendix \ref{superArrhenius}). In our simulation, we have chosen the parameters such that we are in the super-Arrhenius regime (Fig.~\ref{simcomp}). On the other hand, the sub-Arrhenius regime seems to be nearly ideal for the mode-coupling theory of glassy dynamics \cite{pandey2023}. How the same system crosses over from one mechanism to the other for the relaxation dynamics and if it is similar in equilibrium systems remain unclear and outside the scope of the current work.

To conclude, our work provides a simple analytical framework to analyze the effect of motility on the glassy dynamics in epithelial monolayers. The strong correlation between cell shape, a static observable, and dynamics  \cite{Atia2018,sadhukhan2022,arora2024} makes the glassiness in these systems distinctive compared to that in particulate systems \cite{Berthier2011,activereview,pandey2023,bi2014,Bi2016}. We have revealed a novel effect of activity, where the time scale of cell shape change leads to a modified rotational diffusivity, $\Dreff$, of the motile forces. This modification is crucial for the relaxation dynamics of the system. Using new and existing simulation data, we have tested the theoretical predictions against two distinct models of confluent systems, the Vertex model (Fig.~\ref{simcomp}) and the Voronoi model (Fig.~\ref{model1}). Therefore, we believe the theory should apply to a broad class of systems, including particle-based models of confluent monolayers \cite{tarle2015,sarkar2021}. The analytical expressions can be used to predict the fluidity behavior resulting from cellular motility. Thus, our work can pave the way forward for a quantitative understanding of the early stage of epithelial-to-mesenchymal transition and how cancerous cells migrate.

	
	\section*{Acknowledgments}
	We thank Daniel Sussman for sharing the data of $\tau$ for bidisperse Voronoi model from Ref. \cite{Sussman2018}. SS thanks ``Infosys-TIFR Leading Edge Travel Grant'', Ref.No.: TFR/Efund/44/Leading Edge TG(R-3)/09/ for the funding for travel to Israel. MP acknowledges  NextGeneration EU (CUP B63C22000730005) within the Project IR0000029 - Humanities and Cultural Heritage Italian Open Science Cloud (H2IOSC) - M4, C2, Action 3.1.1. We acknowledge the support of the Department of Atomic Energy, Government of India, under Project Identification No. RTI 4007. S.K.N. thanks SERB for grant via SRG/2021/002014.

	\appendix

	\section{Details of the RFOT theory calculation}
	\label{rfotdetails}
	As discussed in the main text, both the configurational entropy, $s_c$, and the surface energy are functions of the interaction potential of the system \cite{parisi2010,Nandi2018}. For our system, we write the total interaction potential as sum of three distinct contributions: $\Phi_p$ for the passive system, $\Phi_s$ for the self-propulsion, and $\Phi_c$  for confluency. Thus, $s_c=s_c[\Phi_p+\Phi_s+\Phi_c]$ and $\Xi=\Xi[\Phi_p+\Phi_s+\Phi_c]$. We assume that self-propulsion, and hence $\Phi_s$, is low and a generalized fluctuation-dissipation relation is valid. For a confluent system, $\Phi_c$ itself cannot be small as the system remains confluent at all times. The interaction potential is parameterized by $p_0$.  We assume a reference state given by $p_0^\text{ref}$ and the corresponding value of potential as $\Phi_c^\text{ref}$. Thus, $\Phi_c=\Phi_c^{\text{ref}}+\delta\Phi_c$, where $\delta \Phi_c$ is small. Therefore, we have
	\begin{align}
		s_c[\Phi,T]=s_c[\Phi_p+&\Phi_c^{\text{ref}}]+\frac{\delta s_c}{\delta\Phi_c}\bigg|_{(\delta\Phi_c,\Phi_s)=0}\delta \Phi_c \nonumber\\
		&+\frac{\delta s_c}{\delta\Phi_s}\bigg|_{(\delta\Phi_c,\Phi_s)=0} \Phi_s+\ldots,
	\end{align}
	where $s_c[\Phi_p+\Phi_c^{\text{ref}}]$ is the configurational entropy of the reference system, ${\delta s_c}/{\delta\Phi_c}\big|_{(\delta\Phi_c,\Phi_s)=0}$ and ${\delta s_c}/{\delta\Phi_s}\big|_{(\delta\Phi_c,\Phi_s)=0}$ are constants. We assume the reference system follows the RFOT theory: the configurational entropy vanishes at the Kauzmann temperature $T_K$. Therefore, we can write $s_c$ close to $T_K$ as 
	$s_c[\Phi_p+\Phi_c] = \frac{\Delta C_p(T-T_K)}{T_K}$, where $\Delta C_p$ is the difference in specific heat between the liquid and the periodic crystalline phase \cite{Kauzmann1948}. 
	
	The interaction energy in confluent systems is given by the perimeter term, and we can write \cite{Sadhukhan2021} $\delta\Phi_c\propto (p_0-p_0^\text{ref})$. Therefore, we have $\delta s_c/\delta\Phi_c\big|_{(\delta\Phi_c,\Phi_s)=0} \delta\Phi_c=\chi_c(p_0-p_0^\text{ref})$, where $\chi_c$ is a constant. Then, we can write $s_c[\Phi,T]$ as 
	\begin{align}
		s_c[\Phi,T]\simeq \frac{\Delta C_p(T-T_K)}{T_K}+\chi_c(p_0 - p_0^\text{ref})+\kappa_s \Phi_s,
	\end{align}
	where $\kappa_s={\delta s_c}/{\delta\Phi_s}\big|_{(\delta\Phi_c,\Phi_s)=0}$ is another constant. On the other hand, following Ref. \cite{Nandi2018}, we ignore the contribution of $\Phi_s$ to the surface reconfiguration energy, thus $\Xi=\Xi[\Phi_p,\Phi_c^\text{ref}+\delta\Phi_c]$. Expanding the effect of confluency around the reference state, we obtain
	\begin{equation}
		\Xi[\Phi]\simeq \Xi[\Phi_p,\Phi_c^\text{ref}]+\frac{\delta\Xi}{\delta \Phi_c}\bigg|_{\delta\Phi_c=0}\delta\Phi_c+\ldots.
	\end{equation}
	Using similar arguments as above, we can write $\Xi[\Phi]\simeq B-C(p_0-p_0^\text{ref})$.

	As shown in the main text, the typical length scale of the mosaics is 
	\begin{align}
		\xi = \Big[\frac{\tilde{\theta} S_d \Upsilon}{d \Omega_d T s_c}\Big]^{1/(d-\tilde{\theta})}.
	\end{align}
	Relaxation within RFOT theory is described by the relaxation of the individual mosaics. Considering a barrier-crossing scenario, we have the relaxation time $\tau$ 
	\begin{align}
		\tau = \tau_0 e^{\Delta_0 \xi^{\psi}},
		\label{Eq:tau}
	\end{align}
	where $\Delta_0$ is an energy scale and $\tau_0$ is the high $T$ relaxation time. Substituting the value of $\xi$ above, we get
	\begin{align}
		\ln\Big(\frac{\tau}{\tau_0}\Big) = \frac{\Delta_0}{k_BT} \Big[\frac{\tilde{\theta} S_d \Upsilon}{d \Omega_d T s_c}\Big]^{\frac{\psi}{d-\tilde{\theta}}}.
		\label{Eq:lntau}
	\end{align}
	{Typically, the surface area exponent $\tilde{\theta}$ should be $d-1$. However, the mosaic surface can be rough and fractal-like, and generally, $\tilde{\theta} \leq d-1$. Simulation results for $\tilde{\theta}$ seem to be consistent with both $d-1$ and $d/2$ \cite{xia2000,biroli2009}. On the other hand, the value of $\psi$  is more non-trivial. To the best of our knowledge, there does not exist any analytical results for $\psi$. We have followed the arguments of Wolynes and co-workers and used $\psi = \tilde{\theta} = d/2$ \cite{lubchenko2007,Kirkpatrick1989}.} Using the expressions of $\Upsilon$ and $s_c$, and the values of $\psi$ and $\tilde{\theta}$ \cite{wolynesbook,lubchenko2007,Biroli2012}, we obtain,
	\begin{equation}
		\ln\Big(\frac{\tau}{\tau_0}\Big) =\frac{E[1 - F(p_0 - p_0^\text{ref})]}{(T-T_K)+\chi(p_0 - p_0^\text{ref})+\tilde{\kappa}_s\delta \Phi_s},
		\label{Eq:lntaufinal}
	\end{equation}
	where, $E = k\tilde{\theta} S_d T_K B/k_B d \Omega_d \Delta C_p$, $\tilde{\kappa}_s = \kappa_sT_K/\Delta C_p$, $\chi=\chi_cT_K/\Delta C_p$ and $F=C/B$.

	\begin{figure*}
		\centering
		\includegraphics[width=18cm]{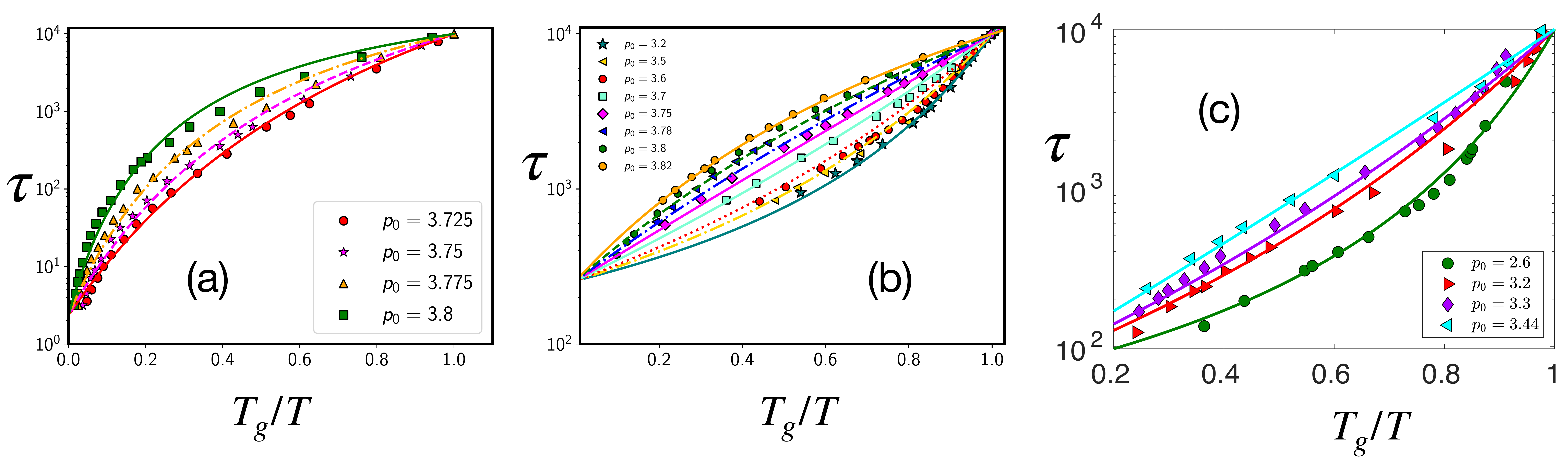}
		\caption{The Angell plot-representation of the relaxation time for various systems. (a) Passive bi-disperse Voronoi model. Simulation data taken from Ref. \cite{Sussman2018}. (b) Data for a mono-disperse Voronoi model with a wide range of $p_0$ values. (c) Simulation results for the passive Vertex model. In all the figures, symbols represent simulation data, and the dotted lines are RFOT theory predictions [Eq.~(\ref{eqn:rfotlowP0})]. See the text for the parameters for each of these plots.}
		\label{fig:passivemodels}
	\end{figure*}

	\section{Consistency check: predictions of super- and sub-Arrhenius regimes in equilibrium}
	\label{superArrhenius}
	One unusual aspect of the glassy dynamics in confluent systems is the readily found sub-Arrhenius relaxation dynamics: In the Angell plot representation \cite{angell1991a,angell1995b}, plotting $\log\tau$ as a function of $T_g/T$, where $T_g$ is the glass transition temperature, a straight line represents the Arrhenius relaxation, that is $\tau$ grows as an exponential of $1/T$. When relaxation is super-Arrhenius, i.e., faster than exponential, the curve appears below the straight line. In contrast, when relaxation is sub-Arrhenius, i.e., slower than exponential, it appears above the straight line. All the models of confluent systems, the Vertex model, the Voronoi model \cite{Sussman2018}, and the CPM \cite{Sadhukhan2021} show sub-Arrhenius behavior  for certain values of the parameters. 
	The extended RFOT theory for the glassy dynamics in confluent systems \cite{Sadhukhan2021} predicts that $T_K^\text{eff}$, the effective Kauzmann temperature, becomes negative in the sub-Arrhenius regime. This result suggests that $\tau$ does not diverge even at zero $T$. On the other hand, Ref. \cite{pandey2023} showed that another theory, the mode-coupling theory (MCT), applies remarkably well in this regime and $\tau$ diverges at a finite $T$ as a power law.  
	
	On the other hand, the extended RFOT theory also predicts the presence of super-Arrhenius relaxation at lower $p_0$ values \cite{Sadhukhan2021}. This prediction has been tested for the CPM \cite{Sadhukhan2021} and the Voronoi model \cite{Li2021}. In the main text, we have shown that the super-Arrhenius regime exists in the self-propelled Vertex model.
	Here we show that this specific prediction is also valid for the equilibrium Vertex model and provide detailed comparison of the RFOT theory for the equilibrium Voronoi and Vertex models..
	
	For the equilibrium systems, the RFOT expression, Eq.~(\ref{rfotgeneral}), becomes
	\begin{equation}\label{eqn:rfotlowP0}
		\ln\left(\f{\tau}{\tau_0}\right)=\frac{k_1-k_2(p_0-p_0^\text{ref})}{T-T_K+\chi(p_0-p_0^\text{ref})},
	\end{equation}
	where $k_1=E$ and $k_2=EF$.
	We first compare the theory with the existing Voronoi model simulation data presented in Ref. \cite{Sussman2018}. We find that $p_0^\text{ref} = 3.53$ gives a good description of the simulation data, and fitting Eq. (\ref{eqn:rfotlowP0}) with one set of data, we find  $\tau_0 = 2.323$, $k_1 = 0.166$, $k_2 = 0.220$, $T_K = 0.000146$ and $\chi = 0.04$.  Fig.~\ref{fig:passivemodels}(a) shows the comparison of the theory (lines) with the simulation data of Ref. \cite{Sussman2018} (symbols). Using these values of the constants, it is easy to verify that $T_K^\text{eff}=T_K-\chi(p_0-\pref)$ becomes negative for these set of curves in Fig.~\ref{fig:passivemodels}. This is consistent with the fact that the curves represent sub-Arrhenius behaviors. 
	
	According to the RFOT theory, we expect the behavior to become super-Arrhenius at lower $p_0$, however, Sussman {\it et al} did not explore this regime \cite{Sussman2018}. Since the detailed values of the parameters depend on the specific system, to test this aspect, we first present simulation data for the equilibrium Voronoi model for a range of $p_0$ values showing both the super- and sub-Arrhenius behavior. For our system with $\lambda_A=1.0$, $\lambda_P=1.0$, and $N=100$, we find $\pref=3.72$, $\tau_0 = 5.545$, $k_1 = 0.064$, $k_2 = 0.112$, $T_K = 0.002$ and $\chi = 0.087$. We show the comparison of our simulation data with Eq.~(\ref{eqn:rfotlowP0}) in Fig.~\ref{fig:passivemodels}(b). Consistent with the results of Ref. \cite{Sussman2018}, the relaxation dynamics is sub-Arrhenius for the range of $p_0=3.75-3.82$, and then the dynamics becomes super-Arrhenius for lower $p_0$, as also found in Ref. \cite{Li2021}.
	
	We finally present our simulation results for the equilibrium Vertex model testing the prediction for the super-Arrhenius behavior at lower $p_0$ values. For these simulations, we have used the open-source software RheoVM \cite{rheovm} to simulate the Vertex model via Brownian dynamics. We have used a system size of $30\times30$ with $N=400$ cells, preferred area $A_0=L^2/N$, $\lambda_A=0.5$, and $\lambda_P=0.05$. We take the friction coefficient for the Brownian dynamics as $0.1$. 
	For the clarity of presentation, we only show the super-Arrhenius behavior for the lower values of $p_0$ in Fig.~\ref{fig:passivemodels}(c). Our analysis suggests the value of $p_0^\text{ref}=3.45$. Fitting one set of data gives the values of the constants in Eq.~\ref{eqn:rfotlowP0} as follows: $\tau_0 = 4.164$, $k_1 = 0.0381$; $k_2 = 0.0056$, $T_K = 0.000312$ and $\chi = 0.0177$. Fig.~\ref{fig:passivemodels}(c) shows the simulation data via symbols and the corresponding RFOT theory plot of Eq.~(\ref{eqn:rfotlowP0}) by lines; the theory agrees remarkably well in this regime with the simulation data.

	\section{Confluency modification of $D_r$}
\begin{figure}
		\includegraphics[width=8.6cm]{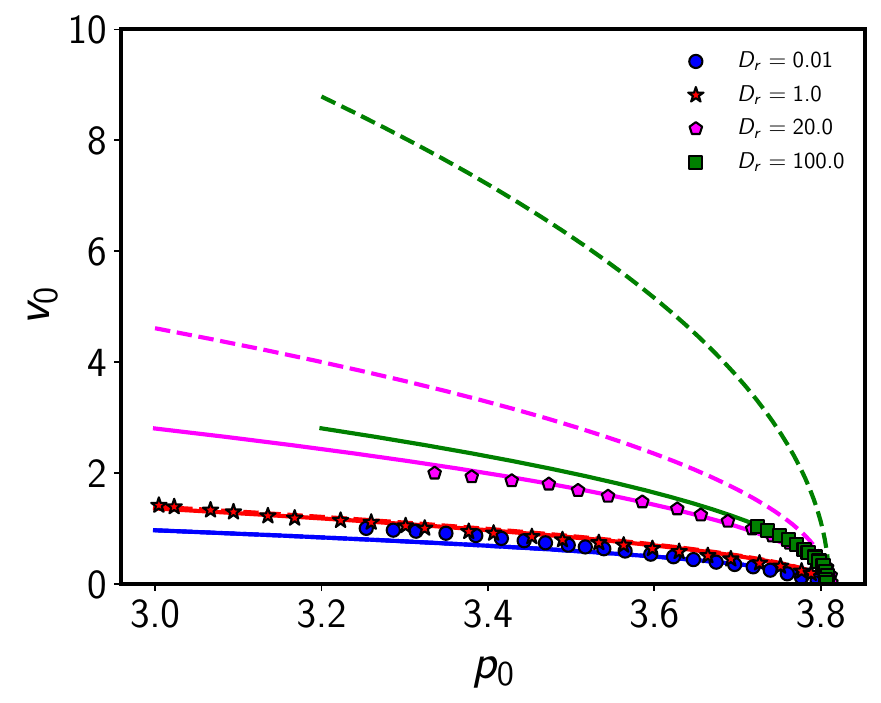}
		\caption{We show the phase diagram in the $v_0-p_0$ plane for various $D_r$ with (solid lines) and without (dashed lines) the confluency modification due to $D_r$ [Eq.~(\ref{effectiveDr})]. The solid and dashed lines overlap for smaller values of $D_r$, but they deviate significantly from each other for higher values. The symbols represent the simulation data from Ref. \cite{Bi2016} for the corresponding values of $D_r$.}
		\label{Drmodification}
	\end{figure}

	We stated in the main text, Sec. \ref{effectiveDrconfluency}, that the confluency modifies the rotational diffusivity of self-propulsion and leads to $D_r^\text{eff}$ [Eq.~(\ref{effectiveDr})]. $D_r^\text{eff}$ enters the RFOT theory, and we have shown the comparison with simulation data in Figs.~\ref{simcomp} and \ref{model1} in the main text. We show in Fig.~\ref{Drmodification} that this modification is crucial for agreement with the simulation data. We present the $v_0-p_0$ phase diagrams at various $D_r$ with this modification and without it (that is, $\kappa=0$ in Eq.~(\ref{effectiveDr})), by solid and dashed lines, respectively. Since $D_r^\text{eff}$ is very close to $D_r$, they nearly agree with each other for $D_r=0.01$ and $1.0$. However, they deviate strongly for the higher values of $D_r$.
	
	\section{An alternative argument for the RFOT theory of confluent systems}
	\label{altarg}
	We have argued, via existing and new simulation results, that the dynamics in confluent systems is sub-Arrhenius at larger values of $p_0$, and it becomes super-Arrhenius as $p_0$ decreases. Furthermore, the application of the RFOT theory to the sub-Arrhenius regime leads to a negative effective Kauzmann temperature, $T_K^\text{eff}$. However, $T_K^\text{eff}$ is positive in the super-Arrhenius regime, and the usual arguments of RFOT theory applies \cite{Biroli2012}. Based on this observation, we can develop a RFOT argument to extend the theory for such systems to understand the effect of $p_0$. Since this argument is specific to the confluent systems, for simplicity, we present it for equilibrium systems alone. The self-propulsion can be included in a similar way as discussed in the main text.
	
	Within RFOT theory, the mosaic length is determined by the competition of an energy cost at the surface and gain in the bulk. The latter is governed by the configurational entropy, $s_c$, which determines the critical behavior. For the confluent systems, $s_c\equiv s_c[\Phi(p_0)]$. We assume that for $p_0<p^*$,
	\begin{equation}
		s_c[\Phi(p_0)]=\begin{cases}
			c(p_0)[T-T_K(p_0)], & \text{ when } T>T_K(p_0)\\
			=0, &\text{ when } T<T_K(p_0)
		\end{cases}
	\end{equation}
	where $c(p_0)$ is a $p_0$-dependent constant. $T_K(p_0)$ here also depends on $p_0$ and we will obtain this dependence below. $p^*$ gives the regime below which the theory is applicable. The simulation results show that $p^*$ should be taken as $p_0^\text{ref}$, where the dynamics is Arrhenius-like. Thus, the theory should be applicable for values of $p_0$ below $p_0^\text{ref}$. We assume that $T_K(p_0)$ is a smooth function of $p_0$ when $p_0$ is close to $p_0^\text{ref}$, therefore, we can write
	\begin{equation}
		T_K(p_0)=-r(p_0-p_0^\text{ref}),
	\end{equation}
	where $r$ is a positive constant. Thus, we have
	\begin{equation}
		s_c[\Phi(p_0)]=\begin{cases}
			c(p_0)[T+r(p_0-p_0^\text{ref}))], & \text{ when } T>T_K(p_0)\\
			=0, &\text{ when } T<T_K(p_0)
		\end{cases}. \nonumber
	\end{equation}
	Using this expression for $s_c$ along with the arguments presented in Sec. \ref{activerfotmodel1}, we obtain the RFOT theory for the confluent system in the absence of self-propulsion as
	\begin{equation}
		\ln\Big(\frac{\tau}{\tau_0}\Big) =\frac{E[1 - F(p_0 - p_0^{ref})]}{T+r(p_0 - p_0^\text{ref})}.
	\end{equation}
	Note that this result is valid for $T>T_K(p_0)$. The system will have dynamics in this regime, and the dynamics freezes in the other regime.

	\section{Simulation details for the Vertex model}
	\label{simulationdetails}
	Here we provide additional details for the self-propelled Vertex Model (VM). To include self-propulsion, we associate each cell with a self-propulsion force $\mathbf{f}_a$ of magnitude $f_0$ (or velocity $v_0$ considering $\mu=1$). $\mathbf{f}_a$ acts along a polarity vector, $\nhat = (\cos \theta_i, \sin \theta_i)$, where $\theta_i$ is an angle measured from the $x$-axis. The total self-propulsion force on each vertex ($k$), ${\bf f}_a^{k}$, is the mean self-propulsion forces of the 3 cells that share vertex $k$. We assume over-damped dynamics and write the equation of motion of vertex $k$ as
	\begin{equation}
		\frac{d{\bf{r}}_k}{dt} = \mu^{-1} \Big({\bf{F}}_k + {\bf{f}}_a^{k} + {\bf F^R_k}\Big),
	\end{equation}
	where ${ \bf F}_k = -{\bf\nabla}_k \mathcal{H}$ is the force arising from the mechanical energy, Eq.~(\ref{energyfunction}), of the tissue.  ${\mathbf{F}^R_k}$ is the random thermal noise with mean zero and standard deviation, $\langle {\mathbf{F}^R}_i(t) {\mathbf{F}^R}_j(t^\prime) \rangle = 2 k_B T \delta(t-t^\prime)\delta_{ij}\mathbf{I}$, with $\mathbf{I}$ being the unit tensor; we have set $k_B$ to unity. ${\bf f}_a^k$ is the active force acting on vertex $k$: ${\bf f}_a^k = \frac{v_0}{3} \sum_{i \in \mathcal{N}(k)} \nhat$, where $\mathcal{N}$ is the list of neighboring cells sharing vertex $k$. $\theta_i(t)$ performs a rotational Brownian motion [Eq.~(\ref{thetaeq})]. We have used the Euler-Murayama integration scheme to update the discretized version of the equations (for each vertex).

	To highlight the effects of activity, we have used $T = 0$ in our simulations. Other parameters are as follows; $\lambda_A = 1.0$, $\lambda_P = 1.0$, $A_{0\alpha} = 1.2$, $A_{0\beta} = 0.8$, $l_{T1} = 0.04$, $\lambda_{T1} = 2.0$, $\delta t = 0.01$, $t_{T1} = 5\delta t$. Our unit of length is $\sqrt{\langle A_0 \rangle}$ and unit of time is $\frac{1}{\mu \lambda_A \langle A_0 \rangle}$.
	Unless otherwise specified, we have equilibrated the system for $3 \times 10^5$ times before collecting the data, performed $100 t_0$  averaging and $32$ ensemble averaging, and simulated $256$ cells.
	
	{\it Self-overlap function, $Q(t)$}: In our simulations, we compute the center of mass, $\mathbf{r}_i$, for the cells from the positions of their vertices. We then define the self-overlap function, $Q(t)$, as 
	\begin{equation}\label{Qoftdef}
		Q(t)=\left\langle\frac{1}{N} \sum_{i=1}^N W\left(a-\left|\mathbf{r}_i(t)-\mathbf{r}_i(0)\right|\right)\right\rangle,
	\end{equation}
	where $W(x)$ is the Heaviside step function:
	\begin{equation}
		W(x)=\begin{cases}
			1 \,\,\, &\text{if }\,\, x>0,\\
			0\,\,\,\ &\text{otherwise}.
		\end{cases}
	\end{equation}
	The parameter $a$ is set to a constant value of $0.346$ throughout the simulation. This value of $a$ corresponds to the caging length scale of vibrational motions revealed via the MSD [inset of Fig.~\ref{simcomp}(b)]. The angular brackets denote initial time and ensemble averages. We define the structural relaxation time $\tau$ as $Q(t=\tau)=0.3$.
	
	{\it Mean-square displacement, $\Delta r^2(t)$}: We define $\Delta r^2(t)$ as
	\begin{align}
		\Delta r^2(t)=\left\langle{\f{1}{N}\sum_{i=1}^N (\mathbf{r}_i(t)-\mathbf{r}_i(0))^2} \right\rangle.
	\end{align}
	
	\bibliography{references.bib}

\begin{thebibliography}{110}%
\makeatletter
\providecommand \@ifxundefined [1]{%
 \@ifx{#1\undefined}
}%
\providecommand \@ifnum [1]{%
 \ifnum #1\expandafter \@firstoftwo
 \else \expandafter \@secondoftwo
 \fi
}%
\providecommand \@ifx [1]{%
 \ifx #1\expandafter \@firstoftwo
 \else \expandafter \@secondoftwo
 \fi
}%
\providecommand \natexlab [1]{#1}%
\providecommand \enquote  [1]{``#1''}%
\providecommand \bibnamefont  [1]{#1}%
\providecommand \bibfnamefont [1]{#1}%
\providecommand \citenamefont [1]{#1}%
\providecommand \href@noop [0]{\@secondoftwo}%
\providecommand \href [0]{\begingroup \@sanitize@url \@href}%
\providecommand \@href[1]{\@@startlink{#1}\@@href}%
\providecommand \@@href[1]{\endgroup#1\@@endlink}%
\providecommand \@sanitize@url [0]{\catcode `\\12\catcode `\$12\catcode
  `\&12\catcode `\#12\catcode `\^12\catcode `\_12\catcode `\%12\relax}%
\providecommand \@@startlink[1]{}%
\providecommand \@@endlink[0]{}%
\providecommand \url  [0]{\begingroup\@sanitize@url \@url }%
\providecommand \@url [1]{\endgroup\@href {#1}{\urlprefix }}%
\providecommand \urlprefix  [0]{URL }%
\providecommand \Eprint [0]{\href }%
\providecommand \doibase [0]{http://dx.doi.org/}%
\providecommand \selectlanguage [0]{\@gobble}%
\providecommand \bibinfo  [0]{\@secondoftwo}%
\providecommand \bibfield  [0]{\@secondoftwo}%
\providecommand \translation [1]{[#1]}%
\providecommand \BibitemOpen [0]{}%
\providecommand \bibitemStop [0]{}%
\providecommand \bibitemNoStop [0]{.\EOS\space}%
\providecommand \EOS [0]{\spacefactor3000\relax}%
\providecommand \BibitemShut  [1]{\csname bibitem#1\endcsname}%
\let\auto@bib@innerbib\@empty
\bibitem [{\citenamefont {Tambe}\ \emph {et~al.}(2011)\citenamefont {Tambe},
  \citenamefont {Hardin}, \citenamefont {Angelini}, \citenamefont {Rajendran},
  \citenamefont {Park}, \citenamefont {Serra-Picamal}, \citenamefont {Zhou},
  \citenamefont {Zaman}, \citenamefont {Butler}, \citenamefont {Weitz},
  \citenamefont {Fredberg},\ and\ \citenamefont {Trepat}}]{Tambe2011}%
  \BibitemOpen
  \bibfield  {author} {\bibinfo {author} {\bibfnamefont {D.~T.}\ \bibnamefont
  {Tambe}}, \bibinfo {author} {\bibfnamefont {C.~C.}\ \bibnamefont {Hardin}},
  \bibinfo {author} {\bibfnamefont {T.~E.}\ \bibnamefont {Angelini}}, \bibinfo
  {author} {\bibfnamefont {K.}~\bibnamefont {Rajendran}}, \bibinfo {author}
  {\bibfnamefont {C.~Y.}\ \bibnamefont {Park}}, \bibinfo {author}
  {\bibfnamefont {X.}~\bibnamefont {Serra-Picamal}}, \bibinfo {author}
  {\bibfnamefont {E.~H.}\ \bibnamefont {Zhou}}, \bibinfo {author}
  {\bibfnamefont {M.~H.}\ \bibnamefont {Zaman}}, \bibinfo {author}
  {\bibfnamefont {J.~P.}\ \bibnamefont {Butler}}, \bibinfo {author}
  {\bibfnamefont {D.~A.}\ \bibnamefont {Weitz}}, \bibinfo {author}
  {\bibfnamefont {J.~J.}\ \bibnamefont {Fredberg}}, \ and\ \bibinfo {author}
  {\bibfnamefont {X.}~\bibnamefont {Trepat}},\ }\href {\doibase
  10.1038/nmat3025} {\bibfield  {journal} {\bibinfo  {journal} {Nat. Mater.}\
  }\textbf {\bibinfo {volume} {10}},\ \bibinfo {pages} {469} (\bibinfo {year}
  {2011})}\BibitemShut {NoStop}%
\bibitem [{\citenamefont {Friedl}\ and\ \citenamefont
  {Gilmour}(2009)}]{Friedl2009b}%
  \BibitemOpen
  \bibfield  {author} {\bibinfo {author} {\bibfnamefont {P.}~\bibnamefont
  {Friedl}}\ and\ \bibinfo {author} {\bibfnamefont {D.}~\bibnamefont
  {Gilmour}},\ }\href {\doibase 10.1038/nrm2720} {\bibfield  {journal}
  {\bibinfo  {journal} {Nat. Rev. Mol. Cell Biol.}\ }\textbf {\bibinfo {volume}
  {10}},\ \bibinfo {pages} {445} (\bibinfo {year} {2009})}\BibitemShut
  {NoStop}%
\bibitem [{\citenamefont {Malmi-Kakkada}\ \emph {et~al.}(2018)\citenamefont
  {Malmi-Kakkada}, \citenamefont {Li}, \citenamefont {Samanta}, \citenamefont
  {Sinha},\ and\ \citenamefont {Thirumalai}}]{Kakkada2018}%
  \BibitemOpen
  \bibfield  {author} {\bibinfo {author} {\bibfnamefont {A.~N.}\ \bibnamefont
  {Malmi-Kakkada}}, \bibinfo {author} {\bibfnamefont {X.}~\bibnamefont {Li}},
  \bibinfo {author} {\bibfnamefont {H.~S.}\ \bibnamefont {Samanta}}, \bibinfo
  {author} {\bibfnamefont {S.}~\bibnamefont {Sinha}}, \ and\ \bibinfo {author}
  {\bibfnamefont {D.}~\bibnamefont {Thirumalai}},\ }\href {\doibase
  10.1103/PhysRevX.8.021025} {\bibfield  {journal} {\bibinfo  {journal} {Phys.
  Rev. X}\ }\textbf {\bibinfo {volume} {8}},\ \bibinfo {pages} {021025}
  (\bibinfo {year} {2018})}\BibitemShut {NoStop}%
\bibitem [{\citenamefont {Sch{\"{o}}tz}\ \emph {et~al.}(2013)\citenamefont
  {Sch{\"{o}}tz}, \citenamefont {Lanio}, \citenamefont {Talbot},\ and\
  \citenamefont {Manning}}]{Schotz2013}%
  \BibitemOpen
  \bibfield  {author} {\bibinfo {author} {\bibfnamefont {E.-M.}\ \bibnamefont
  {Sch{\"{o}}tz}}, \bibinfo {author} {\bibfnamefont {M.}~\bibnamefont {Lanio}},
  \bibinfo {author} {\bibfnamefont {J.~A.}\ \bibnamefont {Talbot}}, \ and\
  \bibinfo {author} {\bibfnamefont {M.~L.}\ \bibnamefont {Manning}},\ }\href
  {\doibase 10.1098/rsif.2013.0726} {\bibfield  {journal} {\bibinfo  {journal}
  {J. R. Soc. Interface}\ }\textbf {\bibinfo {volume} {10}},\ \bibinfo {pages}
  {20130726} (\bibinfo {year} {2013})}\BibitemShut {NoStop}%
\bibitem [{\citenamefont {Poujade}\ \emph {et~al.}(2007)\citenamefont
  {Poujade}, \citenamefont {Grasland-Mongrain}, \citenamefont {Hertzog},
  \citenamefont {Jouanneau}, \citenamefont {Chavrier}, \citenamefont {Ladoux},
  \citenamefont {Buguin},\ and\ \citenamefont {Silberzan}}]{poujade2007}%
  \BibitemOpen
  \bibfield  {author} {\bibinfo {author} {\bibfnamefont {M.}~\bibnamefont
  {Poujade}}, \bibinfo {author} {\bibfnamefont {E.}~\bibnamefont
  {Grasland-Mongrain}}, \bibinfo {author} {\bibfnamefont {A.}~\bibnamefont
  {Hertzog}}, \bibinfo {author} {\bibfnamefont {J.}~\bibnamefont {Jouanneau}},
  \bibinfo {author} {\bibfnamefont {P.}~\bibnamefont {Chavrier}}, \bibinfo
  {author} {\bibfnamefont {B.}~\bibnamefont {Ladoux}}, \bibinfo {author}
  {\bibfnamefont {A.}~\bibnamefont {Buguin}}, \ and\ \bibinfo {author}
  {\bibfnamefont {P.}~\bibnamefont {Silberzan}},\ }\href {\doibase
  10.1073/pnas.0705062104} {\bibfield  {journal} {\bibinfo  {journal} {Proc.
  Natl. Acad. Sci. (USA)}\ }\textbf {\bibinfo {volume} {104}},\ \bibinfo
  {pages} {15988} (\bibinfo {year} {2007})}\BibitemShut {NoStop}%
\bibitem [{\citenamefont {Das}\ \emph {et~al.}(2015)\citenamefont {Das},
  \citenamefont {Safferling}, \citenamefont {Rausch}, \citenamefont {Grabe},
  \citenamefont {Boehm},\ and\ \citenamefont {Spatz}}]{Das2015}%
  \BibitemOpen
  \bibfield  {author} {\bibinfo {author} {\bibfnamefont {T.}~\bibnamefont
  {Das}}, \bibinfo {author} {\bibfnamefont {K.}~\bibnamefont {Safferling}},
  \bibinfo {author} {\bibfnamefont {S.}~\bibnamefont {Rausch}}, \bibinfo
  {author} {\bibfnamefont {N.}~\bibnamefont {Grabe}}, \bibinfo {author}
  {\bibfnamefont {H.}~\bibnamefont {Boehm}}, \ and\ \bibinfo {author}
  {\bibfnamefont {J.~P.}\ \bibnamefont {Spatz}},\ }\href {\doibase
  10.1038/ncb3115} {\bibfield  {journal} {\bibinfo  {journal} {Nat. Cell
  Biol.}\ }\textbf {\bibinfo {volume} {17}},\ \bibinfo {pages} {276} (\bibinfo
  {year} {2015})}\BibitemShut {NoStop}%
\bibitem [{\citenamefont {Brugu{\'{e}}s}\ \emph {et~al.}(2014)\citenamefont
  {Brugu{\'{e}}s}, \citenamefont {Anon}, \citenamefont {Conte}, \citenamefont
  {Veldhuis}, \citenamefont {Gupta}, \citenamefont {Colombelli}, \citenamefont
  {Mu{\~{n}}oz}, \citenamefont {Brodland}, \citenamefont {Ladoux},\ and\
  \citenamefont {Trepat}}]{Brugues2014}%
  \BibitemOpen
  \bibfield  {author} {\bibinfo {author} {\bibfnamefont {A.}~\bibnamefont
  {Brugu{\'{e}}s}}, \bibinfo {author} {\bibfnamefont {E.}~\bibnamefont {Anon}},
  \bibinfo {author} {\bibfnamefont {V.}~\bibnamefont {Conte}}, \bibinfo
  {author} {\bibfnamefont {J.~H.}\ \bibnamefont {Veldhuis}}, \bibinfo {author}
  {\bibfnamefont {M.}~\bibnamefont {Gupta}}, \bibinfo {author} {\bibfnamefont
  {J.}~\bibnamefont {Colombelli}}, \bibinfo {author} {\bibfnamefont {J.~J.}\
  \bibnamefont {Mu{\~{n}}oz}}, \bibinfo {author} {\bibfnamefont {G.~W.}\
  \bibnamefont {Brodland}}, \bibinfo {author} {\bibfnamefont {B.}~\bibnamefont
  {Ladoux}}, \ and\ \bibinfo {author} {\bibfnamefont {X.}~\bibnamefont
  {Trepat}},\ }\href {\doibase 10.1038/NPHYS3040} {\bibfield  {journal}
  {\bibinfo  {journal} {Nat. Phys.}\ }\textbf {\bibinfo {volume} {10}},\
  \bibinfo {pages} {683} (\bibinfo {year} {2014})}\BibitemShut {NoStop}%
\bibitem [{\citenamefont {Malinverno}\ \emph {et~al.}(2017)\citenamefont
  {Malinverno}, \citenamefont {Corallino}, \citenamefont {Giavazzi},
  \citenamefont {Bergert}, \citenamefont {Li}, \citenamefont {Leoni},
  \citenamefont {Disanza}, \citenamefont {Frittoli}, \citenamefont {Oldani},
  \citenamefont {Martini}, \citenamefont {Lendenmann}, \citenamefont
  {Deflorian}, \citenamefont {Beznoussenko}, \citenamefont {Poulikakos},
  \citenamefont {Ong}, \citenamefont {Uroz}, \citenamefont {Trepat},
  \citenamefont {Parazzoli}, \citenamefont {Maiuri}, \citenamefont {Yu},
  \citenamefont {Ferrari}, \citenamefont {Cerbino},\ and\ \citenamefont
  {Scita}}]{malinverno2017}%
  \BibitemOpen
  \bibfield  {author} {\bibinfo {author} {\bibfnamefont {C.}~\bibnamefont
  {Malinverno}}, \bibinfo {author} {\bibfnamefont {S.}~\bibnamefont
  {Corallino}}, \bibinfo {author} {\bibfnamefont {F.}~\bibnamefont {Giavazzi}},
  \bibinfo {author} {\bibfnamefont {M.}~\bibnamefont {Bergert}}, \bibinfo
  {author} {\bibfnamefont {Q.}~\bibnamefont {Li}}, \bibinfo {author}
  {\bibfnamefont {M.}~\bibnamefont {Leoni}}, \bibinfo {author} {\bibfnamefont
  {A.}~\bibnamefont {Disanza}}, \bibinfo {author} {\bibfnamefont
  {E.}~\bibnamefont {Frittoli}}, \bibinfo {author} {\bibfnamefont
  {A.}~\bibnamefont {Oldani}}, \bibinfo {author} {\bibfnamefont
  {E.}~\bibnamefont {Martini}}, \bibinfo {author} {\bibfnamefont
  {T.}~\bibnamefont {Lendenmann}}, \bibinfo {author} {\bibfnamefont
  {G.}~\bibnamefont {Deflorian}}, \bibinfo {author} {\bibfnamefont {G.~V.}\
  \bibnamefont {Beznoussenko}}, \bibinfo {author} {\bibfnamefont
  {D.}~\bibnamefont {Poulikakos}}, \bibinfo {author} {\bibfnamefont {K.~H.}\
  \bibnamefont {Ong}}, \bibinfo {author} {\bibfnamefont {M.}~\bibnamefont
  {Uroz}}, \bibinfo {author} {\bibfnamefont {X.}~\bibnamefont {Trepat}},
  \bibinfo {author} {\bibfnamefont {D.}~\bibnamefont {Parazzoli}}, \bibinfo
  {author} {\bibfnamefont {P.}~\bibnamefont {Maiuri}}, \bibinfo {author}
  {\bibfnamefont {W.}~\bibnamefont {Yu}}, \bibinfo {author} {\bibfnamefont
  {A.}~\bibnamefont {Ferrari}}, \bibinfo {author} {\bibfnamefont
  {R.}~\bibnamefont {Cerbino}}, \ and\ \bibinfo {author} {\bibfnamefont
  {G.}~\bibnamefont {Scita}},\ }\href {\doibase 10.1038/NMAT4848} {\bibfield
  {journal} {\bibinfo  {journal} {Nat. Mater.}\ }\textbf {\bibinfo {volume}
  {16}},\ \bibinfo {pages} {587} (\bibinfo {year} {2017})}\BibitemShut
  {NoStop}%
\bibitem [{\citenamefont {Streitberger}\ \emph {et~al.}(2020)\citenamefont
  {Streitberger}, \citenamefont {Lilaj}, \citenamefont {Schrank}, \citenamefont
  {J{\"{u}}rgen~Braun}, \citenamefont {Reiss-Zimmermann}, \citenamefont
  {K{\"{a}}s},\ and\ \citenamefont {Sack}}]{Streitberger2020}%
  \BibitemOpen
  \bibfield  {author} {\bibinfo {author} {\bibfnamefont {K.-J.}\ \bibnamefont
  {Streitberger}}, \bibinfo {author} {\bibfnamefont {L.}~\bibnamefont {Lilaj}},
  \bibinfo {author} {\bibfnamefont {F.}~\bibnamefont {Schrank}}, \bibinfo
  {author} {\bibfnamefont {a.~K.-T.~H.}\ \bibnamefont {J{\"{u}}rgen~Braun}},
  \bibinfo {author} {\bibfnamefont {M.}~\bibnamefont {Reiss-Zimmermann}},
  \bibinfo {author} {\bibfnamefont {J.~A.}\ \bibnamefont {K{\"{a}}s}}, \ and\
  \bibinfo {author} {\bibfnamefont {I.}~\bibnamefont {Sack}},\ }\href {\doibase
  10.1073/pnas.1913511116} {\bibfield  {journal} {\bibinfo  {journal} {Proc.
  Natl. Acad. Sci. (USA)}\ }\textbf {\bibinfo {volume} {117}},\ \bibinfo
  {pages} {128} (\bibinfo {year} {2020})}\BibitemShut {NoStop}%
\bibitem [{\citenamefont {Friedl}\ and\ \citenamefont
  {Wolf}(2003)}]{Friedl2003a}%
  \BibitemOpen
  \bibfield  {author} {\bibinfo {author} {\bibfnamefont {P.}~\bibnamefont
  {Friedl}}\ and\ \bibinfo {author} {\bibfnamefont {K.}~\bibnamefont {Wolf}},\
  }\href {\doibase 10.1038/nrc1075} {\bibfield  {journal} {\bibinfo  {journal}
  {Nat. Rev. Cancer}\ }\textbf {\bibinfo {volume} {3}},\ \bibinfo {pages} {362}
  (\bibinfo {year} {2003})}\BibitemShut {NoStop}%
\bibitem [{\citenamefont {Berthier}\ and\ \citenamefont
  {Biroli}(2011)}]{Berthier2011}%
  \BibitemOpen
  \bibfield  {author} {\bibinfo {author} {\bibfnamefont {L.}~\bibnamefont
  {Berthier}}\ and\ \bibinfo {author} {\bibfnamefont {G.}~\bibnamefont
  {Biroli}},\ }\href {\doibase 10.1103/RevModPhys.83.587} {\bibfield  {journal}
  {\bibinfo  {journal} {Rev. Mod. Phys.}\ }\textbf {\bibinfo {volume} {83}},\
  \bibinfo {pages} {587} (\bibinfo {year} {2011})}\BibitemShut {NoStop}%
\bibitem [{\citenamefont {Berthier}\ \emph {et~al.}(2019)\citenamefont
  {Berthier}, \citenamefont {Flenner},\ and\ \citenamefont
  {Szamel}}]{Berthier2019c}%
  \BibitemOpen
  \bibfield  {author} {\bibinfo {author} {\bibfnamefont {L.}~\bibnamefont
  {Berthier}}, \bibinfo {author} {\bibfnamefont {E.}~\bibnamefont {Flenner}}, \
  and\ \bibinfo {author} {\bibfnamefont {G.}~\bibnamefont {Szamel}},\ }\href
  {\doibase 10.1063/1.5093240} {\bibfield  {journal} {\bibinfo  {journal} {J.
  Chem. Phys.}\ }\textbf {\bibinfo {volume} {150}},\ \bibinfo {pages} {200901}
  (\bibinfo {year} {2019})}\BibitemShut {NoStop}%
\bibitem [{\citenamefont {Pareek}\ \emph {et~al.}(2023)\citenamefont {Pareek},
  \citenamefont {Adhikari}, \citenamefont {Dasgupta},\ and\ \citenamefont
  {Nandi}}]{pareek2023}%
  \BibitemOpen
  \bibfield  {author} {\bibinfo {author} {\bibfnamefont {P.}~\bibnamefont
  {Pareek}}, \bibinfo {author} {\bibfnamefont {M.}~\bibnamefont {Adhikari}},
  \bibinfo {author} {\bibfnamefont {C.}~\bibnamefont {Dasgupta}}, \ and\
  \bibinfo {author} {\bibfnamefont {S.~K.}\ \bibnamefont {Nandi}},\ }\href
  {\doibase 10.1063/5.0166404} {\bibfield  {journal} {\bibinfo  {journal} {J.
  Chem. Phys.}\ }\textbf {\bibinfo {volume} {159}},\ \bibinfo {pages} {174503}
  (\bibinfo {year} {2023})}\BibitemShut {NoStop}%
\bibitem [{\citenamefont {Atia}\ \emph {et~al.}(2021)\citenamefont {Atia},
  \citenamefont {Fredberg}, \citenamefont {Gov},\ and\ \citenamefont
  {Pegoraro}}]{Atia2021}%
  \BibitemOpen
  \bibfield  {author} {\bibinfo {author} {\bibfnamefont {L.}~\bibnamefont
  {Atia}}, \bibinfo {author} {\bibfnamefont {J.~J.}\ \bibnamefont {Fredberg}},
  \bibinfo {author} {\bibfnamefont {N.~S.}\ \bibnamefont {Gov}}, \ and\
  \bibinfo {author} {\bibfnamefont {A.~F.}\ \bibnamefont {Pegoraro}},\ }\href
  {\doibase https://doi.org/10.1016/j.cdev.2021.203727} {\bibfield  {journal}
  {\bibinfo  {journal} {Cells and Development}\ }\textbf {\bibinfo {volume}
  {168}},\ \bibinfo {pages} {203727} (\bibinfo {year} {2021})}\BibitemShut
  {NoStop}%
\bibitem [{\citenamefont {Sadhukhan}\ \emph {et~al.}(2024)\citenamefont
  {Sadhukhan}, \citenamefont {Dey}, \citenamefont {Karmakar},\ and\
  \citenamefont {Nandi}}]{activereview}%
  \BibitemOpen
  \bibfield  {author} {\bibinfo {author} {\bibfnamefont {S.}~\bibnamefont
  {Sadhukhan}}, \bibinfo {author} {\bibfnamefont {S.}~\bibnamefont {Dey}},
  \bibinfo {author} {\bibfnamefont {S.}~\bibnamefont {Karmakar}}, \ and\
  \bibinfo {author} {\bibfnamefont {S.~K.}\ \bibnamefont {Nandi}},\ }\href
  {https://doi.org/10.1140/epjs/s11734-024-01188-1} {\bibfield  {journal}
  {\bibinfo  {journal} {The European Physical Journal Special Topics}\ }
  (\bibinfo {year} {2024})}\BibitemShut {NoStop}%
\bibitem [{\citenamefont {Angelini}\ \emph {et~al.}(2011)\citenamefont
  {Angelini}, \citenamefont {Hannezo}, \citenamefont {Trepat}, \citenamefont
  {Marquez}, \citenamefont {Fredberg},\ and\ \citenamefont
  {Weitz}}]{Angelini2011}%
  \BibitemOpen
  \bibfield  {author} {\bibinfo {author} {\bibfnamefont {T.~E.}\ \bibnamefont
  {Angelini}}, \bibinfo {author} {\bibfnamefont {E.}~\bibnamefont {Hannezo}},
  \bibinfo {author} {\bibfnamefont {X.}~\bibnamefont {Trepat}}, \bibinfo
  {author} {\bibfnamefont {M.}~\bibnamefont {Marquez}}, \bibinfo {author}
  {\bibfnamefont {J.~J.}\ \bibnamefont {Fredberg}}, \ and\ \bibinfo {author}
  {\bibfnamefont {D.~A.}\ \bibnamefont {Weitz}},\ }\href {\doibase
  10.1073/pnas.1010059108} {\bibfield  {journal} {\bibinfo  {journal} {Proc.
  Natl. Acad. Sci. (USA)}\ }\textbf {\bibinfo {volume} {108}},\ \bibinfo
  {pages} {4717} (\bibinfo {year} {2011})}\BibitemShut {NoStop}%
\bibitem [{\citenamefont {Park}\ \emph {et~al.}(2015)\citenamefont {Park},
  \citenamefont {Kim}, \citenamefont {Bi}, \citenamefont {Mitchel},
  \citenamefont {Qazvini}, \citenamefont {Tantisira}, \citenamefont {Park},
  \citenamefont {McGill}, \citenamefont {Kim}, \citenamefont {Gweon},
  \citenamefont {Notbohm}, \citenamefont {Jr}, \citenamefont {Burger},
  \citenamefont {Randell}, \citenamefont {Kho}, \citenamefont {Tambe},
  \citenamefont {Hardin}, \citenamefont {Shore}, \citenamefont {Israel},
  \citenamefont {Weitz}, \citenamefont {Tschumperlin}, \citenamefont {Henske},
  \citenamefont {Weiss}, \citenamefont {Manning}, \citenamefont {Butler},
  \citenamefont {Drazen},\ and\ \citenamefont {Fredberg}}]{Park2015}%
  \BibitemOpen
  \bibfield  {author} {\bibinfo {author} {\bibfnamefont {J.-A.}\ \bibnamefont
  {Park}}, \bibinfo {author} {\bibfnamefont {J.~H.}\ \bibnamefont {Kim}},
  \bibinfo {author} {\bibfnamefont {D.}~\bibnamefont {Bi}}, \bibinfo {author}
  {\bibfnamefont {J.~A.}\ \bibnamefont {Mitchel}}, \bibinfo {author}
  {\bibfnamefont {N.~T.}\ \bibnamefont {Qazvini}}, \bibinfo {author}
  {\bibfnamefont {K.}~\bibnamefont {Tantisira}}, \bibinfo {author}
  {\bibfnamefont {C.~Y.}\ \bibnamefont {Park}}, \bibinfo {author}
  {\bibfnamefont {M.}~\bibnamefont {McGill}}, \bibinfo {author} {\bibfnamefont
  {S.-H.}\ \bibnamefont {Kim}}, \bibinfo {author} {\bibfnamefont
  {B.}~\bibnamefont {Gweon}}, \bibinfo {author} {\bibfnamefont
  {J.}~\bibnamefont {Notbohm}}, \bibinfo {author} {\bibfnamefont {R.~S.}\
  \bibnamefont {Jr}}, \bibinfo {author} {\bibfnamefont {S.}~\bibnamefont
  {Burger}}, \bibinfo {author} {\bibfnamefont {S.~H.}\ \bibnamefont {Randell}},
  \bibinfo {author} {\bibfnamefont {A.~T.}\ \bibnamefont {Kho}}, \bibinfo
  {author} {\bibfnamefont {D.~T.}\ \bibnamefont {Tambe}}, \bibinfo {author}
  {\bibfnamefont {C.}~\bibnamefont {Hardin}}, \bibinfo {author} {\bibfnamefont
  {S.~A.}\ \bibnamefont {Shore}}, \bibinfo {author} {\bibfnamefont
  {E.}~\bibnamefont {Israel}}, \bibinfo {author} {\bibfnamefont {D.~A.}\
  \bibnamefont {Weitz}}, \bibinfo {author} {\bibfnamefont {D.~J.}\ \bibnamefont
  {Tschumperlin}}, \bibinfo {author} {\bibfnamefont {E.~P.}\ \bibnamefont
  {Henske}}, \bibinfo {author} {\bibfnamefont {S.~T.}\ \bibnamefont {Weiss}},
  \bibinfo {author} {\bibfnamefont {M.~L.}\ \bibnamefont {Manning}}, \bibinfo
  {author} {\bibfnamefont {J.~P.}\ \bibnamefont {Butler}}, \bibinfo {author}
  {\bibfnamefont {J.~M.}\ \bibnamefont {Drazen}}, \ and\ \bibinfo {author}
  {\bibfnamefont {J.~J.}\ \bibnamefont {Fredberg}},\ }\href {\doibase
  10.1038/NMAT4357} {\bibfield  {journal} {\bibinfo  {journal} {Nat. Mater.}\
  }\textbf {\bibinfo {volume} {14}},\ \bibinfo {pages} {1040} (\bibinfo {year}
  {2015})}\BibitemShut {NoStop}%
\bibitem [{\citenamefont {Nnetu}\ \emph {et~al.}(2012)\citenamefont {Nnetu},
  \citenamefont {Knorr}, \citenamefont {Strehle}, \citenamefont {Zink},\ and\
  \citenamefont {K{\"a}s}}]{Nnetu2012}%
  \BibitemOpen
  \bibfield  {author} {\bibinfo {author} {\bibfnamefont {K.~D.}\ \bibnamefont
  {Nnetu}}, \bibinfo {author} {\bibfnamefont {M.}~\bibnamefont {Knorr}},
  \bibinfo {author} {\bibfnamefont {D.}~\bibnamefont {Strehle}}, \bibinfo
  {author} {\bibfnamefont {M.}~\bibnamefont {Zink}}, \ and\ \bibinfo {author}
  {\bibfnamefont {J.~A.}\ \bibnamefont {K{\"a}s}},\ }\href {\doibase
  10.1039/C2SM07208D} {\bibfield  {journal} {\bibinfo  {journal} {Soft Matter}\
  }\textbf {\bibinfo {volume} {8}},\ \bibinfo {pages} {6913} (\bibinfo {year}
  {2012})}\BibitemShut {NoStop}%
\bibitem [{\citenamefont {Szab\'o}\ \emph {et~al.}(2006)\citenamefont
  {Szab\'o}, \citenamefont {Sz\"oll\"osi}, \citenamefont {G\"onci},
  \citenamefont {Jur\'anyi}, \citenamefont {Selmeczi},\ and\ \citenamefont
  {Vicsek}}]{Szabo2006}%
  \BibitemOpen
  \bibfield  {author} {\bibinfo {author} {\bibfnamefont {B.}~\bibnamefont
  {Szab\'o}}, \bibinfo {author} {\bibfnamefont {G.~J.}\ \bibnamefont
  {Sz\"oll\"osi}}, \bibinfo {author} {\bibfnamefont {B.}~\bibnamefont
  {G\"onci}}, \bibinfo {author} {\bibfnamefont {Z.}~\bibnamefont {Jur\'anyi}},
  \bibinfo {author} {\bibfnamefont {D.}~\bibnamefont {Selmeczi}}, \ and\
  \bibinfo {author} {\bibfnamefont {T.}~\bibnamefont {Vicsek}},\ }\href
  {\doibase 10.1103/PhysRevE.74.061908} {\bibfield  {journal} {\bibinfo
  {journal} {Phys. Rev. E}\ }\textbf {\bibinfo {volume} {74}},\ \bibinfo
  {pages} {061908} (\bibinfo {year} {2006})}\BibitemShut {NoStop}%
\bibitem [{\citenamefont {Trepat}\ \emph {et~al.}(2009)\citenamefont {Trepat},
  \citenamefont {Wasserman}, \citenamefont {Angelini}, \citenamefont {Millet},
  \citenamefont {Weitz}, \citenamefont {Butler},\ and\ \citenamefont
  {Fredberg}}]{Trepat2009a}%
  \BibitemOpen
  \bibfield  {author} {\bibinfo {author} {\bibfnamefont {X.}~\bibnamefont
  {Trepat}}, \bibinfo {author} {\bibfnamefont {M.~R.}\ \bibnamefont
  {Wasserman}}, \bibinfo {author} {\bibfnamefont {T.~E.}\ \bibnamefont
  {Angelini}}, \bibinfo {author} {\bibfnamefont {E.}~\bibnamefont {Millet}},
  \bibinfo {author} {\bibfnamefont {D.~A.}\ \bibnamefont {Weitz}}, \bibinfo
  {author} {\bibfnamefont {J.~P.}\ \bibnamefont {Butler}}, \ and\ \bibinfo
  {author} {\bibfnamefont {J.~J.}\ \bibnamefont {Fredberg}},\ }\href {\doibase
  10.1038/nphys1269} {\bibfield  {journal} {\bibinfo  {journal} {Nat. Phys.}\
  }\textbf {\bibinfo {volume} {5}},\ \bibinfo {pages} {426} (\bibinfo {year}
  {2009})}\BibitemShut {NoStop}%
\bibitem [{\citenamefont {Giavazzi}\ \emph {et~al.}(2018)\citenamefont
  {Giavazzi}, \citenamefont {Malinverno}, \citenamefont {Scita},\ and\
  \citenamefont {Cerbino}}]{giavazzi2018}%
  \BibitemOpen
  \bibfield  {author} {\bibinfo {author} {\bibfnamefont {F.}~\bibnamefont
  {Giavazzi}}, \bibinfo {author} {\bibfnamefont {C.}~\bibnamefont
  {Malinverno}}, \bibinfo {author} {\bibfnamefont {G.}~\bibnamefont {Scita}}, \
  and\ \bibinfo {author} {\bibfnamefont {R.}~\bibnamefont {Cerbino}},\ }\href
  {\doibase 10.3389/fphy.2018.00120} {\bibfield  {journal} {\bibinfo  {journal}
  {Front. Phys.}\ }\textbf {\bibinfo {volume} {6}},\ \bibinfo {pages} {120}
  (\bibinfo {year} {2018})}\BibitemShut {NoStop}%
\bibitem [{\citenamefont {Prost}\ \emph {et~al.}(2015)\citenamefont {Prost},
  \citenamefont {J{\"{u}}licher},\ and\ \citenamefont {Joanny}}]{jacques2015}%
  \BibitemOpen
  \bibfield  {author} {\bibinfo {author} {\bibfnamefont {J.}~\bibnamefont
  {Prost}}, \bibinfo {author} {\bibfnamefont {F.}~\bibnamefont
  {J{\"{u}}licher}}, \ and\ \bibinfo {author} {\bibfnamefont {J.~F.}\
  \bibnamefont {Joanny}},\ }\href {\doibase 10.1038/nphys3224} {\bibfield
  {journal} {\bibinfo  {journal} {Nat. Phys.}\ }\textbf {\bibinfo {volume}
  {11}},\ \bibinfo {pages} {111} (\bibinfo {year} {2015})}\BibitemShut
  {NoStop}%
\bibitem [{\citenamefont {Ranft}\ \emph {et~al.}(2010)\citenamefont {Ranft},
  \citenamefont {Basan}, \citenamefont {Elgeti}, \citenamefont {Joanny},
  \citenamefont {Prost},\ and\ \citenamefont {J{\"{u}}licher}}]{ranft2010}%
  \BibitemOpen
  \bibfield  {author} {\bibinfo {author} {\bibfnamefont {J.}~\bibnamefont
  {Ranft}}, \bibinfo {author} {\bibfnamefont {M.}~\bibnamefont {Basan}},
  \bibinfo {author} {\bibfnamefont {J.}~\bibnamefont {Elgeti}}, \bibinfo
  {author} {\bibfnamefont {J.-F.}\ \bibnamefont {Joanny}}, \bibinfo {author}
  {\bibfnamefont {J.}~\bibnamefont {Prost}}, \ and\ \bibinfo {author}
  {\bibfnamefont {F.}~\bibnamefont {J{\"{u}}licher}},\ }\href {\doibase
  10.1073/pnas.101108610} {\bibfield  {journal} {\bibinfo  {journal} {Proc.
  Natl. Acad. Sci. (USA)}\ }\textbf {\bibinfo {volume} {107}},\ \bibinfo
  {pages} {20863} (\bibinfo {year} {2010})}\BibitemShut {NoStop}%
\bibitem [{\citenamefont {Matoz-Fernandez}\ \emph {et~al.}(2017)\citenamefont
  {Matoz-Fernandez}, \citenamefont {Martens}, \citenamefont {Sknepnek},
  \citenamefont {Barrat},\ and\ \citenamefont {Henkes}}]{silke2017}%
  \BibitemOpen
  \bibfield  {author} {\bibinfo {author} {\bibfnamefont {D.~A.}\ \bibnamefont
  {Matoz-Fernandez}}, \bibinfo {author} {\bibfnamefont {K.}~\bibnamefont
  {Martens}}, \bibinfo {author} {\bibfnamefont {R.}~\bibnamefont {Sknepnek}},
  \bibinfo {author} {\bibfnamefont {J.~L.}\ \bibnamefont {Barrat}}, \ and\
  \bibinfo {author} {\bibfnamefont {S.}~\bibnamefont {Henkes}},\ }\href
  {\doibase 10.1039/C6SM02580C} {\bibfield  {journal} {\bibinfo  {journal}
  {Soft Matter}\ }\textbf {\bibinfo {volume} {13}},\ \bibinfo {pages} {3205}
  (\bibinfo {year} {2017})}\BibitemShut {NoStop}%
\bibitem [{\citenamefont {Alvarado}\ and\ \citenamefont
  {Yamanaka}(2014)}]{alvarado2014}%
  \BibitemOpen
  \bibfield  {author} {\bibinfo {author} {\bibfnamefont {A.~S.}\ \bibnamefont
  {Alvarado}}\ and\ \bibinfo {author} {\bibfnamefont {S.}~\bibnamefont
  {Yamanaka}},\ }\href {\doibase 10.1016/j.cell.2014.02.041} {\bibfield
  {journal} {\bibinfo  {journal} {Cell}\ }\textbf {\bibinfo {volume} {157}},\
  \bibinfo {pages} {110} (\bibinfo {year} {2014})}\BibitemShut {NoStop}%
\bibitem [{\citenamefont {Farhadifar}\ \emph {et~al.}(2007)\citenamefont
  {Farhadifar}, \citenamefont {R{\"{o}}per}, \citenamefont {Aigouy},
  \citenamefont {Eaton},\ and\ \citenamefont
  {J{\"{u}}licher}}]{Farhadifar2007}%
  \BibitemOpen
  \bibfield  {author} {\bibinfo {author} {\bibfnamefont {R.}~\bibnamefont
  {Farhadifar}}, \bibinfo {author} {\bibfnamefont {J.-C.}\ \bibnamefont
  {R{\"{o}}per}}, \bibinfo {author} {\bibfnamefont {B.}~\bibnamefont {Aigouy}},
  \bibinfo {author} {\bibfnamefont {S.}~\bibnamefont {Eaton}}, \ and\ \bibinfo
  {author} {\bibfnamefont {F.}~\bibnamefont {J{\"{u}}licher}},\ }\href
  {\doibase 10.1016/j.cub.2007.11.049} {\bibfield  {journal} {\bibinfo
  {journal} {Curr. Biol.}\ }\textbf {\bibinfo {volume} {17}},\ \bibinfo {pages}
  {2095} (\bibinfo {year} {2007})}\BibitemShut {NoStop}%
\bibitem [{\citenamefont {Garcia}\ \emph {et~al.}(2015)\citenamefont {Garcia},
  \citenamefont {Hannezo}, \citenamefont {Elgeti}, \citenamefont {Joanny},
  \citenamefont {Silberzan},\ and\ \citenamefont {Gov}}]{Garcia2015}%
  \BibitemOpen
  \bibfield  {author} {\bibinfo {author} {\bibfnamefont {S.}~\bibnamefont
  {Garcia}}, \bibinfo {author} {\bibfnamefont {E.}~\bibnamefont {Hannezo}},
  \bibinfo {author} {\bibfnamefont {J.}~\bibnamefont {Elgeti}}, \bibinfo
  {author} {\bibfnamefont {J.~F.}\ \bibnamefont {Joanny}}, \bibinfo {author}
  {\bibfnamefont {P.}~\bibnamefont {Silberzan}}, \ and\ \bibinfo {author}
  {\bibfnamefont {N.~S.}\ \bibnamefont {Gov}},\ }\href {\doibase
  10.1073/pnas.1510973112} {\bibfield  {journal} {\bibinfo  {journal} {Proc.
  Natl. Acad. Sci. (USA)}\ }\textbf {\bibinfo {volume} {112}},\ \bibinfo
  {pages} {15314} (\bibinfo {year} {2015})}\BibitemShut {NoStop}%
\bibitem [{\citenamefont {Ramaswamy}(2010)}]{sriramreview}%
  \BibitemOpen
  \bibfield  {author} {\bibinfo {author} {\bibfnamefont {S.}~\bibnamefont
  {Ramaswamy}},\ }\href {\doibase 10.1146/annurev-conmatphys-070909-104101}
  {\bibfield  {journal} {\bibinfo  {journal} {Annu. Rev. Condens. Matter
  Phys.}\ }\textbf {\bibinfo {volume} {1}},\ \bibinfo {pages} {323} (\bibinfo
  {year} {2010})}\BibitemShut {NoStop}%
\bibitem [{\citenamefont {Marchetti}\ \emph {et~al.}(2013)\citenamefont
  {Marchetti}, \citenamefont {Joanny}, \citenamefont {Ramaswamy}, \citenamefont
  {Liverpool}, \citenamefont {Prost}, \citenamefont {Rao},\ and\ \citenamefont
  {Simha}}]{sriramrmp}%
  \BibitemOpen
  \bibfield  {author} {\bibinfo {author} {\bibfnamefont {M.~C.}\ \bibnamefont
  {Marchetti}}, \bibinfo {author} {\bibfnamefont {J.~F.}\ \bibnamefont
  {Joanny}}, \bibinfo {author} {\bibfnamefont {S.}~\bibnamefont {Ramaswamy}},
  \bibinfo {author} {\bibfnamefont {T.~B.}\ \bibnamefont {Liverpool}}, \bibinfo
  {author} {\bibfnamefont {J.}~\bibnamefont {Prost}}, \bibinfo {author}
  {\bibfnamefont {M.}~\bibnamefont {Rao}}, \ and\ \bibinfo {author}
  {\bibfnamefont {R.~A.}\ \bibnamefont {Simha}},\ }\href {\doibase
  10.1103/RevModPhys.85.1143} {\bibfield  {journal} {\bibinfo  {journal} {Rev.
  Mod. Phys.}\ }\textbf {\bibinfo {volume} {85}},\ \bibinfo {pages} {1143}
  (\bibinfo {year} {2013})}\BibitemShut {NoStop}%
\bibitem [{\citenamefont {Park}\ \emph {et~al.}(2016)\citenamefont {Park},
  \citenamefont {Atia}, \citenamefont {Mitchel}, \citenamefont {Fredberg},\
  and\ \citenamefont {Butler}}]{Park2016}%
  \BibitemOpen
  \bibfield  {author} {\bibinfo {author} {\bibfnamefont {J.-A.}\ \bibnamefont
  {Park}}, \bibinfo {author} {\bibfnamefont {L.}~\bibnamefont {Atia}}, \bibinfo
  {author} {\bibfnamefont {J.~A.}\ \bibnamefont {Mitchel}}, \bibinfo {author}
  {\bibfnamefont {J.~J.}\ \bibnamefont {Fredberg}}, \ and\ \bibinfo {author}
  {\bibfnamefont {J.~P.}\ \bibnamefont {Butler}},\ }\href {\doibase
  10.1242/jcs.187922} {\bibfield  {journal} {\bibinfo  {journal} {J. Cell
  Sci.}\ }\textbf {\bibinfo {volume} {129}},\ \bibinfo {pages} {3375} (\bibinfo
  {year} {2016})}\BibitemShut {NoStop}%
\bibitem [{\citenamefont {Thiery}(2002)}]{thiery2002}%
  \BibitemOpen
  \bibfield  {author} {\bibinfo {author} {\bibfnamefont {J.~P.}\ \bibnamefont
  {Thiery}},\ }\href {\doibase 10.1038/nrc822} {\bibfield  {journal} {\bibinfo
  {journal} {Nat. Rev. Cancer}\ }\textbf {\bibinfo {volume} {2}},\ \bibinfo
  {pages} {442} (\bibinfo {year} {2002})}\BibitemShut {NoStop}%
\bibitem [{\citenamefont {Bi}\ \emph {et~al.}(2016)\citenamefont {Bi},
  \citenamefont {Yang}, \citenamefont {Marchetti},\ and\ \citenamefont
  {Manning}}]{Bi2016}%
  \BibitemOpen
  \bibfield  {author} {\bibinfo {author} {\bibfnamefont {D.}~\bibnamefont
  {Bi}}, \bibinfo {author} {\bibfnamefont {X.}~\bibnamefont {Yang}}, \bibinfo
  {author} {\bibfnamefont {M.~C.}\ \bibnamefont {Marchetti}}, \ and\ \bibinfo
  {author} {\bibfnamefont {M.~L.}\ \bibnamefont {Manning}},\ }\href {\doibase
  10.1103/PhysRevX.6.021011} {\bibfield  {journal} {\bibinfo  {journal} {Phys.
  Rev. X}\ }\textbf {\bibinfo {volume} {6}},\ \bibinfo {pages} {021011}
  (\bibinfo {year} {2016})}\BibitemShut {NoStop}%
\bibitem [{\citenamefont {Mitchel}\ \emph {et~al.}(2020)\citenamefont
  {Mitchel}, \citenamefont {Das}, \citenamefont {O'Sullivan}, \citenamefont
  {Stancil}, \citenamefont {DeCamp}, \citenamefont {Koehler}, \citenamefont
  {Oca{\~{a}}}, \citenamefont {Butler}, \citenamefont {Fredberg}, \citenamefont
  {Nieto}, \citenamefont {Bi},\ and\ \citenamefont {Park}}]{mitchel2020}%
  \BibitemOpen
  \bibfield  {author} {\bibinfo {author} {\bibfnamefont {J.~A.}\ \bibnamefont
  {Mitchel}}, \bibinfo {author} {\bibfnamefont {A.}~\bibnamefont {Das}},
  \bibinfo {author} {\bibfnamefont {M.~J.}\ \bibnamefont {O'Sullivan}},
  \bibinfo {author} {\bibfnamefont {I.~T.}\ \bibnamefont {Stancil}}, \bibinfo
  {author} {\bibfnamefont {S.~J.}\ \bibnamefont {DeCamp}}, \bibinfo {author}
  {\bibfnamefont {S.}~\bibnamefont {Koehler}}, \bibinfo {author} {\bibfnamefont
  {O.~H.}\ \bibnamefont {Oca{\~{a}}}}, \bibinfo {author} {\bibfnamefont
  {J.~P.}\ \bibnamefont {Butler}}, \bibinfo {author} {\bibfnamefont {J.~J.}\
  \bibnamefont {Fredberg}}, \bibinfo {author} {\bibfnamefont {M.~A.}\
  \bibnamefont {Nieto}}, \bibinfo {author} {\bibfnamefont {D.}~\bibnamefont
  {Bi}}, \ and\ \bibinfo {author} {\bibfnamefont {J.-A.}\ \bibnamefont
  {Park}},\ }\href {\doibase 10.1038/s41467-020-18841-7} {\bibfield  {journal}
  {\bibinfo  {journal} {Nat. Commun.}\ }\textbf {\bibinfo {volume} {11}},\
  \bibinfo {pages} {5053} (\bibinfo {year} {2020})}\BibitemShut {NoStop}%
\bibitem [{\citenamefont {Janssen}(2019)}]{Janssen2019}%
  \BibitemOpen
  \bibfield  {author} {\bibinfo {author} {\bibfnamefont {L.~M.~C.}\
  \bibnamefont {Janssen}},\ }\href {\doibase 10.1088/1361-648X/ab3e90}
  {\bibfield  {journal} {\bibinfo  {journal} {J. Phys.: Condens. Matter}\
  }\textbf {\bibinfo {volume} {31}},\ \bibinfo {pages} {503002} (\bibinfo
  {year} {2019})}\BibitemShut {NoStop}%
\bibitem [{\citenamefont {Paoluzzi}\ \emph {et~al.}(2022)\citenamefont
  {Paoluzzi}, \citenamefont {Levis},\ and\ \citenamefont
  {Pagonabarraga}}]{paoluzzi2022}%
  \BibitemOpen
  \bibfield  {author} {\bibinfo {author} {\bibfnamefont {M.}~\bibnamefont
  {Paoluzzi}}, \bibinfo {author} {\bibfnamefont {D.}~\bibnamefont {Levis}}, \
  and\ \bibinfo {author} {\bibfnamefont {I.}~\bibnamefont {Pagonabarraga}},\
  }\href {\doibase 10.1038/s42005-022-00886-3} {\bibfield  {journal} {\bibinfo
  {journal} {Commun Phys}\ }\textbf {\bibinfo {volume} {5}},\ \bibinfo {pages}
  {111} (\bibinfo {year} {2022})}\BibitemShut {NoStop}%
\bibitem [{\citenamefont {Palacci}\ \emph {et~al.}(2010)\citenamefont
  {Palacci}, \citenamefont {Cottin-Bizonne}, \citenamefont {Ybert},\ and\
  \citenamefont {Bocquet}}]{Palacci2010}%
  \BibitemOpen
  \bibfield  {author} {\bibinfo {author} {\bibfnamefont {J.}~\bibnamefont
  {Palacci}}, \bibinfo {author} {\bibfnamefont {C.}~\bibnamefont
  {Cottin-Bizonne}}, \bibinfo {author} {\bibfnamefont {C.}~\bibnamefont
  {Ybert}}, \ and\ \bibinfo {author} {\bibfnamefont {L.}~\bibnamefont
  {Bocquet}},\ }\href {\doibase 10.1103/PhysRevLett.105.088304} {\bibfield
  {journal} {\bibinfo  {journal} {Phys. Rev. Lett.}\ }\textbf {\bibinfo
  {volume} {105}},\ \bibinfo {pages} {088304} (\bibinfo {year}
  {2010})}\BibitemShut {NoStop}%
\bibitem [{\citenamefont {Szamel}(2014)}]{Szamel2014}%
  \BibitemOpen
  \bibfield  {author} {\bibinfo {author} {\bibfnamefont {G.}~\bibnamefont
  {Szamel}},\ }\href {\doibase 10.1103/PhysRevE.90.012111} {\bibfield
  {journal} {\bibinfo  {journal} {Phys. Rev. E}\ }\textbf {\bibinfo {volume}
  {90}},\ \bibinfo {pages} {012111} (\bibinfo {year} {2014})}\BibitemShut
  {NoStop}%
\bibitem [{\citenamefont {Han}\ \emph {et~al.}(2017)\citenamefont {Han},
  \citenamefont {Yan}, \citenamefont {Granick},\ and\ \citenamefont
  {Luijten}}]{Han2017}%
  \BibitemOpen
  \bibfield  {author} {\bibinfo {author} {\bibfnamefont {M.}~\bibnamefont
  {Han}}, \bibinfo {author} {\bibfnamefont {J.}~\bibnamefont {Yan}}, \bibinfo
  {author} {\bibfnamefont {S.}~\bibnamefont {Granick}}, \ and\ \bibinfo
  {author} {\bibfnamefont {E.}~\bibnamefont {Luijten}},\ }\href {\doibase
  10.1073/pnas.1706702114} {\bibfield  {journal} {\bibinfo  {journal} {Proc.
  Natl. Acad. Sci. (USA)}\ }\textbf {\bibinfo {volume} {114}},\ \bibinfo
  {pages} {7513} (\bibinfo {year} {2017})}\BibitemShut {NoStop}%
\bibitem [{\citenamefont {Fodor}\ \emph {et~al.}(2016)\citenamefont {Fodor},
  \citenamefont {Nardini}, \citenamefont {Cates}, \citenamefont {Tailleur},
  \citenamefont {Visco},\ and\ \citenamefont {van Wijland}}]{fodor2016}%
  \BibitemOpen
  \bibfield  {author} {\bibinfo {author} {\bibfnamefont {E.}~\bibnamefont
  {Fodor}}, \bibinfo {author} {\bibfnamefont {C.}~\bibnamefont {Nardini}},
  \bibinfo {author} {\bibfnamefont {M.~E.}\ \bibnamefont {Cates}}, \bibinfo
  {author} {\bibfnamefont {J.}~\bibnamefont {Tailleur}}, \bibinfo {author}
  {\bibfnamefont {P.}~\bibnamefont {Visco}}, \ and\ \bibinfo {author}
  {\bibfnamefont {F.}~\bibnamefont {van Wijland}},\ }\href {\doibase
  10.1103/PhysRevLett.117.038103} {\bibfield  {journal} {\bibinfo  {journal}
  {Phys. Rev. Lett.}\ }\textbf {\bibinfo {volume} {117}},\ \bibinfo {pages}
  {038103} (\bibinfo {year} {2016})}\BibitemShut {NoStop}%
\bibitem [{\citenamefont {Parisi}(2005)}]{parisi2005}%
  \BibitemOpen
  \bibfield  {author} {\bibinfo {author} {\bibfnamefont {G.}~\bibnamefont
  {Parisi}},\ }\href@noop {} {\bibfield  {journal} {\bibinfo  {journal}
  {Nature}\ }\textbf {\bibinfo {volume} {433}},\ \bibinfo {pages} {221}
  (\bibinfo {year} {2005})}\BibitemShut {NoStop}%
\bibitem [{\citenamefont {Berthier}(2014)}]{Berthier2014}%
  \BibitemOpen
  \bibfield  {author} {\bibinfo {author} {\bibfnamefont {L.}~\bibnamefont
  {Berthier}},\ }\href {\doibase 10.1103/PhysRevLett.112.220602} {\bibfield
  {journal} {\bibinfo  {journal} {Phys. Rev. Lett.}\ }\textbf {\bibinfo
  {volume} {112}},\ \bibinfo {pages} {220602} (\bibinfo {year}
  {2014})}\BibitemShut {NoStop}%
\bibitem [{\citenamefont {Mandal}\ \emph {et~al.}(2016)\citenamefont {Mandal},
  \citenamefont {Bhuyan}, \citenamefont {Rao},\ and\ \citenamefont
  {Dasgupta}}]{Mandal2016}%
  \BibitemOpen
  \bibfield  {author} {\bibinfo {author} {\bibfnamefont {R.}~\bibnamefont
  {Mandal}}, \bibinfo {author} {\bibfnamefont {P.~J.}\ \bibnamefont {Bhuyan}},
  \bibinfo {author} {\bibfnamefont {M.}~\bibnamefont {Rao}}, \ and\ \bibinfo
  {author} {\bibfnamefont {C.}~\bibnamefont {Dasgupta}},\ }\href {\doibase
  10.1039/c5sm02950c} {\bibfield  {journal} {\bibinfo  {journal} {Soft Matter}\
  }\textbf {\bibinfo {volume} {12}},\ \bibinfo {pages} {6268} (\bibinfo {year}
  {2016})}\BibitemShut {NoStop}%
\bibitem [{\citenamefont {Flenner}\ \emph {et~al.}(2016)\citenamefont
  {Flenner}, \citenamefont {Szamel},\ and\ \citenamefont
  {Berthier}}]{Flenner2016}%
  \BibitemOpen
  \bibfield  {author} {\bibinfo {author} {\bibfnamefont {E.}~\bibnamefont
  {Flenner}}, \bibinfo {author} {\bibfnamefont {G.}~\bibnamefont {Szamel}}, \
  and\ \bibinfo {author} {\bibfnamefont {L.}~\bibnamefont {Berthier}},\ }\href
  {\doibase 10.1039/c6sm01322h} {\bibfield  {journal} {\bibinfo  {journal}
  {Soft Matter}\ }\textbf {\bibinfo {volume} {12}},\ \bibinfo {pages} {7136}
  (\bibinfo {year} {2016})}\BibitemShut {NoStop}%
\bibitem [{\citenamefont {Ni}\ \emph {et~al.}(2013)\citenamefont {Ni},
  \citenamefont {Stuart},\ and\ \citenamefont {Dijkstra}}]{Ni2013}%
  \BibitemOpen
  \bibfield  {author} {\bibinfo {author} {\bibfnamefont {R.}~\bibnamefont
  {Ni}}, \bibinfo {author} {\bibfnamefont {M.~A.~C.}\ \bibnamefont {Stuart}}, \
  and\ \bibinfo {author} {\bibfnamefont {M.}~\bibnamefont {Dijkstra}},\ }\href
  {\doibase 10.1038/ncomms3704} {\bibfield  {journal} {\bibinfo  {journal}
  {Nat. Commun}\ }\textbf {\bibinfo {volume} {4}},\ \bibinfo {pages} {2704}
  (\bibinfo {year} {2013})}\BibitemShut {NoStop}%
\bibitem [{\citenamefont {Klongvessa}\ \emph
  {et~al.}(2019{\natexlab{a}})\citenamefont {Klongvessa}, \citenamefont
  {Ginot}, \citenamefont {Ybert}, \citenamefont {Cottin-Bizonne},\ and\
  \citenamefont {Leocmach}}]{Klongvessa2019a}%
  \BibitemOpen
  \bibfield  {author} {\bibinfo {author} {\bibfnamefont {N.}~\bibnamefont
  {Klongvessa}}, \bibinfo {author} {\bibfnamefont {F.}~\bibnamefont {Ginot}},
  \bibinfo {author} {\bibfnamefont {C.}~\bibnamefont {Ybert}}, \bibinfo
  {author} {\bibfnamefont {C.}~\bibnamefont {Cottin-Bizonne}}, \ and\ \bibinfo
  {author} {\bibfnamefont {M.}~\bibnamefont {Leocmach}},\ }\href {\doibase
  10.1103/PhysRevLett.123.248004} {\bibfield  {journal} {\bibinfo  {journal}
  {Phys. Rev. Lett.}\ }\textbf {\bibinfo {volume} {123}},\ \bibinfo {pages}
  {248004} (\bibinfo {year} {2019}{\natexlab{a}})}\BibitemShut {NoStop}%
\bibitem [{\citenamefont {Klongvessa}\ \emph
  {et~al.}(2019{\natexlab{b}})\citenamefont {Klongvessa}, \citenamefont
  {Ginot}, \citenamefont {Ybert}, \citenamefont {Cottin-Bizonne},\ and\
  \citenamefont {Leocmach}}]{Klongvessa2019b}%
  \BibitemOpen
  \bibfield  {author} {\bibinfo {author} {\bibfnamefont {N.}~\bibnamefont
  {Klongvessa}}, \bibinfo {author} {\bibfnamefont {F.}~\bibnamefont {Ginot}},
  \bibinfo {author} {\bibfnamefont {C.}~\bibnamefont {Ybert}}, \bibinfo
  {author} {\bibfnamefont {C.}~\bibnamefont {Cottin-Bizonne}}, \ and\ \bibinfo
  {author} {\bibfnamefont {M.}~\bibnamefont {Leocmach}},\ }\href {\doibase
  10.1103/PhysRevE.100.062603} {\bibfield  {journal} {\bibinfo  {journal}
  {Phys. Rev. E}\ }\textbf {\bibinfo {volume} {100}},\ \bibinfo {pages}
  {062603} (\bibinfo {year} {2019}{\natexlab{b}})}\BibitemShut {NoStop}%
\bibitem [{\citenamefont {Arora}\ \emph {et~al.}(2022)\citenamefont {Arora},
  \citenamefont {Sood},\ and\ \citenamefont {Ganapathy}}]{Arora2022}%
  \BibitemOpen
  \bibfield  {author} {\bibinfo {author} {\bibfnamefont {P.}~\bibnamefont
  {Arora}}, \bibinfo {author} {\bibfnamefont {A.~K.}\ \bibnamefont {Sood}}, \
  and\ \bibinfo {author} {\bibfnamefont {R.}~\bibnamefont {Ganapathy}},\ }\href
  {\doibase 10.1103/PhysRevLett.128.178002} {\bibfield  {journal} {\bibinfo
  {journal} {Phys. Rev. Lett.}\ }\textbf {\bibinfo {volume} {128}},\ \bibinfo
  {pages} {178002} (\bibinfo {year} {2022})}\BibitemShut {NoStop}%
\bibitem [{\citenamefont {Berthier}\ and\ \citenamefont
  {Kurchan}(2013)}]{Berthier2013}%
  \BibitemOpen
  \bibfield  {author} {\bibinfo {author} {\bibfnamefont {L.}~\bibnamefont
  {Berthier}}\ and\ \bibinfo {author} {\bibfnamefont {J.}~\bibnamefont
  {Kurchan}},\ }\href {\doibase 10.1038/nphys2592} {\bibfield  {journal}
  {\bibinfo  {journal} {Nat. Phys.}\ }\textbf {\bibinfo {volume} {9}},\
  \bibinfo {pages} {310} (\bibinfo {year} {2013})}\BibitemShut {NoStop}%
\bibitem [{\citenamefont {Szamel}(2016)}]{Szamel2016}%
  \BibitemOpen
  \bibfield  {author} {\bibinfo {author} {\bibfnamefont {G.}~\bibnamefont
  {Szamel}},\ }\href {\doibase 10.1103/PhysRevE.93.012603} {\bibfield
  {journal} {\bibinfo  {journal} {Phys. Rev. E}\ }\textbf {\bibinfo {volume}
  {93}},\ \bibinfo {pages} {012603} (\bibinfo {year} {2016})}\BibitemShut
  {NoStop}%
\bibitem [{\citenamefont {Liluashvili}\ \emph {et~al.}(2017)\citenamefont
  {Liluashvili}, \citenamefont {{\'{O}}nody},\ and\ \citenamefont
  {Voigtmann}}]{Liluashvili2017}%
  \BibitemOpen
  \bibfield  {author} {\bibinfo {author} {\bibfnamefont {A.}~\bibnamefont
  {Liluashvili}}, \bibinfo {author} {\bibfnamefont {J.}~\bibnamefont
  {{\'{O}}nody}}, \ and\ \bibinfo {author} {\bibfnamefont {T.}~\bibnamefont
  {Voigtmann}},\ }\href {\doibase 10.1103/PhysRevE.96.062608} {\bibfield
  {journal} {\bibinfo  {journal} {Phys. Rev. E}\ }\textbf {\bibinfo {volume}
  {96}},\ \bibinfo {pages} {062608} (\bibinfo {year} {2017})}\BibitemShut
  {NoStop}%
\bibitem [{\citenamefont {Feng}\ and\ \citenamefont {Hou}(2017)}]{Feng2017}%
  \BibitemOpen
  \bibfield  {author} {\bibinfo {author} {\bibfnamefont {M.}~\bibnamefont
  {Feng}}\ and\ \bibinfo {author} {\bibfnamefont {Z.}~\bibnamefont {Hou}},\
  }\href {\doibase 10.1039/C7SM00852J} {\bibfield  {journal} {\bibinfo
  {journal} {Soft Matter}\ }\textbf {\bibinfo {volume} {13}},\ \bibinfo {pages}
  {4464} (\bibinfo {year} {2017})}\BibitemShut {NoStop}%
\bibitem [{\citenamefont {Nandi}\ and\ \citenamefont {Gov}(2017)}]{Nandi2017}%
  \BibitemOpen
  \bibfield  {author} {\bibinfo {author} {\bibfnamefont {S.~K.}\ \bibnamefont
  {Nandi}}\ and\ \bibinfo {author} {\bibfnamefont {N.~S.}\ \bibnamefont
  {Gov}},\ }\href {\doibase 10.1039/C7SM01648D} {\bibfield  {journal} {\bibinfo
   {journal} {Soft Matter}\ }\textbf {\bibinfo {volume} {13}},\ \bibinfo
  {pages} {7609} (\bibinfo {year} {2017})}\BibitemShut {NoStop}%
\bibitem [{\citenamefont {Nandi}\ \emph {et~al.}(2018)\citenamefont {Nandi},
  \citenamefont {Mandal}, \citenamefont {Bhuyan}, \citenamefont {Dasgupta},
  \citenamefont {Rao},\ and\ \citenamefont {Gov}}]{Nandi2018}%
  \BibitemOpen
  \bibfield  {author} {\bibinfo {author} {\bibfnamefont {S.~K.}\ \bibnamefont
  {Nandi}}, \bibinfo {author} {\bibfnamefont {R.}~\bibnamefont {Mandal}},
  \bibinfo {author} {\bibfnamefont {P.~J.}\ \bibnamefont {Bhuyan}}, \bibinfo
  {author} {\bibfnamefont {C.}~\bibnamefont {Dasgupta}}, \bibinfo {author}
  {\bibfnamefont {M.}~\bibnamefont {Rao}}, \ and\ \bibinfo {author}
  {\bibfnamefont {N.~S.}\ \bibnamefont {Gov}},\ }\href {\doibase
  10.1073/pnas.1721324115} {\bibfield  {journal} {\bibinfo  {journal} {Proc.
  Natl. Acad. Sci. (USA)}\ }\textbf {\bibinfo {volume} {115}},\ \bibinfo
  {pages} {7688} (\bibinfo {year} {2018})}\BibitemShut {NoStop}%
\bibitem [{\citenamefont {Mandal}\ \emph {et~al.}(2022)\citenamefont {Mandal},
  \citenamefont {Nandi}, \citenamefont {Dasgupta}, \citenamefont {Sollich},\
  and\ \citenamefont {Gov}}]{Mandal2022}%
  \BibitemOpen
  \bibfield  {author} {\bibinfo {author} {\bibfnamefont {R.}~\bibnamefont
  {Mandal}}, \bibinfo {author} {\bibfnamefont {S.~K.}\ \bibnamefont {Nandi}},
  \bibinfo {author} {\bibfnamefont {C.}~\bibnamefont {Dasgupta}}, \bibinfo
  {author} {\bibfnamefont {P.}~\bibnamefont {Sollich}}, \ and\ \bibinfo
  {author} {\bibfnamefont {N.~S.}\ \bibnamefont {Gov}},\ }\href {\doibase
  10.1088/2399-6528/ac9c47} {\bibfield  {journal} {\bibinfo  {journal} {J.
  phys. commun.}\ }\textbf {\bibinfo {volume} {6}},\ \bibinfo {pages} {115001}
  (\bibinfo {year} {2022})}\BibitemShut {NoStop}%
\bibitem [{\citenamefont {Graner}\ and\ \citenamefont
  {Glazier}(1992)}]{Graner1992}%
  \BibitemOpen
  \bibfield  {author} {\bibinfo {author} {\bibfnamefont {F.}~\bibnamefont
  {Graner}}\ and\ \bibinfo {author} {\bibfnamefont {J.~A.}\ \bibnamefont
  {Glazier}},\ }\href {\doibase 10.1103/PhysRevLett.69.2013} {\bibfield
  {journal} {\bibinfo  {journal} {Phys. Rev. Lett.}\ }\textbf {\bibinfo
  {volume} {69}},\ \bibinfo {pages} {2013} (\bibinfo {year}
  {1992})}\BibitemShut {NoStop}%
\bibitem [{\citenamefont {Glazier}\ and\ \citenamefont
  {Graner}(1993)}]{Glazier1993}%
  \BibitemOpen
  \bibfield  {author} {\bibinfo {author} {\bibfnamefont {J.~A.}\ \bibnamefont
  {Glazier}}\ and\ \bibinfo {author} {\bibfnamefont {F.}~\bibnamefont
  {Graner}},\ }\href {\doibase 10.1103/PhysRevE.47.2128} {\bibfield  {journal}
  {\bibinfo  {journal} {Phys. Rev. E}\ }\textbf {\bibinfo {volume} {47}},\
  \bibinfo {pages} {2128} (\bibinfo {year} {1993})}\BibitemShut {NoStop}%
\bibitem [{\citenamefont {Hogeweg}(2000)}]{Hogeweg2000}%
  \BibitemOpen
  \bibfield  {author} {\bibinfo {author} {\bibfnamefont {P.}~\bibnamefont
  {Hogeweg}},\ }\href {\doibase 10.1006/jtbi.2000.1087} {\bibfield  {journal}
  {\bibinfo  {journal} {J. Theor. Biol.}\ }\textbf {\bibinfo {volume} {203}},\
  \bibinfo {pages} {317} (\bibinfo {year} {2000})}\BibitemShut {NoStop}%
\bibitem [{\citenamefont {Hirashima}\ \emph {et~al.}(2017)\citenamefont
  {Hirashima}, \citenamefont {Rens},\ and\ \citenamefont
  {Merks}}]{hirashima2017}%
  \BibitemOpen
  \bibfield  {author} {\bibinfo {author} {\bibfnamefont {T.}~\bibnamefont
  {Hirashima}}, \bibinfo {author} {\bibfnamefont {E.~G.}\ \bibnamefont {Rens}},
  \ and\ \bibinfo {author} {\bibfnamefont {R.~M.~H.}\ \bibnamefont {Merks}},\
  }\href {\doibase 10.1111/dgd.12358} {\bibfield  {journal} {\bibinfo
  {journal} {Develop. Growth Differ.}\ }\textbf {\bibinfo {volume} {59}},\
  \bibinfo {pages} {329} (\bibinfo {year} {2017})}\BibitemShut {NoStop}%
\bibitem [{\citenamefont {Chiang}\ and\ \citenamefont
  {Marenduzzo}(2016)}]{Chiang2016}%
  \BibitemOpen
  \bibfield  {author} {\bibinfo {author} {\bibfnamefont {M.}~\bibnamefont
  {Chiang}}\ and\ \bibinfo {author} {\bibfnamefont {D.}~\bibnamefont
  {Marenduzzo}},\ }\href {\doibase 10.1209/0295-5075/116/28009} {\bibfield
  {journal} {\bibinfo  {journal} {{EPL} (Europhysics Letters)}\ }\textbf
  {\bibinfo {volume} {116}},\ \bibinfo {pages} {28009} (\bibinfo {year}
  {2016})}\BibitemShut {NoStop}%
\bibitem [{\citenamefont {Sadhukhan}\ and\ \citenamefont
  {Nandi}(2021)}]{Sadhukhan2021}%
  \BibitemOpen
  \bibfield  {author} {\bibinfo {author} {\bibfnamefont {S.}~\bibnamefont
  {Sadhukhan}}\ and\ \bibinfo {author} {\bibfnamefont {S.~K.}\ \bibnamefont
  {Nandi}},\ }\href {\doibase 10.1103/PhysRevE.103.062403} {\bibfield
  {journal} {\bibinfo  {journal} {Phys. Rev. E}\ }\textbf {\bibinfo {volume}
  {103}},\ \bibinfo {pages} {062403} (\bibinfo {year} {2021})}\BibitemShut
  {NoStop}%
\bibitem [{\citenamefont {Honda}\ and\ \citenamefont
  {Eguchi}(1980)}]{Honda1980}%
  \BibitemOpen
  \bibfield  {author} {\bibinfo {author} {\bibfnamefont {H.}~\bibnamefont
  {Honda}}\ and\ \bibinfo {author} {\bibfnamefont {G.}~\bibnamefont {Eguchi}},\
  }\href {\doibase 10.1016/S0022-5193(80)80021-X} {\bibfield  {journal}
  {\bibinfo  {journal} {J. Theor. Biol.}\ }\textbf {\bibinfo {volume} {84}},\
  \bibinfo {pages} {575} (\bibinfo {year} {1980})}\BibitemShut {NoStop}%
\bibitem [{\citenamefont {Marder}(1987)}]{Marder1987}%
  \BibitemOpen
  \bibfield  {author} {\bibinfo {author} {\bibfnamefont {M.}~\bibnamefont
  {Marder}},\ }\href {\doibase 10.1103/PhysRevA.36.438} {\bibfield  {journal}
  {\bibinfo  {journal} {Phys. Rev. A}\ }\textbf {\bibinfo {volume} {36}},\
  \bibinfo {pages} {438(R)} (\bibinfo {year} {1987})}\BibitemShut {NoStop}%
\bibitem [{\citenamefont {Fletcher}\ \emph {et~al.}(2014)\citenamefont
  {Fletcher}, \citenamefont {Osterfield}, \citenamefont {Baker},\ and\
  \citenamefont {Shvartsman}}]{Fletcher2014}%
  \BibitemOpen
  \bibfield  {author} {\bibinfo {author} {\bibfnamefont {A.~G.}\ \bibnamefont
  {Fletcher}}, \bibinfo {author} {\bibfnamefont {M.}~\bibnamefont
  {Osterfield}}, \bibinfo {author} {\bibfnamefont {R.~E.}\ \bibnamefont
  {Baker}}, \ and\ \bibinfo {author} {\bibfnamefont {S.~Y.}\ \bibnamefont
  {Shvartsman}},\ }\href {\doibase 10.1016/j.bpj.2013.11.4498} {\bibfield
  {journal} {\bibinfo  {journal} {Biophys. J.}\ }\textbf {\bibinfo {volume}
  {106}},\ \bibinfo {pages} {2291} (\bibinfo {year} {2014})}\BibitemShut
  {NoStop}%
\bibitem [{\citenamefont {Li}\ \emph {et~al.}(2021)\citenamefont {Li},
  \citenamefont {Wei}, \citenamefont {Paoluzzi},\ and\ \citenamefont
  {Ciamarra}}]{Li2021}%
  \BibitemOpen
  \bibfield  {author} {\bibinfo {author} {\bibfnamefont {Y.-W.}\ \bibnamefont
  {Li}}, \bibinfo {author} {\bibfnamefont {L.~L.~Y.}\ \bibnamefont {Wei}},
  \bibinfo {author} {\bibfnamefont {M.}~\bibnamefont {Paoluzzi}}, \ and\
  \bibinfo {author} {\bibfnamefont {M.~P.}\ \bibnamefont {Ciamarra}},\ }\href
  {\doibase 10.1103/PhysRevE.103.022607} {\bibfield  {journal} {\bibinfo
  {journal} {Phys. Rev. E}\ }\textbf {\bibinfo {volume} {103}},\ \bibinfo
  {pages} {022607} (\bibinfo {year} {2021})}\BibitemShut {NoStop}%
\bibitem [{\citenamefont {Czajkowski}\ \emph {et~al.}(2019)\citenamefont
  {Czajkowski}, \citenamefont {Sussman}, \citenamefont {Marchetti},\ and\
  \citenamefont {Manning}}]{czajkowski2019}%
  \BibitemOpen
  \bibfield  {author} {\bibinfo {author} {\bibfnamefont {M.}~\bibnamefont
  {Czajkowski}}, \bibinfo {author} {\bibfnamefont {D.~M.}\ \bibnamefont
  {Sussman}}, \bibinfo {author} {\bibfnamefont {M.~C.}\ \bibnamefont
  {Marchetti}}, \ and\ \bibinfo {author} {\bibfnamefont {M.~L.}\ \bibnamefont
  {Manning}},\ }\href {\doibase 10.1039/C9SM00916G} {\bibfield  {journal}
  {\bibinfo  {journal} {Soft Matter}\ }\textbf {\bibinfo {volume} {15}},\
  \bibinfo {pages} {9133} (\bibinfo {year} {2019})}\BibitemShut {NoStop}%
\bibitem [{\citenamefont {Nonomura}(2012)}]{nonomura2012}%
  \BibitemOpen
  \bibfield  {author} {\bibinfo {author} {\bibfnamefont {M.}~\bibnamefont
  {Nonomura}},\ }\href {\doibase 10.1371/journal.pone.0033501} {\bibfield
  {journal} {\bibinfo  {journal} {PLoS ONE}\ }\textbf {\bibinfo {volume} {7}},\
  \bibinfo {pages} {e33501} (\bibinfo {year} {2012})}\BibitemShut {NoStop}%
\bibitem [{\citenamefont {Palmieri}\ \emph {et~al.}(2015)\citenamefont
  {Palmieri}, \citenamefont {Bresler}, \citenamefont {Wirtz},\ and\
  \citenamefont {Grant}}]{palmieri2015}%
  \BibitemOpen
  \bibfield  {author} {\bibinfo {author} {\bibfnamefont {B.}~\bibnamefont
  {Palmieri}}, \bibinfo {author} {\bibfnamefont {Y.}~\bibnamefont {Bresler}},
  \bibinfo {author} {\bibfnamefont {D.}~\bibnamefont {Wirtz}}, \ and\ \bibinfo
  {author} {\bibfnamefont {M.}~\bibnamefont {Grant}},\ }\href {\doibase
  10.1038/srep11745} {\bibfield  {journal} {\bibinfo  {journal} {Sci. Rep.}\
  }\textbf {\bibinfo {volume} {5}},\ \bibinfo {pages} {11745} (\bibinfo {year}
  {2015})}\BibitemShut {NoStop}%
\bibitem [{\citenamefont {Loewe}\ \emph {et~al.}(2020)\citenamefont {Loewe},
  \citenamefont {Chiang}, \citenamefont {Marenduzzo},\ and\ \citenamefont
  {Marchetti}}]{loewe2020}%
  \BibitemOpen
  \bibfield  {author} {\bibinfo {author} {\bibfnamefont {B.}~\bibnamefont
  {Loewe}}, \bibinfo {author} {\bibfnamefont {M.}~\bibnamefont {Chiang}},
  \bibinfo {author} {\bibfnamefont {D.}~\bibnamefont {Marenduzzo}}, \ and\
  \bibinfo {author} {\bibfnamefont {M.~C.}\ \bibnamefont {Marchetti}},\ }\href
  {\doibase 10.1103/PhysRevLett.125.038003} {\bibfield  {journal} {\bibinfo
  {journal} {Phys. Rev. Lett.}\ }\textbf {\bibinfo {volume} {125}},\ \bibinfo
  {pages} {038003} (\bibinfo {year} {2020})}\BibitemShut {NoStop}%
\bibitem [{\citenamefont {Bi}\ \emph {et~al.}(2014)\citenamefont {Bi},
  \citenamefont {Lopez}, \citenamefont {Schwarz},\ and\ \citenamefont
  {Manning}}]{bi2014}%
  \BibitemOpen
  \bibfield  {author} {\bibinfo {author} {\bibfnamefont {D.}~\bibnamefont
  {Bi}}, \bibinfo {author} {\bibfnamefont {J.~H.}\ \bibnamefont {Lopez}},
  \bibinfo {author} {\bibfnamefont {J.~M.}\ \bibnamefont {Schwarz}}, \ and\
  \bibinfo {author} {\bibfnamefont {M.~L.}\ \bibnamefont {Manning}},\ }\href
  {\doibase 10.1039/C3SM52893F} {\bibfield  {journal} {\bibinfo  {journal}
  {Soft Matter}\ }\textbf {\bibinfo {volume} {10}},\ \bibinfo {pages} {1885}
  (\bibinfo {year} {2014})}\BibitemShut {NoStop}%
\bibitem [{\citenamefont {Sussman}\ \emph {et~al.}(2018)\citenamefont
  {Sussman}, \citenamefont {Paoluzzi}, \citenamefont {Marchetti},\ and\
  \citenamefont {Manning}}]{Sussman2018}%
  \BibitemOpen
  \bibfield  {author} {\bibinfo {author} {\bibfnamefont {D.~M.}\ \bibnamefont
  {Sussman}}, \bibinfo {author} {\bibfnamefont {M.}~\bibnamefont {Paoluzzi}},
  \bibinfo {author} {\bibfnamefont {M.~C.}\ \bibnamefont {Marchetti}}, \ and\
  \bibinfo {author} {\bibfnamefont {M.~L.}\ \bibnamefont {Manning}},\ }\href
  {\doibase 10.1209/0295-5075/121/36001} {\bibfield  {journal} {\bibinfo
  {journal} {Europhys. Lett.}\ }\textbf {\bibinfo {volume} {121}},\ \bibinfo
  {pages} {36001} (\bibinfo {year} {2018})}\BibitemShut {NoStop}%
\bibitem [{\citenamefont {Sadhukhan}\ and\ \citenamefont
  {Nandi}(2022)}]{sadhukhan2022}%
  \BibitemOpen
  \bibfield  {author} {\bibinfo {author} {\bibfnamefont {S.}~\bibnamefont
  {Sadhukhan}}\ and\ \bibinfo {author} {\bibfnamefont {S.~K.}\ \bibnamefont
  {Nandi}},\ }\href {\doibase 10.7554/eLife.76406} {\bibfield  {journal}
  {\bibinfo  {journal} {eLife}\ }\textbf {\bibinfo {volume} {11}},\ \bibinfo
  {pages} {e76406} (\bibinfo {year} {2022})}\BibitemShut {NoStop}%
\bibitem [{\citenamefont {Lubchenko}\ and\ \citenamefont
  {Wolynes}(2007)}]{lubchenko2007}%
  \BibitemOpen
  \bibfield  {author} {\bibinfo {author} {\bibfnamefont {V.}~\bibnamefont
  {Lubchenko}}\ and\ \bibinfo {author} {\bibfnamefont {P.~G.}\ \bibnamefont
  {Wolynes}},\ }\href {\doibase 10.1146/annurev.physchem.58.032806.104653}
  {\bibfield  {journal} {\bibinfo  {journal} {Annu. Rev. Phys. Chem.}\ }\textbf
  {\bibinfo {volume} {58}},\ \bibinfo {pages} {235} (\bibinfo {year}
  {2007})}\BibitemShut {NoStop}%
\bibitem [{\citenamefont {Kirkpatrick}\ and\ \citenamefont
  {Thirumalai}(2015)}]{kirkpatrick2015}%
  \BibitemOpen
  \bibfield  {author} {\bibinfo {author} {\bibfnamefont {T.~R.}\ \bibnamefont
  {Kirkpatrick}}\ and\ \bibinfo {author} {\bibfnamefont {D.}~\bibnamefont
  {Thirumalai}},\ }\href {\doibase 10.1103/RevModPhys.87.183} {\bibfield
  {journal} {\bibinfo  {journal} {Rev. Mod. Phys.}\ }\textbf {\bibinfo {volume}
  {87}},\ \bibinfo {pages} {183} (\bibinfo {year} {2015})}\BibitemShut
  {NoStop}%
\bibitem [{\citenamefont {Biroli}\ and\ \citenamefont
  {Bouchaud}(2012)}]{Biroli2012}%
  \BibitemOpen
  \bibfield  {author} {\bibinfo {author} {\bibfnamefont {G.}~\bibnamefont
  {Biroli}}\ and\ \bibinfo {author} {\bibfnamefont {J.~P.}\ \bibnamefont
  {Bouchaud}},\ }in\ \href {\doibase 10.1002/9781118202470.ch2} {\emph
  {\bibinfo {booktitle} {Structural Glasses and Supercooled Liquids: Theory,
  Experiment, and Applications}}},\ \bibinfo {editor} {edited by\ \bibinfo
  {editor} {\bibfnamefont {P.~G.}\ \bibnamefont {Wolynes}}\ and\ \bibinfo
  {editor} {\bibfnamefont {V.}~\bibnamefont {Lubchenko}}}\ (\bibinfo {year}
  {2012})\BibitemShut {NoStop}%
\bibitem [{\citenamefont {Henkes}\ \emph {et~al.}(2020)\citenamefont {Henkes},
  \citenamefont {Kostanjevec}, \citenamefont {Collinson}, \citenamefont
  {Sknepnek},\ and\ \citenamefont {Bertin}}]{henkes2020}%
  \BibitemOpen
  \bibfield  {author} {\bibinfo {author} {\bibfnamefont {S.}~\bibnamefont
  {Henkes}}, \bibinfo {author} {\bibfnamefont {K.}~\bibnamefont {Kostanjevec}},
  \bibinfo {author} {\bibfnamefont {J.~M.}\ \bibnamefont {Collinson}}, \bibinfo
  {author} {\bibfnamefont {R.}~\bibnamefont {Sknepnek}}, \ and\ \bibinfo
  {author} {\bibfnamefont {E.}~\bibnamefont {Bertin}},\ }\href {\doibase
  10.1038/s41467-020-15164-5} {\bibfield  {journal} {\bibinfo  {journal}
  {Nature Communications}\ }\textbf {\bibinfo {volume} {11}},\ \bibinfo {pages}
  {1405} (\bibinfo {year} {2020})}\BibitemShut {NoStop}%
\bibitem [{\citenamefont {Li}\ \emph {et~al.}(2024)\citenamefont {Li},
  \citenamefont {Lei},\ and\ \citenamefont {qiang Ma}}]{li2024}%
  \BibitemOpen
  \bibfield  {author} {\bibinfo {author} {\bibfnamefont {Z.-Q.}\ \bibnamefont
  {Li}}, \bibinfo {author} {\bibfnamefont {Q.-L.}\ \bibnamefont {Lei}}, \ and\
  \bibinfo {author} {\bibfnamefont {Y.}~\bibnamefont {qiang Ma}},\ }\href@noop
  {} {\enquote {\bibinfo {title} {Fluidization and anomalous density
  fluctuations in epithelial tissues with pulsating activity},}\ } (\bibinfo
  {year} {2024}),\ \Eprint {http://arxiv.org/abs/2402.02981} {arXiv:2402.02981
  [cond-mat.soft]} \BibitemShut {NoStop}%
\bibitem [{\citenamefont {Weaire}\ and\ \citenamefont
  {Hutzler}(2001)}]{Weaire2001}%
  \BibitemOpen
  \bibfield  {author} {\bibinfo {author} {\bibfnamefont {D.}~\bibnamefont
  {Weaire}}\ and\ \bibinfo {author} {\bibfnamefont {S.}~\bibnamefont
  {Hutzler}},\ }\href@noop {} {\emph {\bibinfo {title} {Thephysicsof foams}}}\
  (\bibinfo  {publisher} {Oxford University Press},\ \bibinfo {year}
  {2001})\BibitemShut {NoStop}%
\bibitem [{\citenamefont {Albert}\ and\ \citenamefont
  {Schwarz}(2016)}]{albert2016}%
  \BibitemOpen
  \bibfield  {author} {\bibinfo {author} {\bibfnamefont {P.~J.}\ \bibnamefont
  {Albert}}\ and\ \bibinfo {author} {\bibfnamefont {U.~S.}\ \bibnamefont
  {Schwarz}},\ }\href {\doibase 10.1080/19336918.2016.1148864} {\bibfield
  {journal} {\bibinfo  {journal} {Cell Adhesion \& Migration}\ }\textbf
  {\bibinfo {volume} {10}},\ \bibinfo {pages} {516} (\bibinfo {year} {2016})},\
  \bibinfo {note} {pMID: 26838278}\BibitemShut {NoStop}%
\bibitem [{\citenamefont {Barton}\ \emph {et~al.}(2017)\citenamefont {Barton},
  \citenamefont {Henkes}, \citenamefont {Weijer},\ and\ \citenamefont
  {Sknepnek}}]{Barton2017}%
  \BibitemOpen
  \bibfield  {author} {\bibinfo {author} {\bibfnamefont {D.~L.}\ \bibnamefont
  {Barton}}, \bibinfo {author} {\bibfnamefont {S.}~\bibnamefont {Henkes}},
  \bibinfo {author} {\bibfnamefont {C.~J.}\ \bibnamefont {Weijer}}, \ and\
  \bibinfo {author} {\bibfnamefont {R.}~\bibnamefont {Sknepnek}},\ }\href
  {\doibase 10.1371/journal.pcbi.1005569} {\bibfield  {journal} {\bibinfo
  {journal} {Plos Comput. Biol.}\ }\textbf {\bibinfo {volume} {13}},\ \bibinfo
  {pages} {e1005569} (\bibinfo {year} {2017})}\BibitemShut {NoStop}%
\bibitem [{\citenamefont {Yang}\ \emph {et~al.}(2017)\citenamefont {Yang},
  \citenamefont {Bi}, \citenamefont {Czajkowski}, \citenamefont {Merkel},
  \citenamefont {Manning},\ and\ \citenamefont {Marchetti}}]{yang2017}%
  \BibitemOpen
  \bibfield  {author} {\bibinfo {author} {\bibfnamefont {X.}~\bibnamefont
  {Yang}}, \bibinfo {author} {\bibfnamefont {D.}~\bibnamefont {Bi}}, \bibinfo
  {author} {\bibfnamefont {M.}~\bibnamefont {Czajkowski}}, \bibinfo {author}
  {\bibfnamefont {M.}~\bibnamefont {Merkel}}, \bibinfo {author} {\bibfnamefont
  {M.~L.}\ \bibnamefont {Manning}}, \ and\ \bibinfo {author} {\bibfnamefont
  {M.~C.}\ \bibnamefont {Marchetti}},\ }\href {\doibase
  10.1073/pnas.1705921114} {\bibfield  {journal} {\bibinfo  {journal} {Proc.
  Natl. Acad. Sci. (USA)}\ }\textbf {\bibinfo {volume} {114}},\ \bibinfo
  {pages} {12663} (\bibinfo {year} {2017})}\BibitemShut {NoStop}%
\bibitem [{\citenamefont {Wolff}\ \emph {et~al.}(2019)\citenamefont {Wolff},
  \citenamefont {Davidson},\ and\ \citenamefont {Merks}}]{wolff2019}%
  \BibitemOpen
  \bibfield  {author} {\bibinfo {author} {\bibfnamefont {H.~B.}\ \bibnamefont
  {Wolff}}, \bibinfo {author} {\bibfnamefont {L.~A.}\ \bibnamefont {Davidson}},
  \ and\ \bibinfo {author} {\bibfnamefont {R.~M.~H.}\ \bibnamefont {Merks}},\
  }\href {\doibase 10.1007/s11538-019-00599-9} {\bibfield  {journal} {\bibinfo
  {journal} {Bul. Math. Biol.}\ }\textbf {\bibinfo {volume} {81}},\ \bibinfo
  {pages} {3322} (\bibinfo {year} {2019})}\BibitemShut {NoStop}%
\bibitem [{\citenamefont {Paoluzzi}\ \emph {et~al.}(2021)\citenamefont
  {Paoluzzi}, \citenamefont {Angelani}, \citenamefont {Gosti}, \citenamefont
  {Marchetti}, \citenamefont {Pagonabarraga},\ and\ \citenamefont
  {Ruocco}}]{paoluzzi2021}%
  \BibitemOpen
  \bibfield  {author} {\bibinfo {author} {\bibfnamefont {M.}~\bibnamefont
  {Paoluzzi}}, \bibinfo {author} {\bibfnamefont {L.}~\bibnamefont {Angelani}},
  \bibinfo {author} {\bibfnamefont {G.}~\bibnamefont {Gosti}}, \bibinfo
  {author} {\bibfnamefont {M.~C.}\ \bibnamefont {Marchetti}}, \bibinfo {author}
  {\bibfnamefont {I.}~\bibnamefont {Pagonabarraga}}, \ and\ \bibinfo {author}
  {\bibfnamefont {G.}~\bibnamefont {Ruocco}},\ }\href {\doibase
  10.1103/PhysRevE.104.044606} {\bibfield  {journal} {\bibinfo  {journal}
  {Phys. Rev. E}\ }\textbf {\bibinfo {volume} {104}},\ \bibinfo {pages}
  {044606} (\bibinfo {year} {2021})}\BibitemShut {NoStop}%
\bibitem [{\citenamefont {Rivas}\ \emph {et~al.}(2020)\citenamefont {Rivas},
  \citenamefont {Shendruk}, \citenamefont {Henry}, \citenamefont {Reich},\ and\
  \citenamefont {Leheny}}]{rivas2020}%
  \BibitemOpen
  \bibfield  {author} {\bibinfo {author} {\bibfnamefont {D.~P.}\ \bibnamefont
  {Rivas}}, \bibinfo {author} {\bibfnamefont {T.~N.}\ \bibnamefont {Shendruk}},
  \bibinfo {author} {\bibfnamefont {R.~R.}\ \bibnamefont {Henry}}, \bibinfo
  {author} {\bibfnamefont {D.~H.}\ \bibnamefont {Reich}}, \ and\ \bibinfo
  {author} {\bibfnamefont {R.~L.}\ \bibnamefont {Leheny}},\ }\href {\doibase
  10.1039/D0SM00693A} {\bibfield  {journal} {\bibinfo  {journal} {Soft Matter}\
  }\textbf {\bibinfo {volume} {16}},\ \bibinfo {pages} {9331} (\bibinfo {year}
  {2020})}\BibitemShut {NoStop}%
\bibitem [{\citenamefont {Atia}\ \emph {et~al.}(2018)\citenamefont {Atia},
  \citenamefont {Bi}, \citenamefont {Sharma}, \citenamefont {Mitchel},
  \citenamefont {Gweon}, \citenamefont {A.~Koehler}, \citenamefont {DeCamp},
  \citenamefont {Lan}, \citenamefont {Kim}, \citenamefont {Hirsch},
  \citenamefont {Pegoraro}, \citenamefont {Lee}, \citenamefont {Starr},
  \citenamefont {Weitz}, \citenamefont {Martin}, \citenamefont {Park},
  \citenamefont {Butler},\ and\ \citenamefont {Fredberg}}]{Atia2018}%
  \BibitemOpen
  \bibfield  {author} {\bibinfo {author} {\bibfnamefont {L.}~\bibnamefont
  {Atia}}, \bibinfo {author} {\bibfnamefont {D.}~\bibnamefont {Bi}}, \bibinfo
  {author} {\bibfnamefont {Y.}~\bibnamefont {Sharma}}, \bibinfo {author}
  {\bibfnamefont {J.~A.}\ \bibnamefont {Mitchel}}, \bibinfo {author}
  {\bibfnamefont {B.}~\bibnamefont {Gweon}}, \bibinfo {author} {\bibfnamefont
  {S.}~\bibnamefont {A.~Koehler}}, \bibinfo {author} {\bibfnamefont {S.~J.}\
  \bibnamefont {DeCamp}}, \bibinfo {author} {\bibfnamefont {B.}~\bibnamefont
  {Lan}}, \bibinfo {author} {\bibfnamefont {J.~H.}\ \bibnamefont {Kim}},
  \bibinfo {author} {\bibfnamefont {R.}~\bibnamefont {Hirsch}}, \bibinfo
  {author} {\bibfnamefont {A.~F.}\ \bibnamefont {Pegoraro}}, \bibinfo {author}
  {\bibfnamefont {K.~H.}\ \bibnamefont {Lee}}, \bibinfo {author} {\bibfnamefont
  {J.~R.}\ \bibnamefont {Starr}}, \bibinfo {author} {\bibfnamefont {D.~A.}\
  \bibnamefont {Weitz}}, \bibinfo {author} {\bibfnamefont {A.~C.}\ \bibnamefont
  {Martin}}, \bibinfo {author} {\bibfnamefont {J.-A.}\ \bibnamefont {Park}},
  \bibinfo {author} {\bibfnamefont {J.~P.}\ \bibnamefont {Butler}}, \ and\
  \bibinfo {author} {\bibfnamefont {J.~J.}\ \bibnamefont {Fredberg}},\ }\href
  {\doibase 10.1038/s41567-018-0089-9} {\bibfield  {journal} {\bibinfo
  {journal} {Nat. Phys.}\ }\textbf {\bibinfo {volume} {14}},\ \bibinfo {pages}
  {613} (\bibinfo {year} {2018})}\BibitemShut {NoStop}%
\bibitem [{\citenamefont {Mandal}\ \emph {et~al.}(2020)\citenamefont {Mandal},
  \citenamefont {Bhuyan}, \citenamefont {Chaudhuri}, \citenamefont {Dasgupta},\
  and\ \citenamefont {Rao}}]{mandal2020}%
  \BibitemOpen
  \bibfield  {author} {\bibinfo {author} {\bibfnamefont {R.}~\bibnamefont
  {Mandal}}, \bibinfo {author} {\bibfnamefont {P.~J.}\ \bibnamefont {Bhuyan}},
  \bibinfo {author} {\bibfnamefont {P.}~\bibnamefont {Chaudhuri}}, \bibinfo
  {author} {\bibfnamefont {C.}~\bibnamefont {Dasgupta}}, \ and\ \bibinfo
  {author} {\bibfnamefont {M.}~\bibnamefont {Rao}},\ }\href {\doibase
  10.1038/s41467-020-16130-x} {\bibfield  {journal} {\bibinfo  {journal} {Nat.
  Comm.}\ }\textbf {\bibinfo {volume} {11}},\ \bibinfo {pages} {2581} (\bibinfo
  {year} {2020})}\BibitemShut {NoStop}%
\bibitem [{\citenamefont {Keta}\ \emph {et~al.}(2022)\citenamefont {Keta},
  \citenamefont {Jack},\ and\ \citenamefont {Berthier}}]{keta2022}%
  \BibitemOpen
  \bibfield  {author} {\bibinfo {author} {\bibfnamefont {Y.-E.}\ \bibnamefont
  {Keta}}, \bibinfo {author} {\bibfnamefont {R.~L.}\ \bibnamefont {Jack}}, \
  and\ \bibinfo {author} {\bibfnamefont {L.}~\bibnamefont {Berthier}},\ }\href
  {\doibase 10.1103/physrevlett.129.048002} {\bibfield  {journal} {\bibinfo
  {journal} {Physical Review Letters}\ }\textbf {\bibinfo {volume} {129}},\
  \bibinfo {pages} {048002} (\bibinfo {year} {2022})}\BibitemShut {NoStop}%
\bibitem [{\citenamefont {Paul}\ \emph {et~al.}(2023)\citenamefont {Paul},
  \citenamefont {Mutneja}, \citenamefont {Nandi},\ and\ \citenamefont
  {Karmakar}}]{paul2023}%
  \BibitemOpen
  \bibfield  {author} {\bibinfo {author} {\bibfnamefont {K.}~\bibnamefont
  {Paul}}, \bibinfo {author} {\bibfnamefont {A.}~\bibnamefont {Mutneja}},
  \bibinfo {author} {\bibfnamefont {S.~K.}\ \bibnamefont {Nandi}}, \ and\
  \bibinfo {author} {\bibfnamefont {S.}~\bibnamefont {Karmakar}},\ }\href
  {\doibase 10.1073/pnas.221707312} {\bibfield  {journal} {\bibinfo  {journal}
  {Proc. Natl. Acad. Sci. (USA)}\ }\textbf {\bibinfo {volume} {120}},\ \bibinfo
  {pages} {e2217073120} (\bibinfo {year} {2023})}\BibitemShut {NoStop}%
\bibitem [{\citenamefont {Paul}\ \emph {et~al.}(2021)\citenamefont {Paul},
  \citenamefont {Nandi},\ and\ \citenamefont {Karmakar}}]{Paul2021b}%
  \BibitemOpen
  \bibfield  {author} {\bibinfo {author} {\bibfnamefont {K.}~\bibnamefont
  {Paul}}, \bibinfo {author} {\bibfnamefont {S.~K.}\ \bibnamefont {Nandi}}, \
  and\ \bibinfo {author} {\bibfnamefont {S.}~\bibnamefont {Karmakar}},\ }\href
  {\doibase 10.48550/ARXIV.2111.09829} {\bibfield  {journal} {\bibinfo
  {journal} {arXiv}\ ,\ \bibinfo {pages} {2111.09829}} (\bibinfo {year}
  {2021})}\BibitemShut {NoStop}%
\bibitem [{\citenamefont {Kirkpatrick}\ and\ \citenamefont
  {Wolynes}(1987)}]{Kirkpatrick1987}%
  \BibitemOpen
  \bibfield  {author} {\bibinfo {author} {\bibfnamefont {T.~R.}\ \bibnamefont
  {Kirkpatrick}}\ and\ \bibinfo {author} {\bibfnamefont {P.~G.}\ \bibnamefont
  {Wolynes}},\ }\href {\doibase 10.1103/PhysRevA.35.3072} {\bibfield  {journal}
  {\bibinfo  {journal} {Phys. Rev. A}\ }\textbf {\bibinfo {volume} {35}},\
  \bibinfo {pages} {3072} (\bibinfo {year} {1987})}\BibitemShut {NoStop}%
\bibitem [{\citenamefont {Kirkpatrick}\ \emph {et~al.}(1989)\citenamefont
  {Kirkpatrick}, \citenamefont {Thirumalai},\ and\ \citenamefont
  {Wolynes}}]{Kirkpatrick1989}%
  \BibitemOpen
  \bibfield  {author} {\bibinfo {author} {\bibfnamefont {T.~R.}\ \bibnamefont
  {Kirkpatrick}}, \bibinfo {author} {\bibfnamefont {D.}~\bibnamefont
  {Thirumalai}}, \ and\ \bibinfo {author} {\bibfnamefont {P.~G.}\ \bibnamefont
  {Wolynes}},\ }\href {\doibase 10.1103/PhysRevA.40.1045} {\bibfield  {journal}
  {\bibinfo  {journal} {Phys. Rev. A}\ }\textbf {\bibinfo {volume} {40}},\
  \bibinfo {pages} {1045} (\bibinfo {year} {1989})}\BibitemShut {NoStop}%
\bibitem [{\citenamefont {Wolynes}\ and\ \citenamefont
  {Lubchenko}(2012)}]{wolynesbook}%
  \BibitemOpen
  \bibfield  {author} {\bibinfo {author} {\bibfnamefont {P.~G.}\ \bibnamefont
  {Wolynes}}\ and\ \bibinfo {author} {\bibfnamefont {V.}~\bibnamefont
  {Lubchenko}},\ }\href@noop {} {\emph {\bibinfo {title} {Structural Glasses
  and Supercooled Liquids}}}\ (\bibinfo  {publisher} {John Wiley and Sons,
  Inc., Hoboken, New Jersey},\ \bibinfo {year} {2012})\BibitemShut {NoStop}%
\bibitem [{\citenamefont {Parisi}\ and\ \citenamefont
  {Zamponi}(2010)}]{parisi2010}%
  \BibitemOpen
  \bibfield  {author} {\bibinfo {author} {\bibfnamefont {G.}~\bibnamefont
  {Parisi}}\ and\ \bibinfo {author} {\bibfnamefont {F.}~\bibnamefont
  {Zamponi}},\ }\href {\doibase 10.1103/RevModPhys.82.789} {\bibfield
  {journal} {\bibinfo  {journal} {Rev. Mod. Phys.}\ }\textbf {\bibinfo {volume}
  {82}},\ \bibinfo {pages} {789} (\bibinfo {year} {2010})}\BibitemShut
  {NoStop}%
\bibitem [{\citenamefont {Kauzmann}(1948)}]{Kauzmann1948}%
  \BibitemOpen
  \bibfield  {author} {\bibinfo {author} {\bibfnamefont {W.}~\bibnamefont
  {Kauzmann}},\ }\href {\doibase 10.1021/cr60135a002} {\bibfield  {journal}
  {\bibinfo  {journal} {Chemical Reviews}\ }\textbf {\bibinfo {volume} {43}},\
  \bibinfo {pages} {219} (\bibinfo {year} {1948})}\BibitemShut {NoStop}%
\bibitem [{\citenamefont {Angell}(1991)}]{angell1991a}%
  \BibitemOpen
  \bibfield  {author} {\bibinfo {author} {\bibfnamefont {C.~A.}\ \bibnamefont
  {Angell}},\ }\href {\doibase 10.1016/0022-3093(91)90266-9} {\bibfield
  {journal} {\bibinfo  {journal} {J. Non-Cryst. Solids}\ }\textbf {\bibinfo
  {volume} {131-133}},\ \bibinfo {pages} {13} (\bibinfo {year}
  {1991})}\BibitemShut {NoStop}%
\bibitem [{\citenamefont {Angell}(1995)}]{angell1995b}%
  \BibitemOpen
  \bibfield  {author} {\bibinfo {author} {\bibfnamefont {C.~A.}\ \bibnamefont
  {Angell}},\ }\href {\doibase 10.1126/science.267.5206.1924} {\bibfield
  {journal} {\bibinfo  {journal} {Science}\ }\textbf {\bibinfo {volume}
  {267}},\ \bibinfo {pages} {1924} (\bibinfo {year} {1995})}\BibitemShut
  {NoStop}%
\bibitem [{\citenamefont {Zhang}\ \emph {et~al.}(2020)\citenamefont {Zhang},
  \citenamefont {Mueller}, \citenamefont {Doostmohammadi},\ and\ \citenamefont
  {Yeomans}}]{zhang2020}%
  \BibitemOpen
  \bibfield  {author} {\bibinfo {author} {\bibfnamefont {G.}~\bibnamefont
  {Zhang}}, \bibinfo {author} {\bibfnamefont {R.}~\bibnamefont {Mueller}},
  \bibinfo {author} {\bibfnamefont {A.}~\bibnamefont {Doostmohammadi}}, \ and\
  \bibinfo {author} {\bibfnamefont {J.~M.}\ \bibnamefont {Yeomans}},\ }\href
  {\doibase 10.1098/rsif.2020.0312} {\bibfield  {journal} {\bibinfo  {journal}
  {J. Royal Soc. Interface}\ }\textbf {\bibinfo {volume} {17}},\ \bibinfo
  {pages} {20200312} (\bibinfo {year} {2020})}\BibitemShut {NoStop}%
\bibitem [{\citenamefont {Chiang}\ \emph {et~al.}(2023)\citenamefont {Chiang},
  \citenamefont {Hopkins}, \citenamefont {Loewe}, \citenamefont {Marchetti},\
  and\ \citenamefont {Marenduzzo}}]{chiang2023}%
  \BibitemOpen
  \bibfield  {author} {\bibinfo {author} {\bibfnamefont {M.}~\bibnamefont
  {Chiang}}, \bibinfo {author} {\bibfnamefont {A.}~\bibnamefont {Hopkins}},
  \bibinfo {author} {\bibfnamefont {B.}~\bibnamefont {Loewe}}, \bibinfo
  {author} {\bibfnamefont {M.~C.}\ \bibnamefont {Marchetti}}, \ and\ \bibinfo
  {author} {\bibfnamefont {D.}~\bibnamefont {Marenduzzo}},\ }\href {\doibase
  10.48550/arXiv.2310.20465} {\bibfield  {journal} {\bibinfo  {journal}
  {arXiv}\ ,\ \bibinfo {pages} {2310.20465}} (\bibinfo {year}
  {2023})}\BibitemShut {NoStop}%
\bibitem [{\citenamefont {Gysin}\ \emph {et~al.}(2011)\citenamefont {Gysin},
  \citenamefont {Salt}, \citenamefont {Young},\ and\ \citenamefont
  {McCormick}}]{gysin2011}%
  \BibitemOpen
  \bibfield  {author} {\bibinfo {author} {\bibfnamefont {S.}~\bibnamefont
  {Gysin}}, \bibinfo {author} {\bibfnamefont {M.}~\bibnamefont {Salt}},
  \bibinfo {author} {\bibfnamefont {A.}~\bibnamefont {Young}}, \ and\ \bibinfo
  {author} {\bibfnamefont {F.}~\bibnamefont {McCormick}},\ }\href {\doibase
  10.1177/1947601911412376} {\bibfield  {journal} {\bibinfo  {journal} {Genes
  and Cancer}\ }\textbf {\bibinfo {volume} {2}},\ \bibinfo {pages} {359}
  (\bibinfo {year} {2011})}\BibitemShut {NoStop}%
\bibitem [{\citenamefont {Rodrigues}\ \emph {et~al.}(2019)\citenamefont
  {Rodrigues}, \citenamefont {Kosaric}, \citenamefont {Bonham},\ and\
  \citenamefont {Gurtner}}]{rodrigues2019}%
  \BibitemOpen
  \bibfield  {author} {\bibinfo {author} {\bibfnamefont {M.}~\bibnamefont
  {Rodrigues}}, \bibinfo {author} {\bibfnamefont {N.}~\bibnamefont {Kosaric}},
  \bibinfo {author} {\bibfnamefont {C.~A.}\ \bibnamefont {Bonham}}, \ and\
  \bibinfo {author} {\bibfnamefont {G.~C.}\ \bibnamefont {Gurtner}},\ }\href
  {\doibase 10.1152/physrev.00067.2017} {\bibfield  {journal} {\bibinfo
  {journal} {Physiol. Rev.}\ }\textbf {\bibinfo {volume} {99}},\ \bibinfo
  {pages} {665} (\bibinfo {year} {2019})}\BibitemShut {NoStop}%
\bibitem [{\citenamefont {Bi}\ \emph {et~al.}(2015)\citenamefont {Bi},
  \citenamefont {Lopez}, \citenamefont {Schwarz},\ and\ \citenamefont
  {Manning}}]{bi2015}%
  \BibitemOpen
  \bibfield  {author} {\bibinfo {author} {\bibfnamefont {D.}~\bibnamefont
  {Bi}}, \bibinfo {author} {\bibfnamefont {J.~H.}\ \bibnamefont {Lopez}},
  \bibinfo {author} {\bibfnamefont {J.~M.}\ \bibnamefont {Schwarz}}, \ and\
  \bibinfo {author} {\bibfnamefont {M.~L.}\ \bibnamefont {Manning}},\ }\href
  {\doibase 10.1038/NPHYS3471} {\bibfield  {journal} {\bibinfo  {journal} {Nat.
  Phys.}\ }\textbf {\bibinfo {volume} {11}},\ \bibinfo {pages} {1074} (\bibinfo
  {year} {2015})}\BibitemShut {NoStop}%
\bibitem [{\citenamefont {Saraswathibhatla}\ and\ \citenamefont
  {Notbohm}(2020)}]{saraswathibhatla2020}%
  \BibitemOpen
  \bibfield  {author} {\bibinfo {author} {\bibfnamefont {A.}~\bibnamefont
  {Saraswathibhatla}}\ and\ \bibinfo {author} {\bibfnamefont {J.}~\bibnamefont
  {Notbohm}},\ }\href {\doibase 10.1103/PhysRevX.10.011016} {\bibfield
  {journal} {\bibinfo  {journal} {Phys. Rev. X}\ }\textbf {\bibinfo {volume}
  {10}},\ \bibinfo {pages} {011016} (\bibinfo {year} {2020})}\BibitemShut
  {NoStop}%
\bibitem [{\citenamefont {Saraswathibhatla}\ \emph {et~al.}(2021)\citenamefont
  {Saraswathibhatla}, \citenamefont {Henkes}, \citenamefont {Galles},
  \citenamefont {Sknepnek},\ and\ \citenamefont
  {Notbohm}}]{saraswathibhatla2021}%
  \BibitemOpen
  \bibfield  {author} {\bibinfo {author} {\bibfnamefont {A.}~\bibnamefont
  {Saraswathibhatla}}, \bibinfo {author} {\bibfnamefont {S.}~\bibnamefont
  {Henkes}}, \bibinfo {author} {\bibfnamefont {E.~E.}\ \bibnamefont {Galles}},
  \bibinfo {author} {\bibfnamefont {R.}~\bibnamefont {Sknepnek}}, \ and\
  \bibinfo {author} {\bibfnamefont {J.}~\bibnamefont {Notbohm}},\ }\href
  {\doibase https://doi.org/10.1016/j.eml.2021.101438} {\bibfield  {journal}
  {\bibinfo  {journal} {Extreme Mechanics Letters}\ }\textbf {\bibinfo {volume}
  {48}},\ \bibinfo {pages} {101438} (\bibinfo {year} {2021})}\BibitemShut
  {NoStop}%
\bibitem [{\citenamefont {Bazelli{\`{e}}res}\ \emph {et~al.}(2015)\citenamefont
  {Bazelli{\`{e}}res}, \citenamefont {Conte}, \citenamefont {Elosegui-Artola},
  \citenamefont {Serra-Picamal}, \citenamefont {Bintanel-Morcillo},
  \citenamefont {Roca-Cusachs}, \citenamefont {Mu{\~{n}}oz}, \citenamefont
  {Sales-Pardo}, \citenamefont {Guimer{\`{a}}},\ and\ \citenamefont
  {Trepat}}]{bazellieres2015}%
  \BibitemOpen
  \bibfield  {author} {\bibinfo {author} {\bibfnamefont {E.}~\bibnamefont
  {Bazelli{\`{e}}res}}, \bibinfo {author} {\bibfnamefont {V.}~\bibnamefont
  {Conte}}, \bibinfo {author} {\bibfnamefont {A.}~\bibnamefont
  {Elosegui-Artola}}, \bibinfo {author} {\bibfnamefont {X.}~\bibnamefont
  {Serra-Picamal}}, \bibinfo {author} {\bibfnamefont {M.}~\bibnamefont
  {Bintanel-Morcillo}}, \bibinfo {author} {\bibfnamefont {P.}~\bibnamefont
  {Roca-Cusachs}}, \bibinfo {author} {\bibfnamefont {J.~J.}\ \bibnamefont
  {Mu{\~{n}}oz}}, \bibinfo {author} {\bibfnamefont {M.}~\bibnamefont
  {Sales-Pardo}}, \bibinfo {author} {\bibfnamefont {R.}~\bibnamefont
  {Guimer{\`{a}}}}, \ and\ \bibinfo {author} {\bibfnamefont {X.}~\bibnamefont
  {Trepat}},\ }\href {\doibase 10.1038/ncb3135} {\bibfield  {journal} {\bibinfo
   {journal} {Nat. Cell Biol.}\ }\textbf {\bibinfo {volume} {17}},\ \bibinfo
  {pages} {409} (\bibinfo {year} {2015})}\BibitemShut {NoStop}%
\bibitem [{\citenamefont {Pandey}\ \emph {et~al.}(2023)\citenamefont {Pandey},
  \citenamefont {Kolya}, \citenamefont {Sadhukhan},\ and\ \citenamefont
  {Nandi}}]{pandey2023}%
  \BibitemOpen
  \bibfield  {author} {\bibinfo {author} {\bibfnamefont {S.}~\bibnamefont
  {Pandey}}, \bibinfo {author} {\bibfnamefont {S.}~\bibnamefont {Kolya}},
  \bibinfo {author} {\bibfnamefont {S.}~\bibnamefont {Sadhukhan}}, \ and\
  \bibinfo {author} {\bibfnamefont {S.~K.}\ \bibnamefont {Nandi}},\ }\href
  {\doibase 10.48550/arXiv.2306.07250} {\bibfield  {journal} {\bibinfo
  {journal} {arXiv}\ }\textbf {\bibinfo {volume} {v1}},\ \bibinfo {pages}
  {2306.07250} (\bibinfo {year} {2023})}\BibitemShut {NoStop}%
\bibitem [{\citenamefont {Arora}\ \emph {et~al.}(2024)\citenamefont {Arora},
  \citenamefont {Sadhukhan}, \citenamefont {Nandi}, \citenamefont {Bi},
  \citenamefont {Sood},\ and\ \citenamefont {Ganapathy}}]{arora2024}%
  \BibitemOpen
  \bibfield  {author} {\bibinfo {author} {\bibfnamefont {P.}~\bibnamefont
  {Arora}}, \bibinfo {author} {\bibfnamefont {S.}~\bibnamefont {Sadhukhan}},
  \bibinfo {author} {\bibfnamefont {S.~K.}\ \bibnamefont {Nandi}}, \bibinfo
  {author} {\bibfnamefont {D.}~\bibnamefont {Bi}}, \bibinfo {author}
  {\bibfnamefont {A.~K.}\ \bibnamefont {Sood}}, \ and\ \bibinfo {author}
  {\bibfnamefont {R.}~\bibnamefont {Ganapathy}},\ }\href {\doibase
  10.48550/arXiv.2401.13437} {\bibfield  {journal} {\bibinfo  {journal}
  {arXiv}\ ,\ \bibinfo {pages} {2401.13437}} (\bibinfo {year}
  {2024})}\BibitemShut {NoStop}%
\bibitem [{\citenamefont {Tarle}\ \emph {et~al.}(2015)\citenamefont {Tarle},
  \citenamefont {Ravasio}, \citenamefont {Hakim},\ and\ \citenamefont
  {Gov}}]{tarle2015}%
  \BibitemOpen
  \bibfield  {author} {\bibinfo {author} {\bibfnamefont {V.}~\bibnamefont
  {Tarle}}, \bibinfo {author} {\bibfnamefont {A.}~\bibnamefont {Ravasio}},
  \bibinfo {author} {\bibfnamefont {V.}~\bibnamefont {Hakim}}, \ and\ \bibinfo
  {author} {\bibfnamefont {N.~S.}\ \bibnamefont {Gov}},\ }\href {\doibase
  10.1039/c5ib00092k} {\bibfield  {journal} {\bibinfo  {journal} {Int. Biol.}\
  }\textbf {\bibinfo {volume} {7}},\ \bibinfo {pages} {1218} (\bibinfo {year}
  {2015})}\BibitemShut {NoStop}%
\bibitem [{\citenamefont {Sarkar}\ \emph {et~al.}(2021)\citenamefont {Sarkar},
  \citenamefont {Gompper},\ and\ \citenamefont {Elgeti}}]{sarkar2021}%
  \BibitemOpen
  \bibfield  {author} {\bibinfo {author} {\bibfnamefont {D.}~\bibnamefont
  {Sarkar}}, \bibinfo {author} {\bibfnamefont {G.}~\bibnamefont {Gompper}}, \
  and\ \bibinfo {author} {\bibfnamefont {J.}~\bibnamefont {Elgeti}},\ }\href
  {\doibase 10.1038/s42005-020-00515-x} {\bibfield  {journal} {\bibinfo
  {journal} {Comm. Phys.}\ }\textbf {\bibinfo {volume} {4}},\ \bibinfo {pages}
  {36} (\bibinfo {year} {2021})}\BibitemShut {NoStop}%
\bibitem [{\citenamefont {Xia}\ and\ \citenamefont {Wolynes}(2000)}]{xia2000}%
  \BibitemOpen
  \bibfield  {author} {\bibinfo {author} {\bibfnamefont {X.}~\bibnamefont
  {Xia}}\ and\ \bibinfo {author} {\bibfnamefont {P.~G.}\ \bibnamefont
  {Wolynes}},\ }\href {\doibase 10.1073/pnas.97.7.2990} {\bibfield  {journal}
  {\bibinfo  {journal} {Proceedings of the National Academy of Sciences}\
  }\textbf {\bibinfo {volume} {97}},\ \bibinfo {pages} {2990} (\bibinfo {year}
  {2000})},\ \Eprint
  {http://arxiv.org/abs/https://www.pnas.org/doi/pdf/10.1073/pnas.97.7.2990}
  {https://www.pnas.org/doi/pdf/10.1073/pnas.97.7.2990} \BibitemShut {NoStop}%
\bibitem [{\citenamefont {Biroli}\ and\ \citenamefont
  {Bouchaud}(2009)}]{biroli2009}%
  \BibitemOpen
  \bibfield  {author} {\bibinfo {author} {\bibfnamefont {G.}~\bibnamefont
  {Biroli}}\ and\ \bibinfo {author} {\bibfnamefont {J.~P.}\ \bibnamefont
  {Bouchaud}},\ }\href {https://arxiv.org/abs/0912.2542} {\enquote {\bibinfo
  {title} {The random first-order transition theory of glasses: a critical
  assessment},}\ } (\bibinfo {year} {2009}),\ \Eprint
  {http://arxiv.org/abs/0912.2542} {arXiv:0912.2542 [cond-mat.dis-nn]}
  \BibitemShut {NoStop}%
\bibitem [{\citenamefont {Tong}\ \emph {et~al.}(2022)\citenamefont {Tong},
  \citenamefont {Singh}, \citenamefont {Sknepnek},\ and\ \citenamefont
  {Košmrlj}}]{rheovm}%
  \BibitemOpen
  \bibfield  {author} {\bibinfo {author} {\bibfnamefont {S.}~\bibnamefont
  {Tong}}, \bibinfo {author} {\bibfnamefont {N.~K.}\ \bibnamefont {Singh}},
  \bibinfo {author} {\bibfnamefont {R.}~\bibnamefont {Sknepnek}}, \ and\
  \bibinfo {author} {\bibfnamefont {A.}~\bibnamefont {Košmrlj}},\ }\href
  {\doibase 10.1371/journal.pcbi.1010135} {\bibfield  {journal} {\bibinfo
  {journal} {PLOS Comput. Biol.}\ }\textbf {\bibinfo {volume} {18}},\ \bibinfo
  {pages} {1} (\bibinfo {year} {2022})}\BibitemShut {NoStop}%
\end{thebibliography}%

\end{document}